\documentclass[a4paper,11pt]{article}
\pdfoutput=1 
\usepackage{jcappub}

\usepackage{amsmath}
\usepackage{bm}
\usepackage{graphicx}
\usepackage{enumitem}
\usepackage{geometry}
\usepackage{xspace}
\usepackage{pdflscape}
\usepackage{placeins}
\usepackage{epstopdf}

\makeatletter
\gdef\@fpheader{}
\g@addto@macro\bfseries{\boldmath}
\makeatother

\newcommand{\ie}{{i.e.~}}

\newcommand{\eg}{e.g.~}

\newcommand{\OmegaPBH}{\Omega_{\mathrm{PBH}}}

\let\oldsqrt\sqrt
\def\sqrt{\mathpalette\DHLhksqrt}
\def\DHLhksqrt#1#2{%
\setbox0=\hbox{$#1\oldsqrt{#2\,}$}\dimen0=\ht0
\advance\dimen0-0.2\ht0
\setbox2=\hbox{\vrule height\ht0 depth -\dimen0}%
{\box0\lower0.4pt\box2}}

\DeclareMathOperator{\erfc}{erfc}

\newcommand{\dd}{\mathrm{d}}

\newcommand{\sss}[1]{{\scriptscriptstyle{#1}}}

\newcommand{\uinstab}{\mathrm{instab}}
\newcommand{\uPl}{\mathrm{Pl}}

\newcommand{\umin}{\mathrm{min}}
\newcommand{\umax}{\mathrm{max}}
\newcommand{\uend}{\mathrm{end}}

\newcommand{\uini}{\mathrm{ini}}

\newcommand{\uinf}{\mathrm{inf}}
\newcommand{\utot}{\mathrm{tot}}
\newcommand{\ureh}{\mathrm{reh}}
\newcommand{\urad}{\mathrm{rad}}

\newcommand{\uc}{\mathrm{c}}

\newcommand{\usssPl}{\sss{\uPl}}

\newcommand{\calP}{\mathcal{P}}

\newcommand{\MeV}{\mathrm{MeV}}
\newcommand{\GeV}{\mathrm{GeV}}
\newcommand{\TeV}{\mathrm{TeV}}

\newcommand{\Mp}{M_\usssPl}

\newcommand{\beq}{\begin{eqnarray}}
\newcommand{\eeq}{\end{eqnarray}}
\newcommand{\bea}{\begin{equation}\begin{aligned}}
\newcommand{\eea}{\end{aligned}\end{equation}}

\newlength{\wsingfig}
\newlength{\wdblefig}
\newlength{\wquadfig}
\newlength{\wtriplefig}
\newcommand{\Eq}[1]{Eq.~(\ref{#1})}
\newcommand{\Eqs}[1]{Eqs.~(\ref{#1})}
\newcommand{\Fig}[1]{Fig.~{\ref{#1}}}
\newcommand{\Figs}[1]{Figs.~{\ref{#1}}}
\newcommand{\Ref}[1]{Ref.~{\cite{#1}}}
\newcommand{\Refs}[1]{Refs.~{\cite{#1}}}
\newcommand{\Sec}[1]{Sec.~\ref{#1}}

\newcommand{\App}[1]{Appendix~\ref{#1}}
\newcommand{\Apps}[1]{Appendixes~\ref{#1}}

\subheader{}

\title{Primordial black holes from the preheating instability in single-field inflation}

\author[a]{J\'er\^ ome Martin,}
\author[b]{Theodoros Papanikolaou,}
\author[b,a]{Vincent Vennin}

\affiliation[a]{Institut d'Astrophysique de Paris, UMR 7095-CNRS,
  Universit\'e Pierre et Marie Curie, 98bis boulevard Arago, 75014
  Paris, France}
\affiliation[b]{Laboratoire Astroparticule et
  Cosmologie, Universit\'e Denis Diderot Paris 7, 75013 Paris, France}

\emailAdd{jmartin@iap.fr}
\emailAdd{theodoros.papanikolaou@apc.univ-paris7.fr}
\emailAdd{vincent.vennin@apc.univ-paris7.fr}

\date{today}

\begin{document}

\sloppy

\abstract{After the end of inflation, the inflaton field oscillates
  around a local minimum of its potential and decays into ordinary
  matter. 
  These oscillations trigger a resonant instability for
  cosmological perturbations with wavelengths that exit the Hubble radius close to the end of
  inflation. In this paper, we study the formation of Primordial Black
  Holes (PBHs) at these enhanced scales. We find that the production
  mechanism can be so efficient that PBHs subsequently dominate the content of the universe and
  reheating proceeds from their evaporation.
  Observational constraints on the PBH abundance also restrict the
  duration of the resonant instability phase, leading to tight limits on the reheating temperature
  that we derive. We conclude that the production of PBHs during reheating is a generic and inevitable
  property of the simplest inflationary single-field models, and does not require any fine tuning of the inflationary potential.
}



\maketitle

\section{Introduction}
\label{sec:intro}

The reheating
stage~\cite{Turner:1983he,Shtanov:1994ce,Kofman:1994rk,Kofman:1997yn}
is a crucial part of the inflationary
scenario~\cite{Starobinsky:1980te,Guth:1980zm,Linde:1981mu,Albrecht:1982wi,Linde:1983gd}. It
allows inflation to come to an end,
and describes how the inflaton field decays and
produces ordinary matter. Although reheating appears to be a rather complicated process, as far as the large scales probed by the Cosmic Microwave Background (CMB) anisotropies are concerned~\cite{Akrami:2018vks, Akrami:2018odb}, the influence of this epoch on the
predictions of inflation is simple, at least in single-field models. This is due to the fact that, on large scales,
the curvature perturbation is
conserved~\cite{Mukhanov:1981xt,Kodama:1985bj}, which implies that the
details of the reheating process do not affect the inflationary
predictions. In fact, those predictions are sensitive to a single
parameter, the so-called reheating
parameter~\cite{Martin:2006rs}, which is a combination
of the reheating temperature and of the mean equation-of-state parameter, and which determines the location of the
observational window along the inflationary potential. Given the restrictions on the shape
of the potential now available~\cite{Martin:2010hh,Martin:2013tda,Martin:2013nzq,Martin:2015dha,Vennin:2015eaa}, 
this can be used to constrain reheating~\cite{Martin:2010kz,Martin:2014nya,Martin:2016oyk,Hardwick:2016whe}.

On small scales however, the situation is different. It was indeed
shown in \Ref{Jedamzik:2010dq} (see also \Ref{Easther:2010mr}) that, for
scales leaving the Hubble radius during the last $\sim 10$
e-folds of inflation (if the energy scale of inflation
is not tuned to extremely low values), there is a parametric
instability that can lead to an enormous growth of
perturbations. This can cause early structure formation and/or
gravitational waves production~\cite{Jedamzik:2010dq,Easther:2010mr,Jedamzik:2010hq},
and may open a new observational window on inflation and reheating.

In the present paper, we study yet another possible consequence of the
presence of this instability, namely the production of Primordial
Black Holes (PBHs)~\cite{Carr:1974nx,Carr:1975qj}. The motivation is twofold. First, this may lead to a new inflationary
mechanism for black hole production which is completely natural and
generic. Usually, it is necessary to consider very specific potentials
in order for this production to be efficient. In this work, the only
assumption is that the potential can be approximated by a parabola
around its minimum. Except for fine-tuned situations (where, for
instance, a symmetry prevents the presence of a quadratic term in the
Taylor expansion of the potential about its minimum), this is
always the case. Second, tight constraints on the abundance of PBHs have been placed in various mass ranges (for a review, see \eg \Refs{Carr:2009jm, Carr:2017jsz}), and this can be used to obtain extra
information about the reheating epoch. Note that PBH formation during preheating has been studied in \Refs{Green:2000he, Bassett:2000ha, Suyama:2004mz, Cai:2018tuh}, although in a different context.

The paper is organised as follows. In the next section,
\Sec{sec:methods}, we briefly review \Ref{Jedamzik:2010dq} and the
physical mechanism that leads to the instability mentioned
above. Then, in \Sec{sec:pbhformation}, we study under which physical
conditions PBHs are formed. In \Sec{subsec:criterion}, based on
\Ref{Goncalves:2000nz}, we derive the critical
density contrast from the requirement that the
instability must last long enough, before reheating is completed, to
allow the initial scalar field overdensity to form a black hole. In
\Sec{subsec:criterionrefined}, the corresponding criterion is
refined by taking Hawking evaporation into account. We then calculate the mass fraction at the end of the
instability phase in
\Sec{subsec:beta}. Due to the high efficiency of the instability, we
find that the corresponding values for the fraction of the universe comprised in PBHs
can be
larger than one, which is not possible. The mass fraction must
therefore be renormalised, which is done in \Sec{subsec:renorma}. We
propose two ways to carry out this procedure, one which accounts for the possible inclusion of PBHs within larger ones (\Sec{subsubsec:inclusion}), and one which accounts for the premature termination of the instability phase by the backreaction of PBHs (\Sec{subsubsec:premature}). Having calculated the abundance of PBHs at
the end of the instability, in \Sec{subsec:evolving:beta}, we
proceed with calculating their abundance in the subsequent radiation-dominated epoch. In some cases, we find that PBHs are so abundant
that the radiation-dominated era is
delayed and we discuss under which conditions this occurs in
\Sec{subsec:Reh:PBH:evap}. In \Sec{subsec:Planckian:Relics}, we also
consider the case where black holes do not entirely evaporate but leave Planckian relics behind. In \Sec{sec:results}, we derive the observational consequences
of the above-described mechanism. In \Sec{subsec:onset}, we establish
restrictions on the energy density at the onset of the radiation dominated era (the reheating temperature). From current constraints on PBHs (\Sec{subsec:PBH:abundance}) and Planckian relics (\Sec{subsec:result:relics}) abundances, we then derive
constraints on the energy scale of inflation and the reheating temperature. In
\Sec{sec:conclusion}, we summarise our main results and present our
conclusions. Finally, the paper ends with two appendices. In
\App{sec:ltb}, we explain how a scalar field (here, the
inflaton field) can collapse and form a black hole and, in
Appendix~\ref{sec:deltacri}, we use these considerations to derive the
expression of the critical density contrast used in the rest of
the paper.
\section{Inflation and the preheating instability}
\label{sec:methods}
We consider scenarios where inflation is realised by a single scalar
field $\phi$ (the inflaton), which slowly rolls down its potential
$V(\phi)$ and then oscillates at
the bottom of it. In flat
Friedmann-Lema\^itre-Robertson-Walker space-times, the dynamics
of the homogeneous inflaton field is driven by the Klein-Gordon and
the Friedmann equations,
\begin{align}
\label{eq:KG&F}
& \ddot{\phi}+3H\dot{\phi}+V'\left(\phi\right) =0\, ,\quad
H^2 = \dfrac{V\left(\phi\right) + \dfrac{\dot{\phi}^2}{2}}{3\Mp^2}\, .
\end{align}
Hereafter, $H=\dot{a}/a$ is the Hubble
parameter, $a(t)$ is the scale factor, a
dot denotes derivative with respect to cosmic time, and $\Mp$ is the reduced Planck mass. These equations can be
solved numerically or with the help of the slow-roll approximation,
and the solution is insensitive to the choice of initial conditions
due to the presence of the slow-roll
attractor~\cite{Salopek:1990jq, Liddle:1994dx, Vennin:2014xta, Grain:2017dqa, Chowdhury:2019otk}. Inflation ends when the first
slow-roll parameter $\epsilon_1 \equiv -\dot{H}/H^2$ reaches one;
then, starts the reheating/preheating phase.

Close to its minimum, we assume the potential to be approximated by a
quadratic function,\footnote{As the amplitude of the oscillations get
  damped, the leading order in a Taylor expansion of the function
  $V(\phi)$ around its minimum quickly dominates, which is of
  quadratic order unless there is an exact cancellation at that
  order. The validity of this approximation is further discussed below.\label{footnote:appr:quad}}
\begin{align}
\label{eq:pot}
V\left(\phi\right) = \frac{m^2}{2}\phi^2.
\end{align}
When, after the end of inflation, the inflaton field explores this
part of the potential, $H\ll m$ and $\phi$ behaves as
\begin{align}
  \phi(t)\simeq\phi_0\left(\frac{a_0}{a}\right)^{3/2}
  \sin\left(mt\right) .
\end{align}
This implies that the energy density stored in $\phi$
redshifts on average as matter~\cite{Turner:1983he}, $\rho_\phi\propto a^{-3}$, and that the
oscillations have a frequency given by the mass $m$. Here, the
subscript ``$0$'' just denotes a reference time that might be taken at
the end of inflation. Let us stress that we only assume the
inflationary potential to be of the quadratic form towards the end of
inflation, see footnote~\ref{footnote:appr:quad}. No restriction on its shape is imposed
at the scales where the cosmological
perturbations observed in the CMB are produced, where the potential
can \eg be of the plateau type, and provide a good fit to
observations. This means that the parameter $m$ in \Eq{eq:pot} should
not be fixed to match the CMB power spectrum amplitude as usually
done, but should be left free in order to scan different values of
$H_\uend$, namely different energy scales at the end of inflation. In
practice, this can be done as follows. Inflation ends when\footnote{The value obtained for
  $\phi_\uend$ is independent of the mass parameter $m$, which can be
  seen with writing \Eqs{eq:KG&F} as a single equation for $\phi$ in
  terms of the number of e-folds $N=\ln a$,
  \begin{align}
    \frac{\dd^2\phi}{\dd
      N}+\left[3-\frac{1}{2\Mp^2}\left(\frac{\dd\phi}{\dd
        N}\right)^2\right]\left(\frac{\dd\phi}{\dd
      N}+\Mp^2\frac{V'}{V}\right)=0\, .
  \end{align}
  In this equation, the potential only appears through the combination
  $V'/V$, in which the mass parameter $m$ cancels out. Since the first
  slow-roll parameter can be written as $\epsilon_1 = (\dd\phi/\dd N)^2/(2\Mp^2)$, the value of $\phi$ at which it crosses one
  does not depend on $m$.} $\phi_\uend\simeq 1.0092 \Mp $. Given that
$\epsilon_1 = 3 \dot{\phi}^2/2 /(V+\dot{\phi}^2/2)$, at the end of
inflation, $\dot{\phi}^2 = V$, and one can relate $H_\uend$ to $m$
according to
\begin{align}
  m = 2 H_\uend \frac{\Mp}{\phi_\uend}\, .
\end{align}
In this way, by varying $m$ one can vary the value of the Hubble
parameter at the end of inflation, $H_\uend$. 

For the cosmological perturbations, there is a single gauge-invariant
scalar degree of freedom that can be described with the
Mukhanov-Sasaki variable~\cite{Mukhanov:1981xt,Kodama:1985bj} $v$,
which is a combination of the perturbed inflaton field and of the
Bardeen potential, the latter being a generalisation of the
gravitational Newtonian potential~\cite{Bardeen:1980kt}. Its Fourrier
component $v_{\bm k}$ evolves according to~\cite{Mukhanov:1990me}
\begin{align}
\label{eq:MS}
  v_{\bm k}''+\left(k^2-\frac{z''}{z}\right)v_{\bm k} = 0,
  \end{align} where a prime
denotes a derivative with respect to conformal time $\eta$, defined as
$\dd t = a \dd \eta$. In this expression, $z\equiv
\sqrt{2\epsilon_1}a\Mp$ and is such that
\begin{align}
\frac{z''}{z}= a^2 H^2 \left[ \left(1+\frac{\epsilon_2}{2}\right)
  \left(2-\epsilon_1+\frac{\epsilon_2}{2}\right)
  +\frac{\epsilon_2\epsilon_3}{2}\right]\, ,
\end{align}
where $\epsilon_2 \equiv \dd\ln\epsilon_1/\dd N$ and $\epsilon_3
\equiv \dd\ln\epsilon_2/\dd N$ are the second and the third slow-roll
parameters respectively. The initial condition is taken in the
Bunch-Davies vacuum, \ie such that $v_{\bm k}\rightarrow e^{-i k
  \eta}/\sqrt{2k}$ when $k\gg aH$, and the function $z''/z$ is
evaluated on the background dynamics that has been numerically
integrated as explained above. In this way, one can compute the
amplitude of $v_{\bm k}$ at the end of inflation for each mode $k$.

\begin{figure}[t]
\begin{center}
\includegraphics[width=0.6\textwidth]{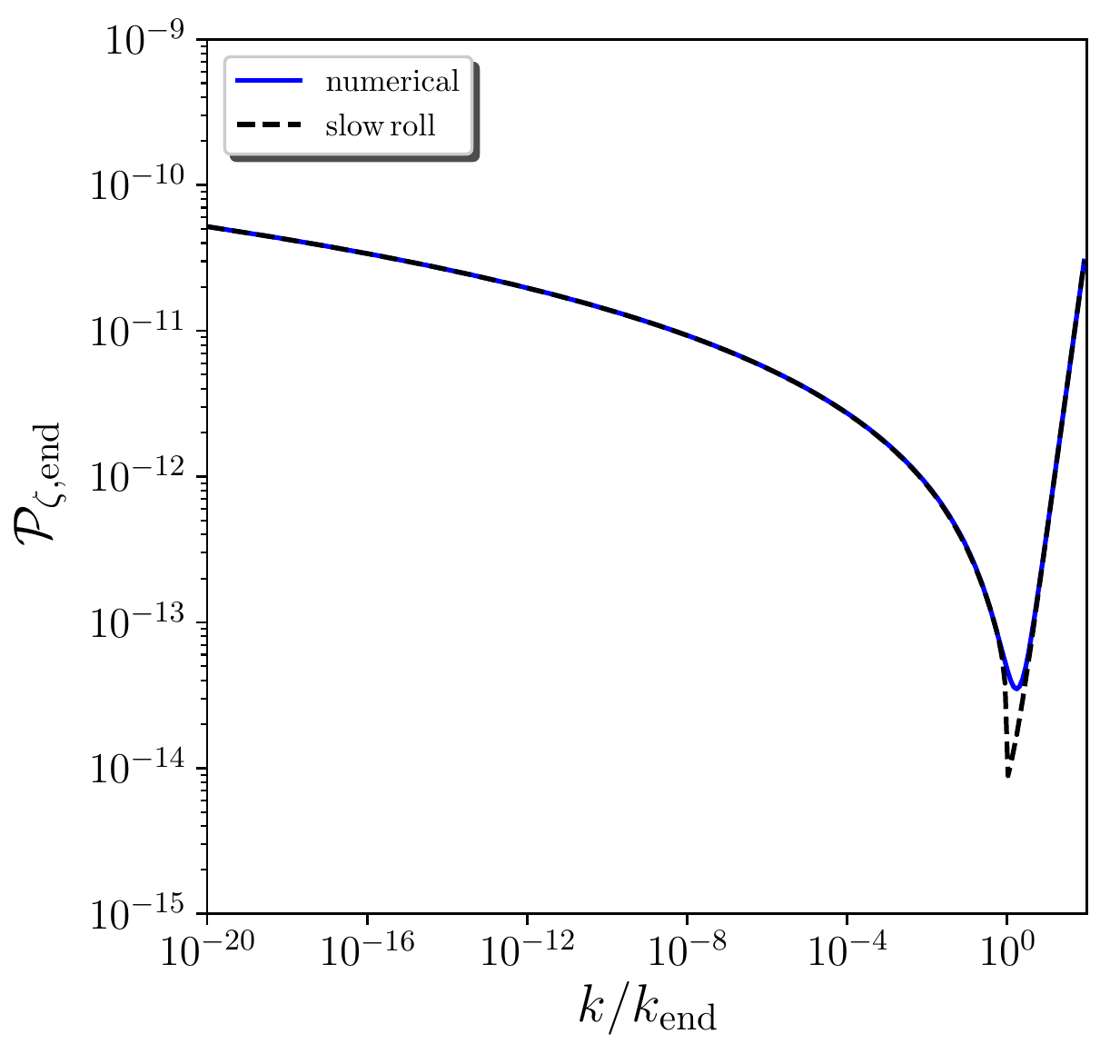}
\caption{Power spectrum of the curvature perturbation at the end of
  inflation for $m\simeq 1.14\times10^{-6}\, \Mp$, corresponding to
  $\rho_\uinf \equiv 3H_\uend^2\Mp^2= 10^{-12}\Mp^4\simeq \left(2.43
  \times 10^{15}\GeV\right)^4$, as a function of $k/k_\uend$, where
  $k_\uend$ is the scale that exits the Hubble radius at the end of
  inflation. The blue solid line corresponds to the numerical solution
  of \Eq{eq:MS} while the black dashed line stands for the slow-roll
  approximation~(\ref{eq:Pzeta:end:SR}).}
\label{fig:Pzeta:end}
\end{center}
\end{figure}

\begin{figure}[t]
\begin{center}
\includegraphics[width=0.8\textwidth]{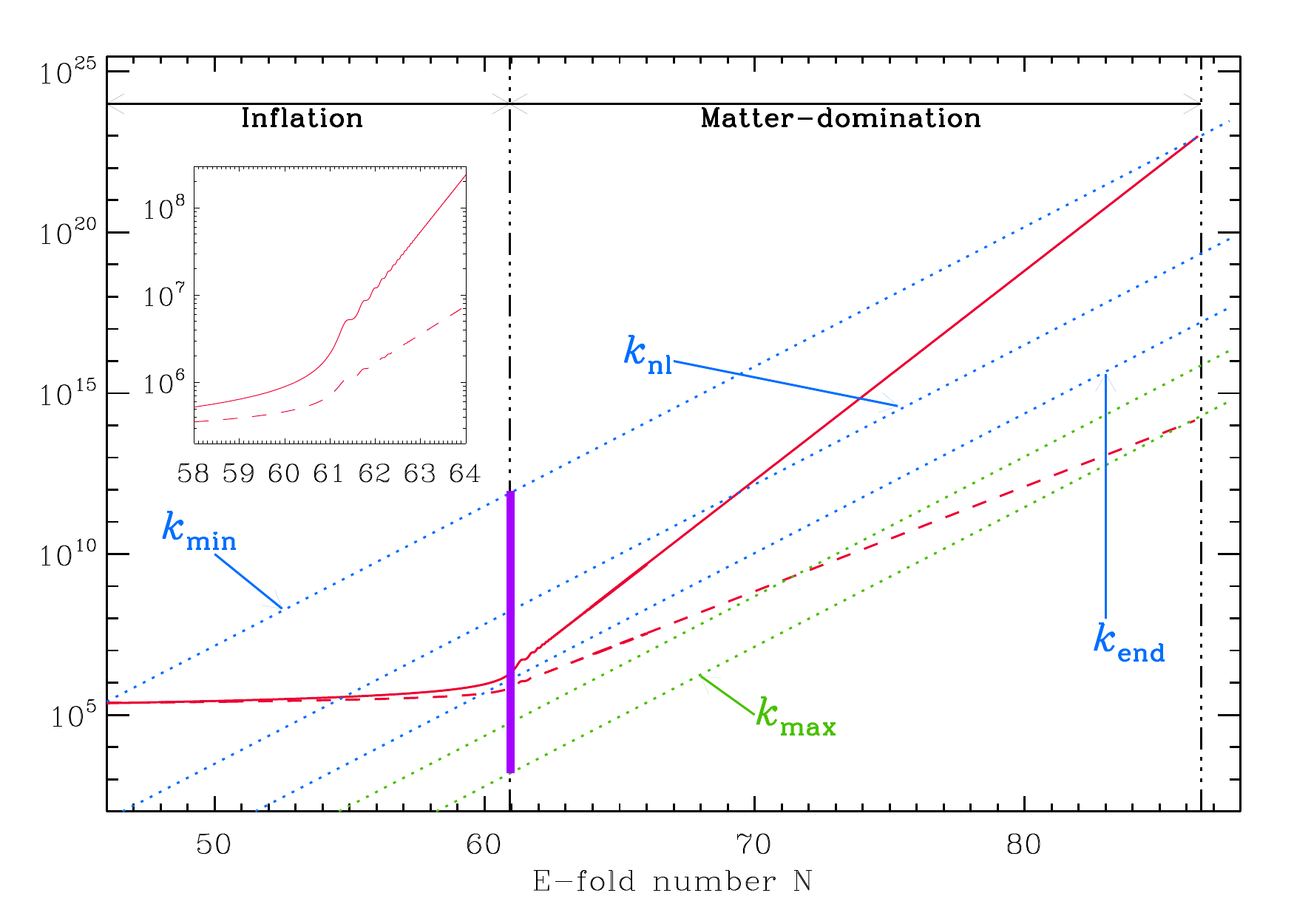}
\caption{Evolution of the relevant physical scales versus the e-folds
  number. The continuous red line denotes the Hubble radius, which is
  also the upper bound of the instability band, while the dashed red
  line represents the scale $\sqrt{3Hm}$ which corresponds to the
  lower bound of the resonance band. The dotted lines represent the
  physical wavelengths of different Fourier modes: the ``green modes''
  enter the instability mode from below while the ``blue modes'' enter
  it from above. The inset shows the detailed behaviours of the Hubble
  radius and $\sqrt{3Hm}$ at the transition between inflation and
  reheating. Figure taken from \Ref{Jedamzik:2010dq}.}
\label{fig:scaleinf}
\end{center}
\end{figure}

It is convenient to introduce the curvature perturbation $\zeta$
defined as $\zeta=v/z$ since this quantity is conserved on
super-Hubble scales, and to compute the power spectrum $\calP_\zeta =
k^3 \vert \zeta_{\bm k} \vert^2/(2\pi^2)$ of that quantity at the end
of inflation. It is displayed in \Fig{fig:Pzeta:end} for the value of
$m$ corresponding to $\rho_\uinf\equiv 3 H_\uend^2\Mp^2 =
10^{-12}\Mp^4\simeq \left(2.43 \times 10^{15}\GeV\right)^4$, as a
function of $k/k_\uend$, where $k_\uend = a_\uend H_\uend$ is the
scale that exits the Hubble radius at the end of inflation, see also
\Fig{fig:scaleinf}. The blue solid line corresponds to the numerical
solution of \Eq{eq:MS}, while the black dashed line stands for the
slow-roll approximated solution of that equation,
namely~\cite{Schwarz:2001vv, Gong:2001he} \bea
\label{eq:Pzeta:end:SR}
\calP_{\zeta,\uend} = \begin{cases}
  \dfrac{H_*^2\left(k\right)}{8\pi^2\Mp^2\epsilon_{1*}
    \left(k\right)}\left[1+\left(\dfrac{k}{k_\uend}\right)^2\right]
  \left[1-2\left(C+1\right)\epsilon_{1*}\left(k\right)
    -C\epsilon_{2*}\left(k\right)\right]\ \text{if}\ k<k_\uend\\
  \dfrac{H_\uend^2}{8\pi^2\Mp^2}\left[1
    +\left(\dfrac{k}{k_\uend}\right)^2\right]\ \text{if}\ k>k_\uend
\end{cases} .
\eea In this expression, for the modes that cross out the Hubble
radius before the end of inflation, $k<k_\uend$, the functions
$H_*(k)$, $\epsilon_{1*}(k)$ and $\epsilon_{2*}(k)$ respectively
denote the values of $H$, $\epsilon_1$ and $\epsilon_2$ at the time
when the mode $k$ exits the Hubble radius, and are evaluated in the
numerical solution of \Eqs{eq:KG&F}. The parameter
$C\simeq-0.7296$ is a numerical constant. One can check in
\Fig{fig:Pzeta:end} that, except for the very few modes that are close
to the Hubble scale at the end of inflation and for which the amount
of power is underestimated, \Eq{eq:Pzeta:end:SR} provides a very good
fit to the numerical solution.

As already mentioned, after the end of inflation, the inflaton
oscillates at the bottom of its quadratic potential and the evolution
of the perturbations through this epoch strongly depends on the scales
considered. On large scales (for instance, CMB scales), the
conservation of curvature perturbation is sufficient to establish that
the power spectrum~(\ref{eq:Pzeta:end:SR}) calculated at the end of
inflation propagates through the reheating epoch without being
distorted. However, on small scales, things can be very
different. As shown in \Ref{Jedamzik:2010dq}, for modes satisfying
\begin{align}
\label{eq:instability:band}
a H < k < a \sqrt{3 H m},
\end{align}
see \Fig{fig:scaleinf},
the oscillations source a parametric resonance (in the narrow
resonance regime). The reason is that, thanks to these oscillations,
\Eq{eq:MS} becomes a Mathieu equation and the
condition~(\ref{eq:instability:band}) is in fact equivalent to being
in the first instability band of that equation. We see that the
instability occurs if the physical wavelength of a mode is smaller
than the Hubble radius (continuous red line in \Fig{fig:scaleinf})
during reheating and larger than a new scale given by $\sqrt{3Hm}$
(dashed red line in \Fig{fig:scaleinf}). Moreover, two types of mode
can be distinguished. The ``blue modes'' in \Fig{fig:scaleinf} exit
the Hubble radius during inflation and re-enter it during reheating;
these modes therefore enter the instability band from above. On the
other hand, the ``green modes'' never exit the Hubble radius and enter
the instability band from below by crossing the new scale
$\sqrt{3Hm}$. Once within the instability band, as described in
\Ref{Jedamzik:2010dq}, the fluctuations get strongly amplified, such
that the density contrast grows linearly with the scale factor. Effectively, they thus behave as pressureless matter
perturbations in a pressureless matter universe. In what follows, this
epoch is referred to as the ``instability phase''. As explained in
\Sec{sec:intro}, during this epoch, cosmological perturbations at the
amplified scales may collapse into PBHs. When the
inflaton decays into other degrees of freedom (or when the PBHs take the inflaton over, see below), the instability stops,
and the density of black holes evolves under various physical effects
(cosmic expansion, Hawking evaporation, accretion, merging,
{\it{etc.}}).

Let us further discuss the quadratic approximation for the inflationary potential. The largest scales amenable to parametric resonance during the
instability phase are such that $k=a_{\uinstab} H_{\uinstab}$, where
the time $t_{\uinstab}$ denotes the end of the instability phase (the
corresponding Fourier mode is denoted ``$k_\mathrm{min}$'' in
\Fig{fig:scaleinf}). During inflation, they cross out the Hubble
radius at a number of e-folds $\sim \ln(H_\uend/H_\uinstab)/3$ before
the end of inflation, where we recall that $H_\uend$ is the value of
the Hubble parameter at the end of inflation and where we have used
that, during the instability, the universe is matter dominated at the
background level. Since observational bounds on the tensor-to-scalar
ratio impose~\cite{Akrami:2018odb} $H_\uend<8\times 10^{13}\, \GeV$,
and given that $H_\uinstab>H_{\mathrm{BBN}}\sim (10\,
\MeV)^2/\sqrt{3\Mp^2} \sim 10^{-23}\, \GeV$, where hereafter ``BBN'' stands for big-bang nucleosynthesis, this number of e-folds
needs to be smaller than $\sim 28$.\footnote{Strictly speaking the
  tensor-to-scalar ratio $r$ is related to $H_*$, the energy scale of
  inflation at the time the CMB modes left the Hubble radius during
  inflation, which is a different quantity that $H_\uend$, the energy
  scale at the end of inflation. Here, we neglect the difference
  between those two quantities. This approximation is especially
  accurate for plateau models, namely for the models favoured by the
  most recent astrophysical data.} All the scales of interest for the
problem at hand are therefore generated in the last $28$ e-folds of
inflation, where we assume the potential to be well approximated by
the quadratic form~(\ref{eq:pot}). Although one may be suspicious that this approximation holds for $28$
e-folds, let us stress that this value is in fact an
extreme upper bound that comes from saturating the condition
$H_\uinstab> H_{\mathrm{BBN}}$, while we will see below that most of
the relevant parameter space is such that $H_\uinstab$ and
$H_\mathrm{BBN}$ are separated by many orders of magnitude and this
number of e-folds is in fact much smaller. In practice, potentials favoured by the data (such as plateau ones) tend to be shallower than the quadratic one away from the end of inflation, and we have explicitly checked that this approximation only slightly underestimates the amplitude of scalar perturbations in such potentials, leading to conservative statements regarding the amount of PBHs.\footnote{Hereafter, ``conservative'' refers to the fact that the approximations performed in this work tend to underestimate the amount of PBHs, such that our results can be viewed as lower bounds on their abundance, and the regions of parameter space that are excluded because they produce
too many PBHs might extend beyond what is obtained below. \label{footnote:conservative}} It is nonetheless clear
that the calculational program laid out below can easily be performed
for any given potential, such that the approximation~(\ref{eq:pot})
for the last e-folds of inflation is released. In this work, it however
allows us to carry out a full parameter-space analysis, where the
energy scale of inflation can be varied without relying on a specific
potential. We will see that this provides an overall picture where several interesting regions are identified, in which a more
detailed analysis can always be carried out.

\section{PBH formation during reheating}
\label{sec:pbhformation}

We have just seen that the modes in the resonance
band~(\ref{eq:instability:band}) behave as pressureless matter
fluctuations in a pressureless matter universe. In
Ref.~\cite{Goncalves:2000nz} and in the two appendices, see
\Eq{eq:collapsetime}, it is shown that they collapse into PBHs after a time~\cite{Goncalves:2000nz}\footnote{Here, we
  correct an error of a factor $2$ in Eq.~(84) of
  \Ref{Goncalves:2000nz}.}
\begin{align}
\label{eq:t:coll}
\Delta t_{\mathrm{collapse}}
= \frac{\pi}{H\left[t_{\mathrm{bc}}(k)\right]
  \delta_{\bm k}^{3/2}\left[t_{\mathrm{bc}}(k)\right]},
\end{align}
where $t_{\mathrm{bc}}(k)$ denotes the ``band-crossing'' time, \ie the time at which the mode $k$
crosses in the instability band~(\ref{eq:instability:band}).

Let us note that, in a matter-dominated universe, $aH$ decreases as
$a^{-1/2}$ while $a\sqrt{H}$ increases as $a^{1/4}$, so the bounds
defining the instability band~(\ref{eq:instability:band}) are such
that, when a mode crosses in the band, it remains in the band (in
other words, modes cannot cross out the band).

This instability stops when the coherent oscillations of the inflaton
are over. This can happen \eg when the inflaton decays into other
fields. In the case of perturbative preheating, this occurs when the
Hubble parameter drops below the decay rate $\Gamma$ of the inflaton,
and for this reason, hereafter this time is referred to as
$t_\Gamma$. One should however note that the results derived below
are independent of the precise way in which the phase of coherent
oscillations stop, since the time at which this happens (regardless of
the way it happens) is simply one of the parameters in the present
scenario.\footnote{As one approaches the point where $H\sim\Gamma$,
  the averaged background equation-of-state parameter becomes progressively finite and
  this could lead to shutting off the instability before the time of
  perturbative decay~\cite{Carr:2018nkm}. In this case
  $H_\Gamma>\Gamma$, but again, $H_\Gamma$ is simply used as a
  parameter to describe the time at which the instability stops, and
  ``$\Gamma$'' is no more than a convenient notation.}

Let us also stress that, for later convenience, we have introduced the two notations
$t_\uinstab$ and $t_\Gamma$. As mentioned above, $t_\uinstab$ denotes
the end of the instability while $t_\Gamma$ denotes the time at which
the field decays. Although they are identical in the standard picture, we will see below that there are cases where they differ (for instance if PBHs come to dominate the universe content before the inflaton decays), which explains the need for two distinct notations.
\subsection{Formation criterion}
\label{subsec:criterion}
Let us now determine under which conditions PBHs form. The last mode to enter the band~(\ref{eq:instability:band})
``from above'' is such that $k=a_\Gamma H_\Gamma$, which leads to
$k/k_\uend=(\rho_\Gamma/\rho_\uinf)^{1/6}$. The last mode that enters
the band ``from below'' is, on the other hand, such that
$k=a_\Gamma\sqrt{3 H_\Gamma m}$. In this paper, however, we restrict
ourselves to modes that enter the instability band from above. Indeed,
as already noticed, the modes that enter the band from below have
never crossed out the Hubble radius and their status is unclear: in practice, one should derive the full real-space profile of the over-densities produced by the instability band~\cite{Musco:2018rwt, Muia:2019coe}, which is beyond the scope of the present work.
We therefore restrict our
analysis to a subset of the instability
band~(\ref{eq:instability:band}) only, namely to modes such that
\begin{align}
\label{eq:instability:new}
\left(\frac{\rho_\Gamma}{\rho_\uinf}\right)^{1/6}<\frac{k}{k_\uend}<1\, .
\end{align}
Obviously, the incorporation of the modes that enter the instability
band ``from below'' could lead to further PBHs
production, and the results presented below are therefore conservative
in the sense of footnote~\ref{footnote:conservative}.

Let us now determine under which condition the time spent in the
instability band~(\ref{eq:instability:band}) is enough for PBHs to form. Since the background energy density decays as
pressureless matter during the instability, one has
\begin{align}
\label{eq:treh}
t_{\Gamma} - t_{\mathrm{bc}} = \frac{2}{3H_{\mathrm{bc}}}
\left[\left(\frac{a_\Gamma}{a_{\mathrm{bc}}}\right)^{3/2}-1\right]\, .
\end{align}
Requiring that this is larger than the time~(\ref{eq:t:coll}) required
for PBHs to form, one obtains the following condition,
\begin{align}
\label{eq:cond:delta}
\left(\frac{3\pi}{2}\right)^{2/3}
\left[\left(\frac{k}{k_\uend}\right)^3 \sqrt{\frac{\rho_\uinf}{\rho_\Gamma}}
  -1\right]^{-2/3}<\delta_{\bm k}[t_{\mathrm{bc}}(k)]<1\, ,
\end{align}
where the upper bound comes from the requirement that PBHs form in the
perturbative regime (the enforcement of this condition is again conservative with regards to the PBH abundance).
\subsection{Refined formation criterion: Hawking evaporation}
\label{subsec:criterionrefined}
The mass $M$ of the PBH
associated to the scale $k$ is given by some fraction $\xi$ of the
mass contained within a Hubble radius at the time $t_{\mathrm{bc}}$ when $k$ re-enters
the Hubble radius. Making use of the fact that the background energy
density decays as pressureless matter during the instability, one
obtains
\begin{align}
\label{eq:M(k)}
M(k) =
\xi \frac{\left(3\Mp^2\right)^{3/2}}{ \sqrt{\rho_\uinf}}
\left(\frac{k}{k_\uend}\right)^{-3}\, .
\end{align}
These masses are typically very small and
can be such that they disappear by Hawking evaporation before the end
of the instability. Since the evaporated black holes should be removed
from the mass fraction, let us determine under which conditions this
happens. The time of evaporation of a black hole with mass $M$ is
given by~\cite{Hawking:1974rv}
\begin{align}
\label{eq:Hawking:time}
\Delta t_{\mathrm{evap}}(M) = \frac{10240}{g}\frac{M^3}{\Mp^4}\, ,
\end{align}
where $g$ is the effective number of degrees of freedom. For the black
hole to survive until the end of the instability, one should therefore
check that $\Delta t_{\mathrm{evap}} > t_\Gamma -
t_{\mathrm{collapse}} = t_\Gamma - t_{\mathrm{bc}} -
(t_{\mathrm{collapse}} - t_{\mathrm{bc}})$, where $t_\Gamma -
t_{\mathrm{bc}}$ is given in \Eq{eq:treh} and $t_{\mathrm{collapse}} -
t_{\mathrm{bc}}$ is given in \Eq{eq:t:coll}. This imposes the
condition
\begin{align}
\label{eq:delta:max:Hawking}
\delta_{\bm k}[t_{\mathrm{bc}}(k)]<\left[\frac{2}{3\pi}
  \left(\frac{k}{k_\uend}\right)^3\sqrt{\frac{\rho_\uinf}{\rho_\Gamma}}
  -\frac{2}{3\pi}-\frac{10240}{g}\frac{\xi^3}{\pi}
  \frac{\left(3\Mp\right)^4}{\rho_\uinf}
  \left(\frac{k}{k_\uend}\right)^{-6}\right]^{-2/3}\, .
\end{align}
When the quantity inside the square brackets is negative, Hawking
evaporation cannot proceed before the end of the instability phase and
this does not need to be taken into account. Otherwise, the value 
for $\delta_\umax(k)$ now needs to be taken as the minimum value between the right-hand side of \Eq{eq:cond:delta}
and the right-hand side of \Eq{eq:delta:max:Hawking}. Let us
note that, comparing \Eqs{eq:cond:delta}
and~(\ref{eq:delta:max:Hawking}), one always has
$\delta_\umax(k)>\delta_\uc(k)$, unless $\delta_{\mathrm{c}}>1$, in
which case we simply take the mass fraction to vanish.

\subsection{Mass fraction}
\label{subsec:beta}

Assuming Gaussian statistics $P$ for the density contrast perturbation
at the band-crossing time, with a variance given
by the power spectrum $\calP_\delta$, the mass fraction of
PBHs can be expressed as~\cite{Harada:2013epa}
\begin{align}
\label{eq:beta:erf}
\beta\left(M,t_\Gamma\right) \equiv
\frac{\dd \Omega_{\mathrm{PBH}}(k,t_\Gamma)}{\dd \ln M}
= 2 \int_{\delta_\uc(k)}^{\delta_{\umax}(k)} P(\delta) \dd\delta 
=\erfc\left[\frac{\delta_\uc(k)}{\sqrt{2 \calP_\delta (k)}}\right]
-\erfc\left[\frac{\delta_{\umax}\left(k\right)}
  {\sqrt{2 \calP_\delta (k)}}\right],
\end{align}
where $\erfc$ is the complementary error function and we have followed
the usual Press-Schechter practice of multiplying by a factor $2$. In
this expression, we recall that $M$ and $k$ are related through \Eq{eq:M(k)}, that the minimum value of the density
contrast, $\delta_\uc(k)$, is given by the left-hand side of
\Eq{eq:cond:delta}, and that the maximum value, $\delta_\umax(k)$, is given
by the considerations presented in \Sec{subsec:criterionrefined}. On
the other hand $\calP_\delta (k)$ [where the argument $
  t_\mathrm{bc}(k)$ has been dropped for notational convenience] can
be obtained from the following considerations. Since the modes
belonging to \Eq{eq:instability:new} are super Hubble between the end
of inflation and the time at which they enter the instability
band~(\ref{eq:instability:band}) from above, the curvature
perturbation $\zeta_{\bm k}$ is conserved, hence $\zeta_{\bm
  k}\left[t_{\mathrm{bc}}(k)\right] = \zeta_{{\bm k},\uend}$. As
explained in \Ref{Jedamzik:2010dq}, for the modes inside the
instability band, one has
\begin{align}
  \delta_{\bm k} =-\frac{2}{5}\left(3+\frac{k^2}{a^2
    H^2}\right)\zeta_{\bm k}\, ,
\end{align}
which allows us to relate the power spectrum of the density contrast
at the band-crossing time to the one of the
curvature perturbation at the end of inflation,
\begin{align}
\label{eq:calP:delta:bc}
\calP_{\delta}\left[{ k}, t_\mathrm{bc}(k)\right]
= \left(\frac{6}{5}\right)^2 \calP_{\zeta,\uend}(k)\, .
\end{align}
\begin{figure}[t]
\begin{center}
\includegraphics[width=0.6\textwidth]{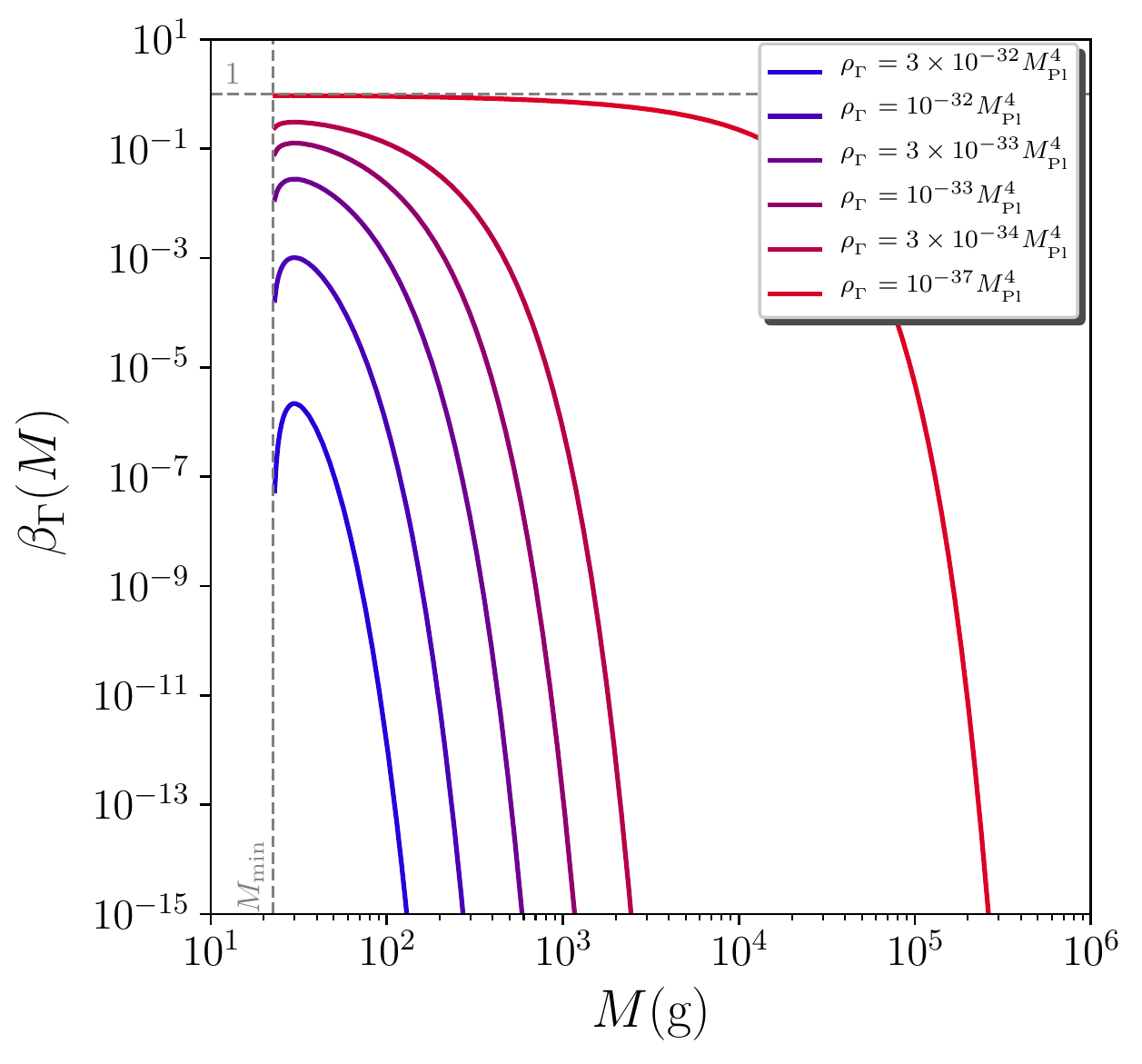}
\caption{Mass fraction of PBHs at the end of the
  instability phase, as a function of the mass at which they form. The
  energy density at the end of inflation is set to $\rho_{\uinf} =
  10^{-12}\Mp^4\simeq (2.43 \times 10^{15}\GeV)^4$, and the result is
  displayed for a few values of $\rho_\Gamma$, namely
  $\rho_\Gamma=3\times 10^{-32}\Mp^4\simeq (3.2\times 10^{10}\GeV)^4$,
  $\rho_\Gamma=10^{-32}\Mp^4\simeq (2.4\times 10^{10}\GeV)^4$,
  $\rho_\Gamma=3\times 10^{-33}\Mp^4\simeq (1.8\times 10^{10}\GeV)^4$,
  $\rho_\Gamma=10^{-33}\Mp^4\simeq (1.4\times 10^{10}\GeV)^4$,
  $\rho_\Gamma=3\times 10^{-34}\Mp^4\simeq (10^{10}\GeV)^4$ and
  $\rho_\Gamma= 10^{-37}\Mp^4\simeq (1.4\times 10^{9}\GeV)^4$. The
  vertical grey dashed line stands for the minimum mass corresponding
  to the scale that matches the Hubble radius at the end of inflation,
  while the horizontal grey dashed line corresponds to $\beta=1$,
  which is the maximum possible value attained in the limit
  $\delta_\uc \ll \sqrt{\calP_\delta}$.}
\label{fig:beta:reh:few:rho:reh}
\end{center}
\end{figure}

The mass fraction at the end of the instability phase can be computed
using the above relations, and is displayed as a function of the mass
in \Fig{fig:beta:reh:few:rho:reh}, for $\rho_\uinf = 10^{-12}
\Mp^4\simeq \left(2.43 \times 10^{15}\GeV\right)^4$ and a few values
of $\rho_\Gamma$. We also take $10240/g=100$ and $\xi =1$. The
vertical grey dashed line stands for the minimum mass $M_\umin$,
corresponding to the scale that matches the Hubble radius at the end
of inflation, and which can be obtained by setting $k/k_\uend = 1$ in
\Eq{eq:M(k)}. For the value of $\rho_\uinf$ used in the figure, one
has $M_\umin\simeq 22.5 \, \mbox{g}\simeq 1.1\times
10^{-32}M_\odot$, where $M_\odot$ denotes the mass of the sun. Since the result depends only on $\rho_\uinf$, and
given that the same value of $\rho_\uinf$ is used for all curves, this
explains why the same value for the minimum mass is found. One can also check that, the lower $\rho_\Gamma$ is, the longer the instability phase is, hence the more amplified the fluctuations are and the more black holes are produced.

The dependence of $\beta(M,t_\Gamma)$ in terms of the mass $M$ can also be understood in simple terms. 
The dominant trend is that the
mass fraction mostly decreases with the value of the mass. This is
because, the larger the mass, the smaller the wavenumber $k$ [see
  \Eq{eq:M(k)}], hence the later the mode enters the instability band,
so the less amplified the perturbation and the larger $\delta_\uc$ [see
  \Eq{eq:cond:delta}]. More precisely, for $\delta_\umax=1$, from \Eq{eq:beta:erf}, $\beta$ decreases with $\delta_\uc/\sqrt{2\calP_\delta}$. Since $\delta_\uc \propto k^{-2}$, see \Eq{eq:cond:delta}, $\beta$ decreases with $M$ (hence increases with $k$) if $\dd\ln\calP_\zeta/\dd\ln k>-4$, \ie if the spectral index is larger than $-3$. This is of course the case away from the end of inflation, where the power spectrum is close to scale invariance, but might not be true for modes that exit the Hubble radius close to the end of inflation, \ie for values of $M$ close to $M_\umin$. In fact, one can check that the spectral index corresponding to the ``numerical'' power spectrum in \Fig{fig:Pzeta:end} (blue curve) is always larger than $-3$, and the reason why $\beta$ increases with $M$ at small masses in some of the curves displayed in  \Fig{fig:beta:reh:few:rho:reh} is because we make use of the slow-roll approximation~\eqref{eq:Pzeta:end:SR} corresponding to the black dashed curve in \Fig{fig:Pzeta:end}, for which the spectral index drops below $-3$ at the very end of inflation. However, as stressed above, although this approximation is necessary to limit the numerical cost of the parameter space exploration performed below, it only affects a tiny range of modes that exit the Hubble radius at the very end of inflation, and is conservative in the sense of footnote~\ref{footnote:conservative}.
\subsection{Renormalising the mass fraction at the end of the instability}
\label{subsec:renorma}
\begin{figure}[t]
\begin{center}
\includegraphics[width=0.6\textwidth]{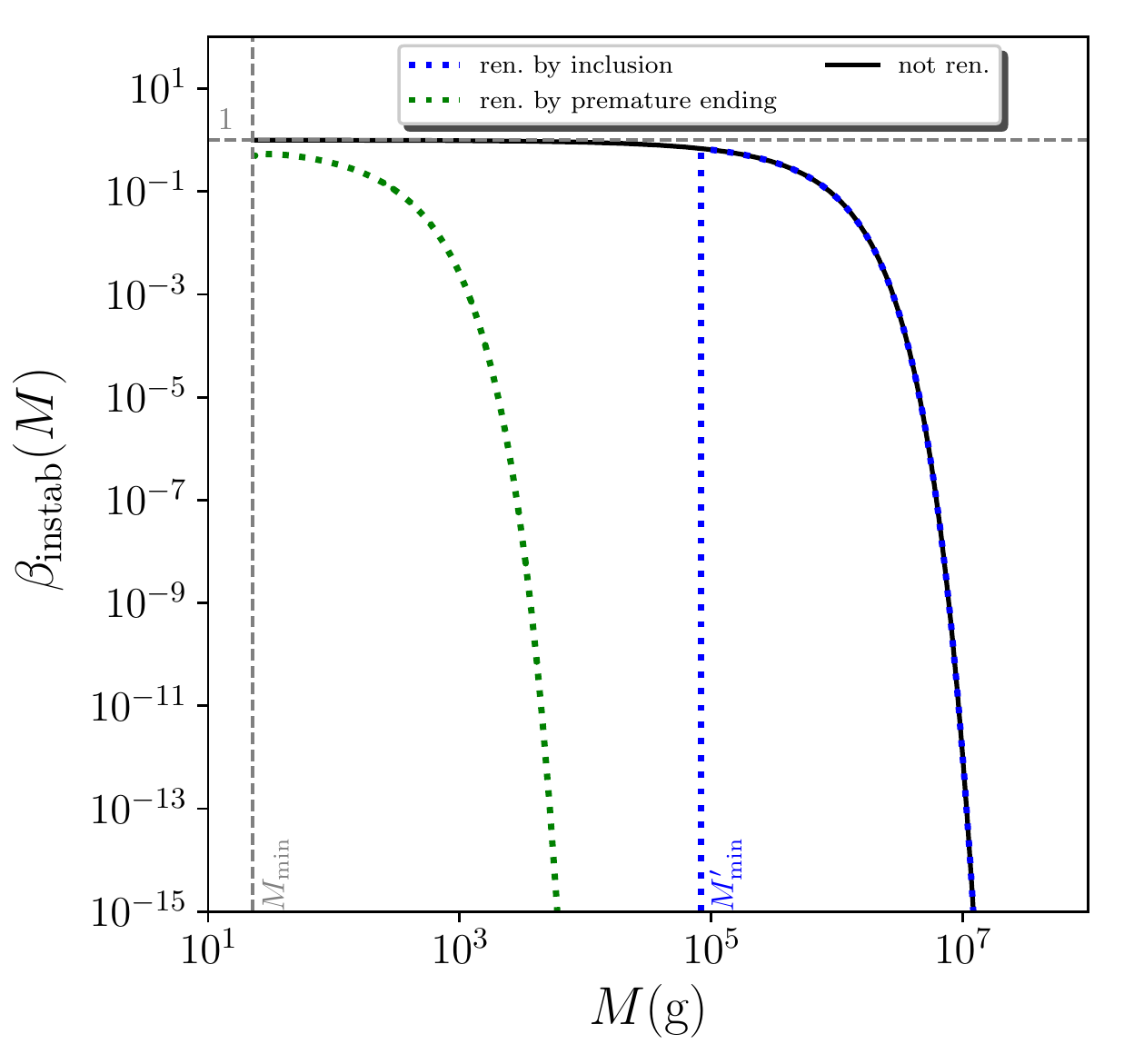}
\caption{Mass fraction of PBHs at the end of the
  instability phase, as a function of the mass at which the black
  holes form, for $\rho_{\uinf} = 10^{-12}\Mp^4\simeq \left(2.43
  \times 10^{15}\GeV\right)^4$ and $\rho_\Gamma = 10^{-40}\Mp^4\simeq
  \left(2.43 \times 10^{8}\GeV\right)^4$. The black line corresponds
  to the result obtained before renormalisation and
  leads to $\Omega_{\mathrm{PBH}}(t_\Gamma) = 8.54>1$, which is not
  physical. The blue dotted line is obtained after renormalisation by
  inclusion, \ie when increasing $M_\umin$ to $M_\umin'$ such that the
  integrated mass fraction $\Omega_{\mathrm{PBH}}(t_\Gamma)$ becomes
  one. This accounts for the absorption of small-black holes into
  larger-mass black holes when the regions that collapse into these
  large-mass black holes already contain smaller ones. The green
  dotted line stands for renormalisation by premature ending, \ie
  by stopping the instability phase before $t_\Gamma$, at the time
  when $\Omega_{\mathrm{PBH}}$ reaches one. This accounts for the fact
  that if the universe becomes dominated by black holes, the
  parametric resonance effect stops.}
\label{fig:beta:reh:renormalisation:procedures}
\end{center}
\end{figure}
The fraction of the energy density of the universe contained within
PBHs at the end of the instability phase is, by
definition, given by
\begin{align}
\label{eq:Omega:PBH:def}
\Omega_{\mathrm{PBH}}\left(t_\Gamma\right)
= \int_{M_\umin}^{M_\umax} \beta\left(M,t_\Gamma\right) \dd \ln M\, .
\end{align}
One can compute its value for the parameters displayed in
\Fig{fig:beta:reh:few:rho:reh} and one finds
$\Omega_{\mathrm{PBH}}(t_\Gamma)\simeq 8.68\times 10^{-7},\ 5.63\times
10^{-4},\ 2.13\times 10^{-2},\ 0.129,\ 0.412,\ 4.58 $ for $\rho_\Gamma
= 3\times 10^{-32}\Mp^3,\cdots,10^{-37}\Mp^4$, respectively. The fact
that $\Omega_{\mathrm{PBH}}$ decreases with $\rho_\Gamma$ is
consistent with what precedes, but the reader should be struck by the
last value, which is above one. This is of course not possible given
that we assume the spatial curvature to vanish, and entails that when
$\rho_\Gamma$ decreases, the production of PBHs is
so efficient that they overtake the energy density stored in the
inflaton field. When this happens, the above approach breaks
down. Below, we propose two procedures to model what may physically
prevent $\Omega_{\mathrm{PBH}}$ to grow larger than one.
\subsubsection{Renormalisation by inclusion}
\label{subsubsec:inclusion}
When $\Omega_{\mathrm{PBH}}$ increases and reaches sizeable values,
PBHs are densely distributed in the universe, and
when a fluctuation at a given scale gets amplified above the
threshold, the region of space that collapses and forms a black hole
may already contain smaller black holes. If this happens, when black
holes with larger masses form, black holes with smaller masses may
be absorbed and disappear from the mass fraction, and we dub this effect ``inclusion''.  In
\Ref{MoradinezhadDizgah:2019wjf}, this is also called the
``could-in-cloud'' phenomenon.

In that case, we proceed as follows: if
$\Omega_{\mathrm{PBH}}(t_\Gamma)$ is found to be larger than one, we
increase the value of $M_\umin$ in \Eq{eq:Omega:PBH:def},
\begin{align}
\label{eq:ren:inclusion}
M_\umin \rightarrow M_\umin'\, ,
\end{align}
in such a way that $\Omega_{\mathrm{PBH}}(t_\Gamma)$ becomes one. We
therefore remove the small mass tail of the distribution that is
responsible for having $\Omega_{\mathrm{PBH}}>1$, accounting for their
absorption into larger-mass black holes.

One should note that this inclusion effect might, in practice, prevent
$\Omega_{\mathrm{PBH}}$ to grow larger than some intermediate value
that is smaller than one, but this would have only very little impact
on the results derived below as long as that value is of order
one (which is expected for the inclusion phenomenon to be significant~\cite{MoradinezhadDizgah:2019wjf}). Another possibility is that small-black holes are indeed removed
from the distribution, but that the decrease in $\beta$ at small $M$
is smoother than a sharp cutoff imposed at $M_\umin'$. In the absence
of a clear way to model the formation of
PBHs and the inclusion dynamics in the dense regime, it seems difficult to go beyond the sharp
cutoff procedure, which can however be seen as a limit bounding
the range of possible renormalisation procedures (the other bounding
procedure being introduced below).

In \Fig{fig:beta:reh:renormalisation:procedures}, we have represented
the mass fraction at the end of the instability phase for
$\rho_{\uinf} = 10^{-12}\Mp^4\simeq \left(2.43 \times
10^{15}\GeV\right)^4$ and $\rho_\Gamma = 10^{-40}\Mp^4\simeq
\left(2.43 \times 10^{8}\GeV\right)^4$. The black solid line
corresponds to what is obtained before renormalisation and leads to
$\Omega_{\mathrm{PBH}}(t_\Gamma) = 8.54$, which is not possible. The
blue dotted line represents the result after renormalisation by
inclusion~(\ref{eq:ren:inclusion}), \ie by removing the low mass part
of the distribution to bring $\Omega_{\mathrm{PBH}}(t_\Gamma)$ back to
one.
\subsubsection{Renormalisation by premature ending}
\label{subsubsec:premature}
Another possibility is that, as $\Omega_{\mathrm{PBH}}$ increases,
PBHs backreact on the dynamics of the universe,
which is no longer dominated by the coherent oscillations of the
inflaton field, and the instability stops. The precise value of
$\Omega_{\mathrm{PBH}}$ at which this premature termination occurs is
difficult to assess, and for simplicity we will assume it to be one,
since our final results mildly depend on it.

In that case, if $\Omega_{\mathrm{PBH}}(t_\Gamma)$ is found to be
larger than one, we change the time at which the instability stops,
\begin{align}
t_\Gamma \rightarrow t_{\mathrm{instab}},
\end{align}
where $t_{\mathrm{instab}}$ is the time at which
$\Omega_{\mathrm{PBH}}$ reaches one. Therefore, as announced before,
there are situations for which $t_\uinstab\neq t_\Gamma$. The result
is displayed in \Fig{fig:beta:reh:renormalisation:procedures} with the
dotted green line.  One can check that the large-mass black holes are
removed from the mass fraction distribution, since those black holes
correspond to scales that enter the instability band towards the end
of the instability phase, at which point the instability is now no
longer on.

Since, as explained above, the procedure of renormalisation by
inclusion removes the small-mass end of the distribution, these two
approaches can therefore be viewed as complementary, and by studying the
results obtained with both one can assess how much the conclusions
depend on the way the mass fraction is renormalised.

The actual renormalisation procedure might lie in between these two
schemes: for instance, it could happen that, as
$\Omega_{\mathrm{PBH}}$ increases, inclusion starts to be important,
which slows down the increase of $\Omega_{\mathrm{PBH}}$ but does not
prevent it from further growing, until the point where premature
ending occurs. In such a case, a distribution that is intermediate
between the blue and the green curves of
\Fig{fig:beta:reh:renormalisation:procedures} would be obtained. As we
will show below, some common conclusions can be drawn with both
renormalisation schemes, which motivates the statement that such
conclusions are mildly dependent on the renormalisation approach.
\subsection{Evolving the mass fraction}
\label{subsec:evolving:beta}
\begin{figure}[t]
\begin{center}
\includegraphics[width=0.6\textwidth]{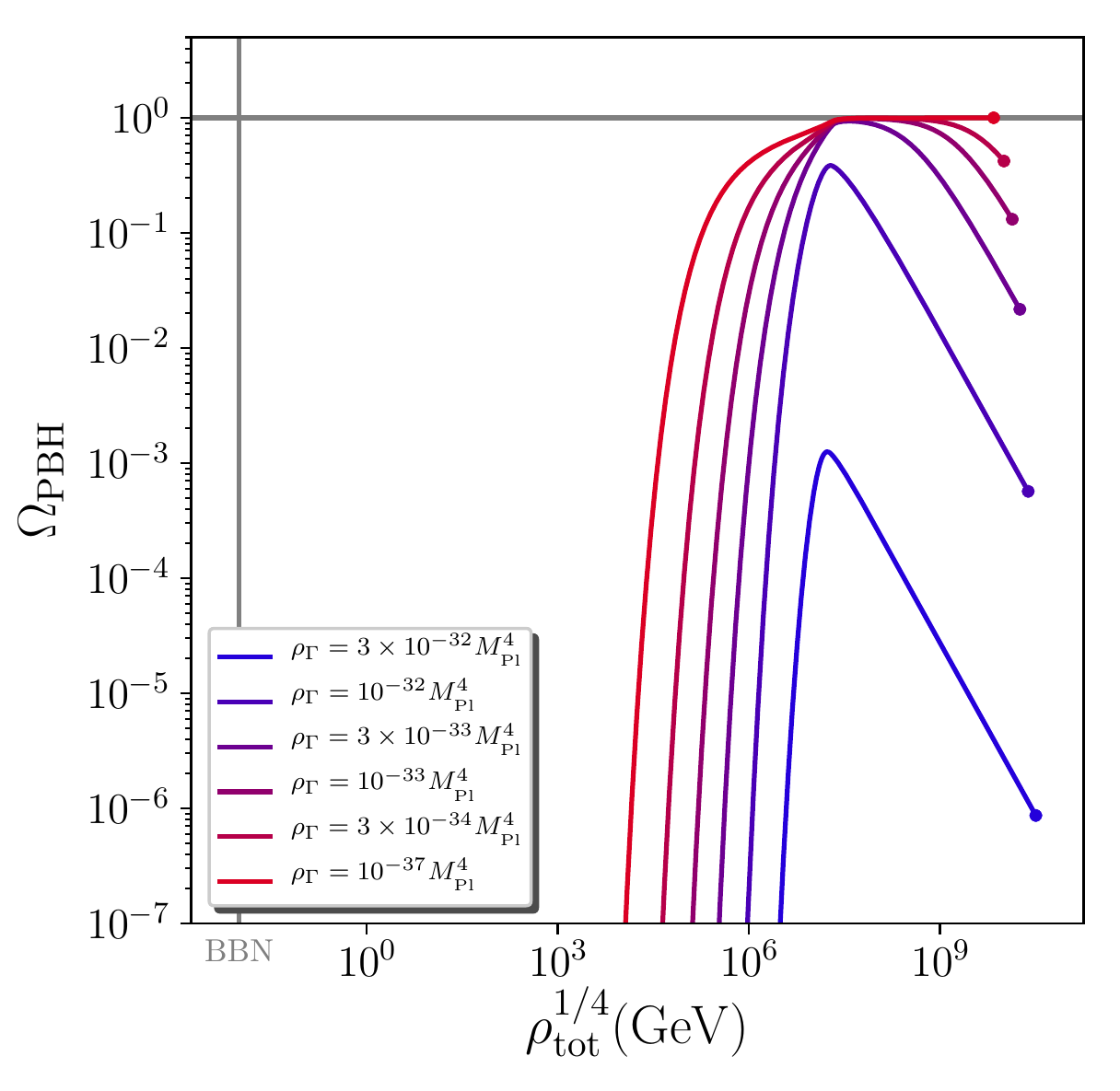}
\caption{Integrated mass fraction $\Omega_{\mathrm{PBH}}$ as a
  function of time, here parametrised by the total energy density
  $\rho_\utot$, from the end of the instability phase
  $t_{\mathrm{instab}}$ until BBN, for the same values of $\rho_\uinf$
  and $\rho_\Gamma$ as the ones displayed in
  \Fig{fig:beta:reh:few:rho:reh}. For $\rho_\Gamma =
  10^{-37}\Mp^4\simeq (1.4\times 10^{9}\GeV)^4$, the mass fraction
  needs to be renormalised, which for illustration here is done using the
  premature-ending procedure.}
\label{fig:Omega:time}
\end{center}
\end{figure}
After the instability stops, the density of black holes evolves under
different physical effects, such as Hawking evaporation, accretion and
merging. In what follows we neglect the two latter and only account
for the former. The reason is that accretion and merging are
technically difficult to model (see \eg \Refs{Ali-Haimoud:2016mbv,
  Ali-Haimoud:2017rtz}), and only contribute to enhancing the final
value of $\Omega_{\mathrm{PBH}}$. The reason why this is the case for
accretion is obvious, and for merging, this is because the Hawking
evaporation time~(\ref{eq:Hawking:time}) cubicly depends on the
mass. Therefore, when two black holes (say of the same mass) merge,
they loose some fraction of their mass through the emission of
gravitational waves, but their evaporation time is multiplied by $8$,
allowing them to live much longer. As a consequence, by only
considering Hawking evaporation, we again derive conservative bounds,
which underestimate the density of black holes at the epochs where
they are observationally constrained.

The mass of a black hole decreases under Hawking evaporation according
to~\cite{Hawking:1974rv}
\begin{align}
M(t,k) = M\left(t_{\mathrm{instab}},k\right)\left\lbrace
1-\frac{t-t_{\mathrm{instab}}}{\Delta t_{\mathrm{evap}}\left[M\left(
    t_{\mathrm{instab}},k\right)\right]}\right\rbrace^{1/3}\, ,
\end{align}
where $\Delta t_{\mathrm{evap}}$ was given in
\Eq{eq:Hawking:time}. This expression should be understood as coming
with a Heaviside function such that, when $t-t_{\mathrm{instab}} >
\Delta t_{\mathrm{evap}}$, $M$ is set to zero. We do not write it
explicitly here for notational convenience. If $\bar{\beta}$ denotes
the mass fraction in the absence of Hawking evaporation, one then has
\begin{align}
\label{eq:OmegaPBH:continuous}
 \Omega_{\mathrm{PBH}}(t)  =
 \int_{M_\umin' }^{M_\umax} \bar{\beta}
 \left(M,t\right)\left[1-\frac{t-t_{\mathrm{instab}}}
   {\Delta t_{\mathrm{evap}}\left(M_\mathrm{instab}\right)}\right]^{1/3} \dd\ln M\, ,
\end{align}
where $M_\mathrm{instab}$ is a short-hand notation for $M(
t_{\mathrm{instab}},k)$, and where one should recall that $M$ and $k$
are related through \Eq{eq:M(k)}. Let us see how $\bar{\beta}$ can be
calculated (in what follows, quantities with a bar denote their values
in the absence of Hawking evaporation). The energy density of
PBHs contained in an infinitesimal range of scales
$\delta (\ln M)$ is given by $ \delta \bar{ \rho} =
\rho_{\mathrm{tot}} \bar{\beta}\left(M,t\right) \delta (\ln M)
$. Since PBHs behave as pressureless matter, in the absence of Hawking
evaporation one would have $\dot{\delta\bar{\rho}}+3H\delta\bar{\rho}
= 0$. Plugging the former expression into the latter, one obtains $
\left(\dot{\rho}_{\mathrm{tot}}+3 H {\rho}_{\mathrm{tot}}
\right)\bar{\beta}\left(M,t\right) + \rho_{\mathrm{tot}}
\dot{\bar{\beta}}\left(M,t\right) =0$. After the end of the
instability phase, we assume that the inflaton instantaneously decays
into a radiation fluid, so $\bar{\rho}_\utot =
\bar{\rho}_{\mathrm{PBH}} + \bar{\rho}_\urad$. In the absence of
Hawking evaporation, one then has $\dot{\bar{\rho}}_\utot = - 3 H
\bar{\rho}_{\mathrm{PBH}} - 4 H \bar{\rho}_\urad = -3H
\Omega_{\mathrm{PBH}}\bar{\rho}_\utot - 4 H (1-\Omega_{\mathrm{PBH}})
\bar{\rho}_\utot = H
\bar{\rho}_{\mathrm{tot}}(\Omega_{\mathrm{PBH}}-4)$. This gives rise
to
\begin{align}
\label{eq:ode:beta:bar}
\dot{\bar{\beta}}(M,t)+H\left(\Omega_{\mathrm{PBH}}-1\right)\bar{\beta}(M,t)=0\, .
\end{align}
A priori, this equation has to be solved for each mass independently,
with the corresponding initial condition at
$t_{\mathrm{instab}}$. However, since the equation is linear and does
not depend explicitly on the mass, a simpler solution to the problem
can be found by introducing the function $\mathfrak{b}$ that satisfies
\begin{align}
\label{eq:ode:frakb}
\dot{\mathfrak{b}}+H\left(\Omega_{\mathrm{PBH}}-1\right)\mathfrak{b}=0
\quad\text{with}\quad
\mathfrak{b}\left(t_{\mathrm{instab}}\right)=1\, ,
\end{align}
and such that
\begin{align}
\label{eq:frakb:beta}
\bar{\beta}\left(M,t\right)
=\bar{ \beta}\left(M,t_{\mathrm{instab}}\right)\mathfrak{b}\left(t\right)
\end{align}
satisfies \Eq{eq:ode:beta:bar}, with the correct initial
condition. The set of equations~(\ref{eq:OmegaPBH:continuous}),
(\ref{eq:ode:frakb}) and~(\ref{eq:frakb:beta}) then defines a
differential system that one can integrate numerically. Finally, let
us note that, in practice, we would like to integrate the differential
system until a time defined by its energy density rather than its
cosmic time (for instance, until BBN defined by
$\rho^{1/4}=\rho_{\mathrm{BBN}}^{1/4}\sim 10\, \MeV$). For this reason
it is more convenient to use $\ln \rho_\utot$ as the time variable
(the log being used for numerical convenience), and \Eq{eq:ode:frakb}
becomes
\begin{align}
\frac{\dd\mathfrak{b}}{\dd\ln\rho_{\mathrm{tot}}} +
\frac{\Omega_{\mathrm{PBH}}-1}{\Omega_{\mathrm{PBH}}-4}
\mathfrak{b}=0\,.
\end{align}
The value of cosmic time is still necessary in order to evaluate the
Hawking suppression term in \Eq{eq:OmegaPBH:continuous}, which can be
tracked solving
\begin{align}
\label{eq:dt:dlnrho}
\frac{\dd (t-t_{\mathrm{instab}})}{\dd\ln\rho_{\mathrm{tot}}}
= \frac{\sqrt{3}\Mp}{\left(\Omega_{\mathrm{PBH}}-4\right)\sqrt{\rho_{\mathrm{tot}}}}
\end{align}
together with the above system. 

In \Fig{fig:Omega:time}, the solution one obtains for
$\Omega_{\mathrm{PBH}}$ as a function of time is displayed for the
same parameter values as the ones used in
\Fig{fig:beta:reh:few:rho:reh}. At early time, the effect of Hawking
evaporation is negligible, and $\rho_{\mathrm{PBH}} \propto
a^{-3}$. If $\Omega_{\mathrm{PBH}}\ll 1$, $\rho_\utot\simeq
\rho_\urad\propto a^{-4}$ and $\Omega_{\mathrm{PBH}} \propto a$,
otherwise $\rho_\utot \simeq \rho_{\mathrm{PBH}}\propto a^{-3}$ and
$\Omega_{\mathrm{PBH}}$ remains equal to one. Let us see when the
black holes complete their evaporation. If a PBH forms from a scale that crosses in the instability band at
$\rho_{\mathrm{bc}}$, its mass is given by setting $k/k_\uend =
(\rho_{\mathrm{bc}}/\rho_\uinf)^{1/6}$ in \Eq{eq:M(k)}. Inserting
the corresponding expression of $M$ into \Eq{eq:Hawking:time}, the time
$t_\mathrm{evap}-t_\uinstab$ at which it evaporates can be derived.
If $\Omega_{\mathrm{PBH}}\ll 1$ until this point, \Eq{eq:dt:dlnrho}
can be integrated and gives $\rho=\rho_{\mathrm{instab}}
[1+2\sqrt{\rho_{\mathrm{instab}}/3}(t-t_{\mathrm{instab}})/\Mp]^{-2}$,
which means that the black hole evaporates at the energy density
\begin{align}
\label{eq:rho:evap:estimate}
\rho_{\mathrm{evap}} \sim \frac{1}{26244 \xi^6}
\left(\frac{g}{10240}\right)^2\frac{\rho_{\mathrm{bc}}^3}{\Mp^{8}}\, .
\end{align}
Notice that, in order to obtain this estimate, we have neglected the
fact that Hawking evaporation starts before the end of the
instability (which was however taken into account for PBHs that entirely evaporate during the instability, see \Sec{subsec:criterionrefined}). Indeed, given than the collapsing time decays with the initial density contrast, see \Eq{eq:t:coll}, and since PBHs form in the Gaussian tail of the distribution function where the smaller the density contrast, the more likely it is, most PBHs form close to the end of the instability phase, and for them Hawking evaporation during the instability can be neglected.

If $\Omega_{\mathrm{PBH}}$ takes sizeable values before the
evaporation of the first black holes, the estimate~(\ref{eq:rho:evap:estimate}) needs only to be
corrected by factors of order one (If $\Omega_{\mathrm{PBH}}=1$, the
corrective factor is $8/9$). The first black holes to evaporate are
the ones with the smallest mass $M_\umin$, \ie such that
$\rho_{\mathrm{bc}} = \rho_\uinf$. In \Fig{fig:Omega:time}, one can
check that the evaporation of these PBHs indeed corresponds to the
turning point of all curves [for $\rho_{\uinf} = 10^{-12}\Mp^4\simeq
  \left(2.43 \times 10^{15}\GeV\right)^4$, \Eq{eq:rho:evap:estimate}
  gives $\rho_{\mathrm{evap}}\sim 3\times 10^{-45} \Mp^4 \simeq
  \left(1.8\times 10^{7}\GeV\right)^4$]. Below this point, Hawking
evaporation is efficient and $\Omega_{\mathrm{PBH}}$ quickly
decreases.
\subsection{Reheating through PBH evaporation}
\label{subsec:Reh:PBH:evap}
\begin{figure}[t]
\begin{center}
\includegraphics[width=0.6\textwidth]{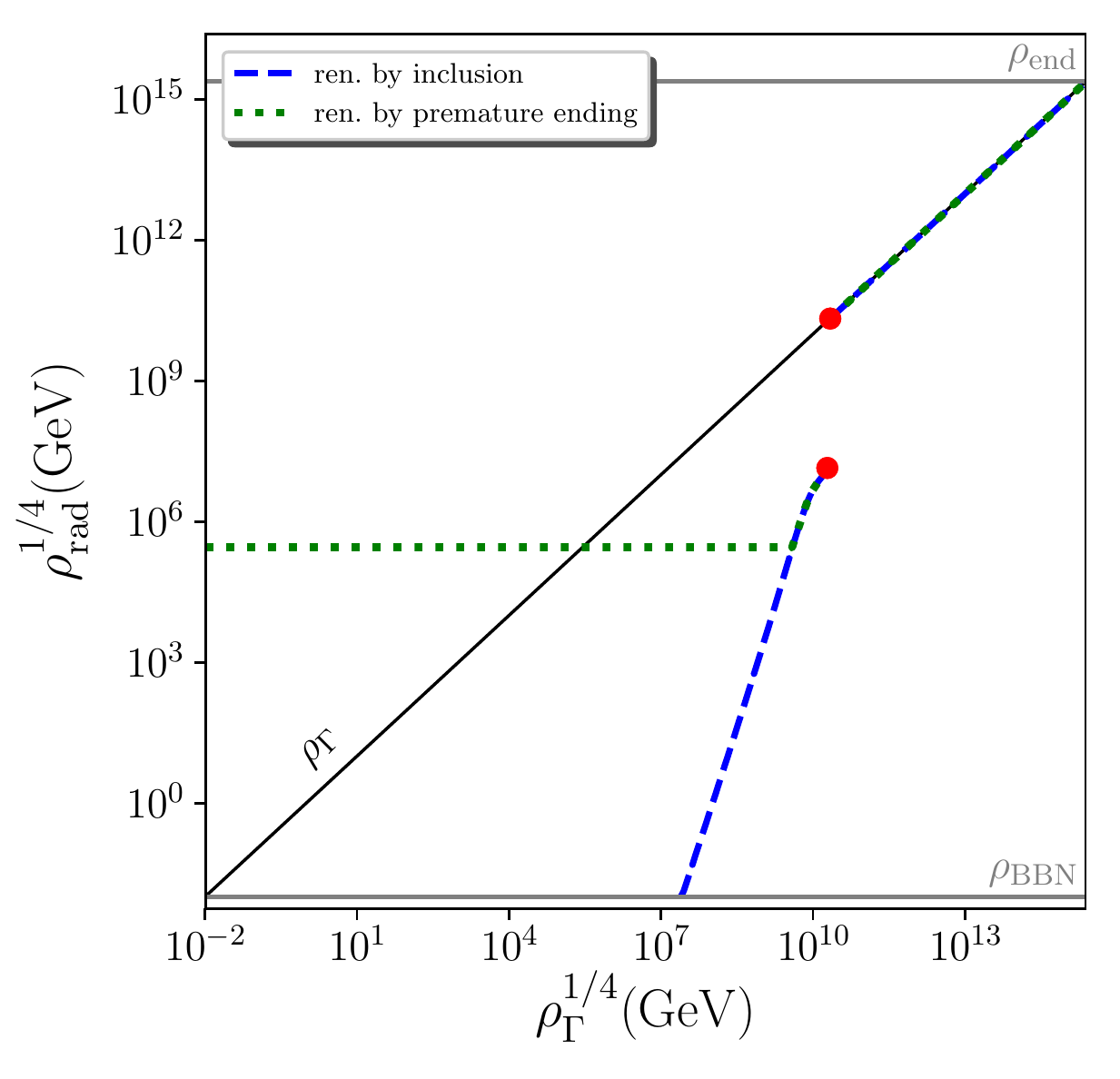}
\caption{Energy density at the onset of the radiation era,
  $\rho_\urad$, as a function of $\rho_\Gamma$, for $\rho_\uinf =
  10^{-12}\Mp^4\simeq \left(2.43 \times 10^{15}\GeV\right)^4$ (which
  is the value used in all previous figures). The blue curve
  corresponds to the renormalisation procedure by inclusion, while the
  green one stands for renormalisation by premature ending. The red
  circles indicate the location of the discontinuity, \ie values of
  $\rho_\urad$ comprised between the two circles are never realised,
  see main text.}
\label{fig:rho:rad}
\end{center}
\end{figure}
The onset of the radiation era, defined as being the time, after the
instability phase, after which $\Omega_{\mathrm{PBH}}$ remains below
$1/2$, does not necessarily coincide with
$t_{\mathrm{instab}}$. Indeed, if the universe is dominated by PBHs at
the end of the instability, as is the case for the curve with
$\rho_\Gamma = 10^{-37}\Mp^4\simeq (1.4\times 10^{9}\GeV)^4$ in
\Fig{fig:Omega:time}, the radiation era only starts with the
evaporation of the first black holes around $\rho\sim
10^{-45}\Mp^4\simeq (10^{7}\GeV)^4$ as explained above. In fact, even
if PBHs do not dominate the universe's content at the end of the
instability phase, they may later do so, see the curve with
$\rho_\Gamma = 3\times 10^{-33}\Mp^4\simeq (1.8\times 10^{10}\GeV)^4$
in \Fig{fig:Omega:time} for instance, in which case the onset of the
radiation epoch is also delayed.

In such cases, let us point out that the reheating of the universe
proceeds from the Hawking evaporation of the PBHs
that dominate the energy budget for a transient period after the
instability phase.\footnote{This possibility has been discussed, in a
  different context, in \Refs{GarciaBellido:1996qt, Hidalgo:2011fj,
    Suyama:2014vga, Zagorac:2019ekv}.} If it completes long before BBN,
such a mechanism is a priori allowed, and we discuss several of its
implications in \Sec{sec:conclusion}. It is then interesting to
extract the energy density at the onset of the radiation period,
$\rho_\urad$, from our computational pipeline. Let us notice that
$\rho_\urad$ is the quantity which is related to what would be defined
as the reheating temperature, $T_\ureh$, through
$\rho_\urad=g_*\pi^2T_\ureh^4/30$, where $g_*$ is the number of
relativistic degrees of freedom.

The quantity $\rho_\urad$ is displayed in \Fig{fig:rho:rad} for
$\rho_\uinf = 10^{-12}\Mp^4\simeq (2.43 \times 10^{15}\GeV)^4$ (which
is the same value employed in all previous figures, in particular in
\Fig{fig:Omega:time}) and as a function of $\rho_\Gamma$, which varies
between $\rho_{\mathrm{BBN}}$ and $\rho_\uinf$. This allows us to identify several relevant regions in parameter space. When
$\rho_\Gamma$ is large, the instability phase is too short to produce
a substantial amount of PBHs and they never dominate the energy content of
the universe. This corresponds \eg to the curve with $\rho_{\Gamma} =
3\times 10^{-32}\Mp^4\simeq (3.2\times 10^{10}\GeV)^4$ in
\Fig{fig:Omega:time}. In this case, the radiation era starts when the
inflaton decays into radiation, and $\rho_{\urad} = \rho_\Gamma$.

When $\rho_\Gamma$ decreases, one first notices in \Fig{fig:rho:rad}
the presence of a discontinuity, that we will explain shortly. In a
small range below the discontinuity, $\rho_\urad$ is different from
$\rho_\Gamma$, denoting the presence of a phase where PBHs dominate
the universe, but does not depend on the renormalisation procedure,
revealing that PBHs do not dominate at the end of the instability
phase. This corresponds \eg to the curve with $\rho_\Gamma = 3\times
10^{-33}\Mp^4\simeq (1.8\times 10^{10}\GeV)^4$ in
\Fig{fig:Omega:time}. In this case, after the instability phase, there
is a first radiation epoch, then PBHs take over and drive a matter
epoch, before they evaporate and reheat the universe, which finally enters a
second radiation epoch. One then finds $\rho_\urad < \rho_\Gamma$.

The discontinuity can be explained as follows: let us consider the case where radiation dominates at $t_\Gamma$, namely
$\Omega_\mathrm{PBH}<1/2$ at $t_\Gamma$. Clearly, in this situation,
no renormalisation is needed since $\Omega_\mathrm{PBH}<1$ at
$t_\Gamma$. Then, as already explained, $\Omega_\mathrm{PBH}$
grows proportionally to the scale factor until Hawking evaporation
becomes efficient and makes $\Omega_\mathrm{PBH}$ decrease, see \Fig{fig:Omega:time}. Assume that the maximum value $\Omega_\mathrm{PBH}$ reaches is slightly smaller than
$1/2$.
In this situation, the start of
the radiation epoch is $t_\Gamma$ and $\rho_\mathrm{rad}=\rho_\Gamma$
since the radiation era is never interrupted. This case
corresponds to the upper red dot in \Fig{fig:rho:rad}. Consider now the situation where at the end of the
instability, the value of $\Omega _\mathrm{PBH}$ is
infinitesimally larger than in the previous case (and, therefore, still
smaller than $1/2$ at $t_\Gamma$). This means that we now start with a
value of $\rho_\Gamma$ that is slightly smaller than before (and
the instability lasts slightly longer). This gives rise
to the same behaviour as described above except that, now, the value at the
maximum is slightly larger than before, and above
$1/2$. This means that the radiation epoch comes to an end and that a
matter dominated era starts. Of course, since this is also the time at
which Hawking radiation starts to become important, this matter-dominated era lasts a very short amount of time and very soon a
new radiation dominated era (the ``real'' one) starts. The
important point, however, is that $\rho_\mathrm{rad}$ is now very
different from $\rho_\Gamma$ and is close to $\rho_\mathrm{evap}$, and this second case corresponds to the lower red dot in \Fig{fig:rho:rad}. 

This explains the discontinuity in the curve $\rho_\mathrm{rad}$ versus
$\rho_\Gamma$. Let us note that an important consequence of this behaviour is the fact that none of the values for $\rho_\urad$ comprised between the two red circles can be physically realised. We therefore identify regions in parameter space that are forbidden, not by the observations, but by self-consistency of the scenario itself.

Finally, when $\rho_\Gamma$ takes small values, PBHs are very
abundantly produced and the mass fraction needs to be renormalised at
the end of the instability phase. If renormalisation is carried out by
inclusion, by keeping only the heavy black holes in the distribution,
Hawking evaporation proceeds at later times when $\rho_\Gamma$
decreases, and the radiation epoch is more and more delayed. There is
even a point where the radiation era has not started yet by BBN, which
is obviously excluded and which explains why the blue curve is not
plotted in \Fig{fig:rho:rad} below that point. If renormalisation is
performed by premature ending on the other hand, the result does not
depend on $\rho_\Gamma$ since $\rho_{\mathrm{instab}}$ becomes
independent of that parameter and, from there, the
value of $\rho_\urad$ is only controlled by the evaporation process. In
that case, for $\rho_\Gamma^{1/4}\gtrsim 286 \TeV$, the onset of the
radiation epoch is delayed compared to what it would have been if
sourced by inflaton decay. 
This also implies that the inflaton could decay ``inside'' the black holes, although due to the no hair theorem, this should not leave any physical imprint.
On the other hand, if $\rho_\Gamma^{1/4}\lesssim 286 \TeV$, reheating occurs earlier than it would have with pure inflaton decay. 

To conclude this section, let us stress again that, for
$\rho_\Gamma\lesssim 10^{10}\GeV$ and $\rho_\uinf =
10^{-12}\Mp^4\simeq \left(2.43 \times 10^{15}\GeV\right)^4$ (a full
scan of the parameter space is presented in the following), namely
below the lower red point in \Fig{fig:rho:rad}, the radiation in our
universe no longer comes from inflaton decay but from the evaporation
of PBHs formed during preheating. Given the generic character of the
situation considered here (single-field inflation with quadratic minimum), this is clearly one of the main conclusions of the present
paper.

\subsection{Planckian relics}
\label{subsec:Planckian:Relics}

The previous considerations show that the universe may have gone
through a phase where PBHs are numerous, and can
even dominate the energy budget of the universe, but that these black
holes can also well have all disappeared before BBN, through Hawking
evaporation. In such a case, there is no direct way to constrain
them, unless they do not fully evaporate and leave some relics behind.

This possibility has been discussed~\cite{Markov:1984xd,
  Coleman:1991ku} in the context of quantum-gravity inspired
scenarios, where it has been suggested that black hole evaporation
might stop when the mass of the black hole reaches the Planck mass. In
this case, the number density of black hole can be computed at the end
of the instability phase according to
\begin{align}
  n_{\mathrm{PBH}}\left(t_\mathrm{instab}\right) = \rho_\utot
  \int_{M_\umin}^{M_\umax}\frac{\tilde{\beta}
    \left(M,t_\mathrm{instab}\right)}{M} \dd \ln M\, .
\end{align}
In this expression, $\tilde{\beta}\left(M,t_\mathrm{instab}\right)$
corresponds to \Eq{eq:beta:erf} (with $t_\Gamma$ replaced with
$t_\mathrm{instab}$) where, instead of taking $\delta_\umax$ as being
the minimum value between one and the right-hand side of
\Eq{eq:delta:max:Hawking}, one simply takes $\delta_\umax = 1$. This
ensures that the black holes that evaporate before the end of the
instability phase are also accounted for in the calculation of relics.

Since this number density is not affected by Hawking evaporation, it
then evolves according to the function $\frak{b}(t)$ introduced in
\Sec{subsec:evolving:beta}, \ie solely under the effect of cosmic
expansion. The fractional energy density of relics at subsequent times
is thus given by
\begin{align}
\label{eq:Omega_Relics}
\Omega_{\mathrm{relics}}(t) = \mathfrak{b}(t) \int_{M_\umin}^{M_\umax}
      {\tilde{\beta}\left(M,t_\mathrm{instab}\right)}\frac{\Mp}{M} \dd \ln M\, .
\end{align}
Let us note that this expression assigns one Planckian relic to each
black hole, whether it has already evaporated or not. It therefore
gives the density of ``naked'' relics only in the late-time limit,
when all black holes have evaporated. It however always provides a
lower bound on the contribution to dark matter (DM) originating from
black holes and their relics, and as such, should be checked to be
smaller than $\Omega_{\mathrm{DM}}$, which will be done in
\Sec{subsec:result:relics}.

\section{Observational consequences}
\label{sec:results}
Having described the physical setup and the methods employed to model
it, let us now turn to the results and discuss their physical
implications.
\subsection{The onset of the radiation era}
\label{subsec:onset}
In \Sec{subsec:Reh:PBH:evap}, it was found that in some cases, the
production of PBHs is so efficient that they may come to dominate the
energy budget of the universe, either before the end of the
instability phase or afterwards. In that case, the onset of the
radiation era does not correspond to the time when the inflaton
decays, \ie when $\rho=\rho_\Gamma$, but rather occurs when the PBHs
evaporate. The corresponding energy density, $\rho_\urad$, has been
displayed as a function of $\rho_\Gamma$ and for a fixed value of
$\rho_\uinf$ in \Fig{fig:rho:rad}.

\begin{figure}[t]
\begin{center}
  \includegraphics[width=0.496\textwidth, clip=true]
                  {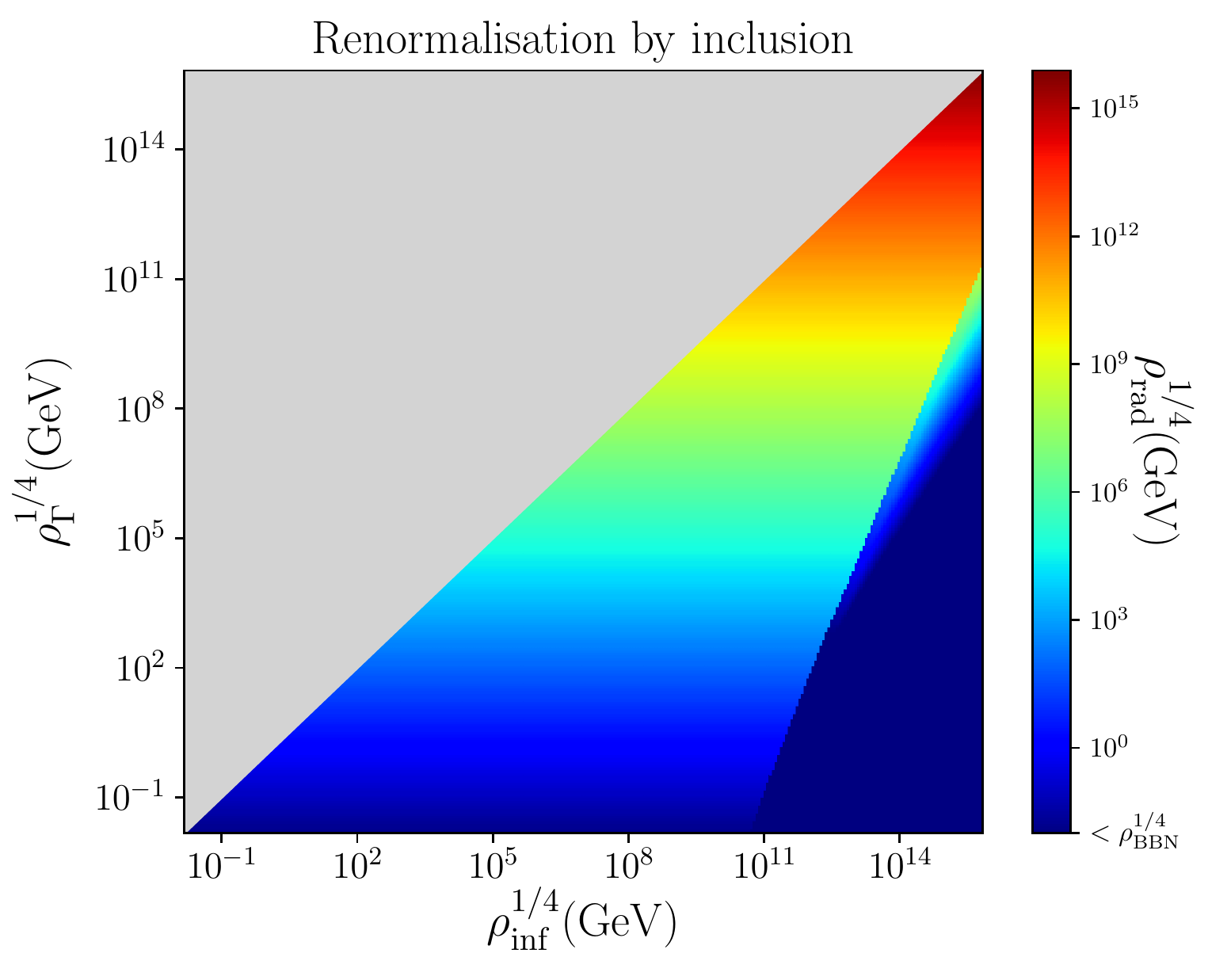}
                  \includegraphics[width=0.496\textwidth, clip=true]
                                  {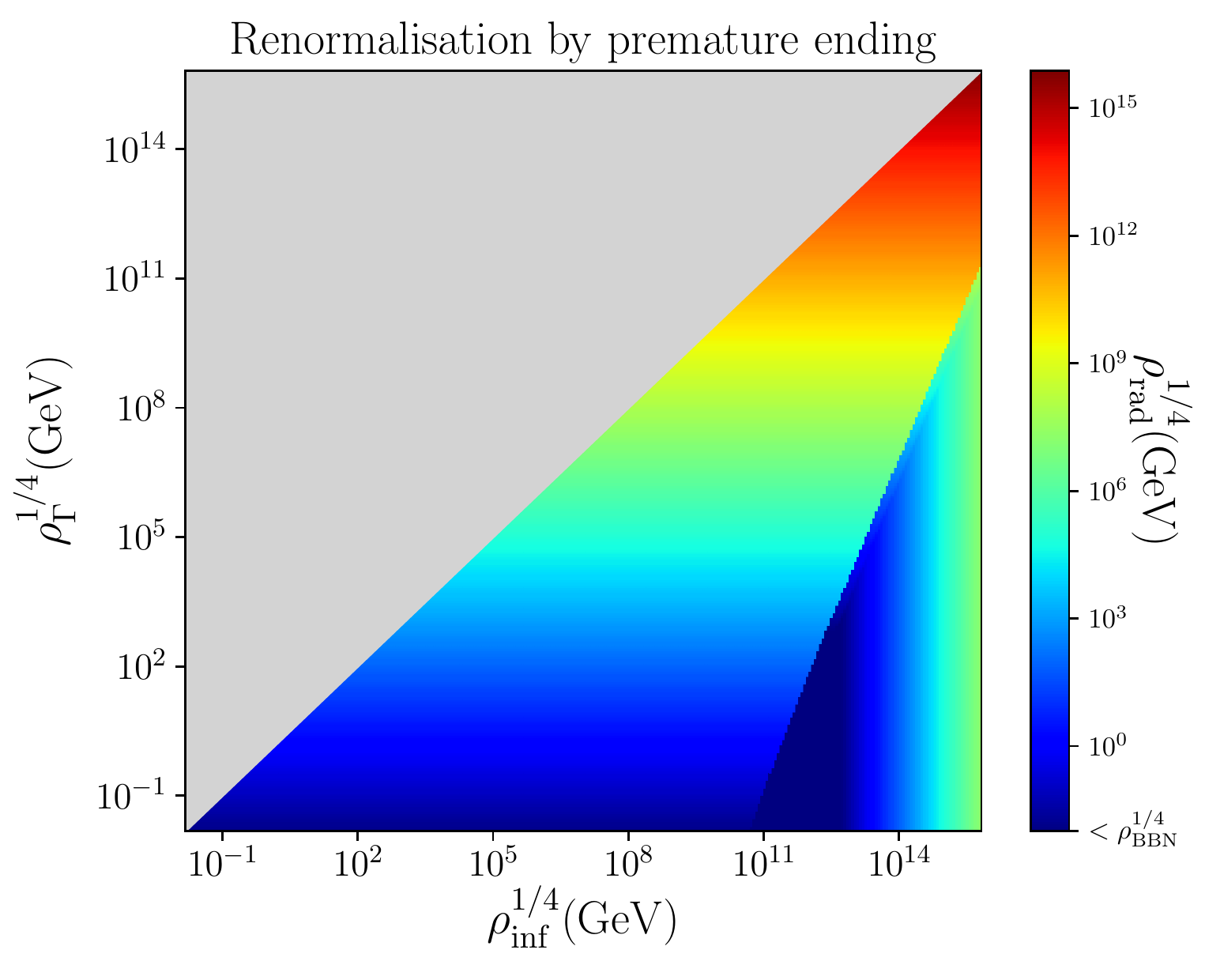}
\caption{The energy density at the onset of the radiation era as a
  function of $\rho_\uinf$ and $\rho_\Gamma$. The grey region is
  excluded since it corresponds to $\rho_\uinf<\rho_\Gamma$. Left
  panel: renormalisation by inclusion. Right panel: renormalisation by
  premature ending.}
\label{fig:rho_rad_maps}
\end{center}
\end{figure}

In \Fig{fig:rho_rad_maps}, the same quantity is shown, but as a
function of both $\rho_\uinf$ and $\rho_\Gamma$. Thus
\Fig{fig:rho:rad} is a vertical slice of \Fig{fig:rho_rad_maps}. The
left panel corresponds to renormalisation by inclusion,
see \Sec{subsubsec:inclusion}, while the right panel stands for
renormalisation by premature ending, see \Sec{subsubsec:premature}. The
grey region is excluded since it corresponds to
$\rho_\Gamma>\rho_\uinf$. In the region where $\rho_\urad =
\rho_\Gamma$, PBHs never dominate and reheating occurs at the end of
the instability phase, through decay and thermalisation of the
inflaton. In both figures, the lower right triangular regions, in
which $\rho_\urad \neq \rho_\Gamma$, are such that reheating proceeds
by PBH evaporation. Notice that, there, the darkest blue region
corresponds to parameter values for which the universe is still not
dominated by radiation at BBN, which is excluded. This allows us to
generalise the remarks made around \Fig{fig:rho:rad}: when
$\rho_\Gamma$ is large, the instability phase is short, PBHs never
dominate the universe, so $\rho_\urad=\rho_\Gamma$ and reheating
proceeds in the standard way; when $\rho_\Gamma$ is sufficiently
small, PBHs can dominate the universe, which results into either
delaying or anticipating the universe reheating. For $\rho_\uinf^{1/4}\simeq 10^{15}\GeV$, which corresponds to a tensor-to-scalar ratio of $r\simeq 10^{-3}$, reheating occurs from PBHs evaporation when $\rho_\Gamma^{1/4}\lesssim 2
\times 10^9 \mathrm{GeV}$.

More generally, the boundary of the lower-right triangles, \ie the condition for
reheating the universe via PBH evaporation, can be worked out as
follows. Clearly, reheating proceeds through PBHs evaporation if the
PBHs are formed in a substantial way. This is the case if the
critical density contrast given in \Eq{eq:cond:delta}, $\delta_\uc \sim ({3\pi}/{2})^{2/3}
({k}/{k_\uend})^{-2} ({\rho_\uinf}/{\rho_{\mathrm{instab}}})^{-1/3}
=({3\pi}/{2})^{2/3}(\rho_\mathrm{instab}/\rho_\mathrm{bc})^{1/3}$
[where we have used $k/k_\uend =
  (\rho_{\mathrm{bc}}/\rho_\uinf)^{1/6}$] is much smaller than
$\sqrt{2 \calP_\delta}$. Moreover, the modes that get the more
amplified are the ones that enter the instability band the earlier,
and thus exit the Hubble radius not long before the end of
inflation. For them, one can take $\calP_{\zeta,\uend} \sim
H_\uend^2/(8\pi^2\Mp^2)$, see \Eq{eq:Pzeta:end:SR}, hence
$\calP_{\delta,\mathrm{bc}} \sim 3 \rho_\uinf/(50 \pi^2 \Mp^4)$, see
\Eq{eq:calP:delta:bc}. As a consequence, the condition
$\delta_\uc/\sqrt{2\calP_\delta}\ll 1$ leads an upper bound on
$\rho_\mathrm{instab}$, namely
$\rho_\mathrm{instab}<4(125\sqrt{3}\pi^5)^{-1}(\rho_\uinf/\Mp^4)^{3/2}
\rho_\mathrm{bc}$. This makes sense since, in order to have sizeable
PBHs production, the instability must last long enough and, therefore,
$\rho_\mathrm{instab}$ must be small enough. Since $\rho_{\mathrm{bc}}<\rho_\uinf$ and
$\rho_\Gamma<\rho_{\mathrm{instab}}$ by
construction, this gives rise to
\begin{align}
\label{eq:reheating:via:PBH:evap:criterion}
\frac{\rho_\Gamma}{\Mp^4}<\frac{4}{125\sqrt{3}\pi^{5}}
\left(\frac{\rho_\uinf}{\Mp^4}\right)^{5/2}\, .
\end{align}
One can check that this expression provides a good fit to the boundary
of the lower right triangular regions in \Fig{fig:rho_rad_maps}, hence
it gives a simple criterion to check whether or not reheating proceeds
via PBH evaporation.


%
\subsection{Constraints from the abundance of PBHs}
\label{subsec:PBH:abundance}
\begin{figure}[t!]
\begin{center}
\includegraphics[width=0.496\textwidth,
  clip=true]{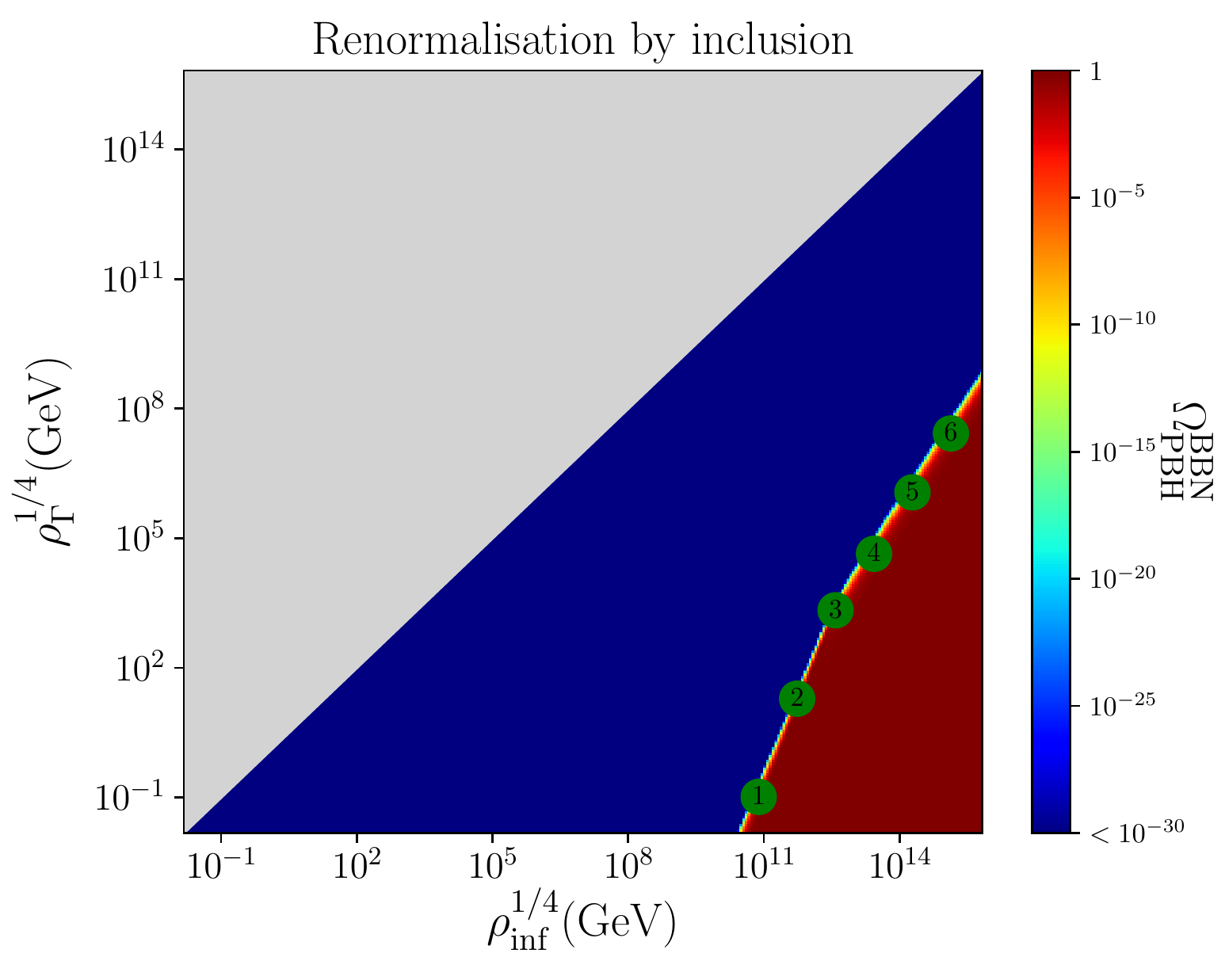}
\includegraphics[height=0.4\textwidth,width=0.45\textwidth,
  clip=true]{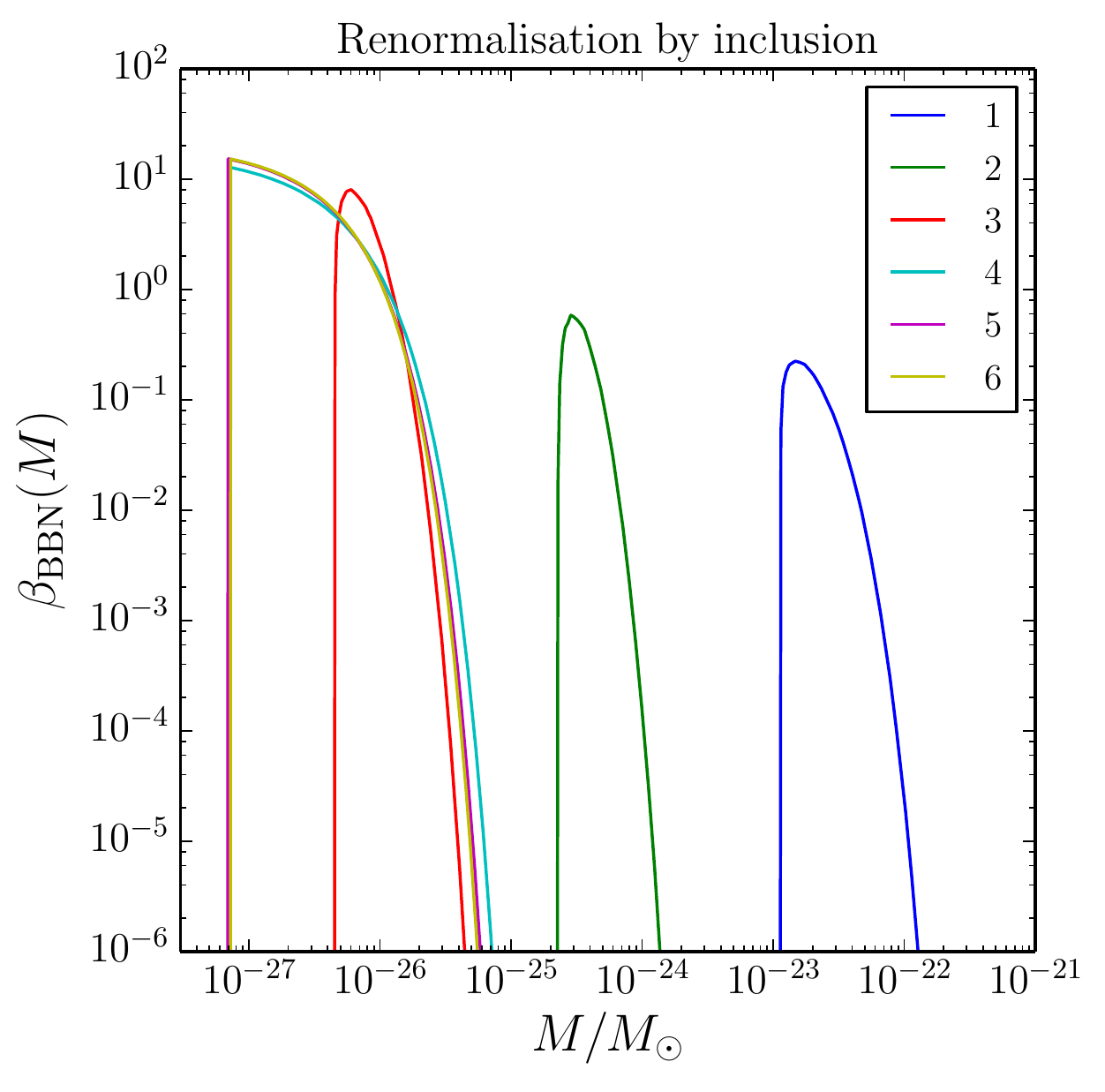}
\caption{In the left panel, the fraction of the universe made of PBHs
  at BBN is displayed as a function of $\rho_\uinf$ and $\rho_\Gamma$,
  when the mass fraction is renormalised by inclusion. The grey region
  corresponds to $\rho_\Gamma>\rho_\uinf$ and is therefore
  forbidden. In the blue region,
  $\Omega_{\mathrm{PBH}}^\mathrm{BBN}<10^{-30}$, which leaves the
  parameters unconstrained. In the dark red region,
  $\Omega_{\mathrm{PBH}}^\mathrm{BBN}\simeq 1$, which is excluded. In
  between, there is a fine-tuned region where
  $\Omega_{\mathrm{PBH}}^\mathrm{BBN}$ takes fractional values, and
  where the details of the mass fraction matter. For that reason, $6$
  points are labeled across that region, for which
  $\OmegaPBH^{\mathrm{BBN}}=10^{-2}$, and their mass fraction is shown
  in the right panel.}
\label{fig:BBN_constraints_inclusion}
\end{center}
\end{figure} 

\begin{figure}[t!]
\begin{center}
\includegraphics[width=0.496\textwidth, clip=true]{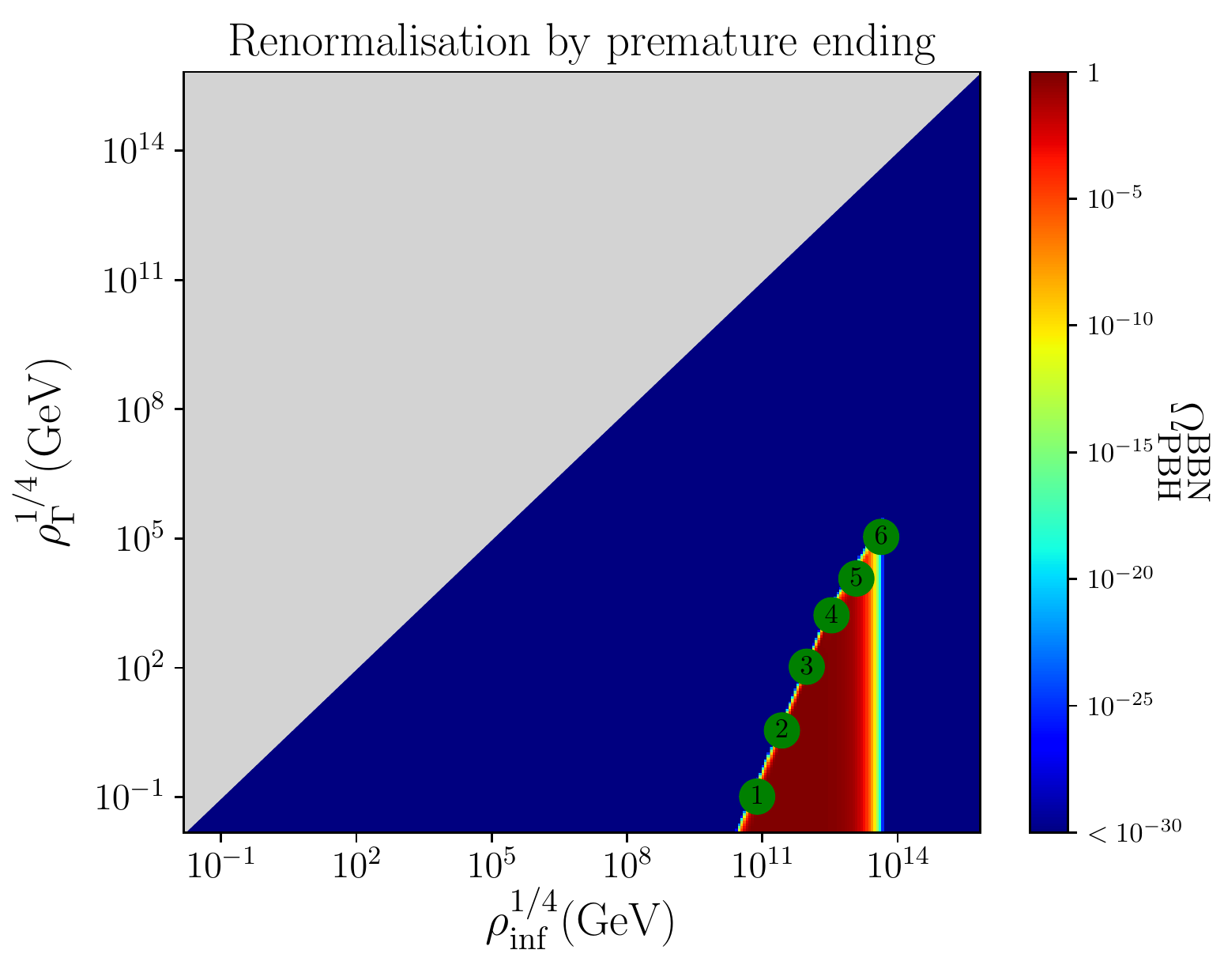}
\includegraphics[height=0.4\textwidth, width=0.45\textwidth, clip=true]{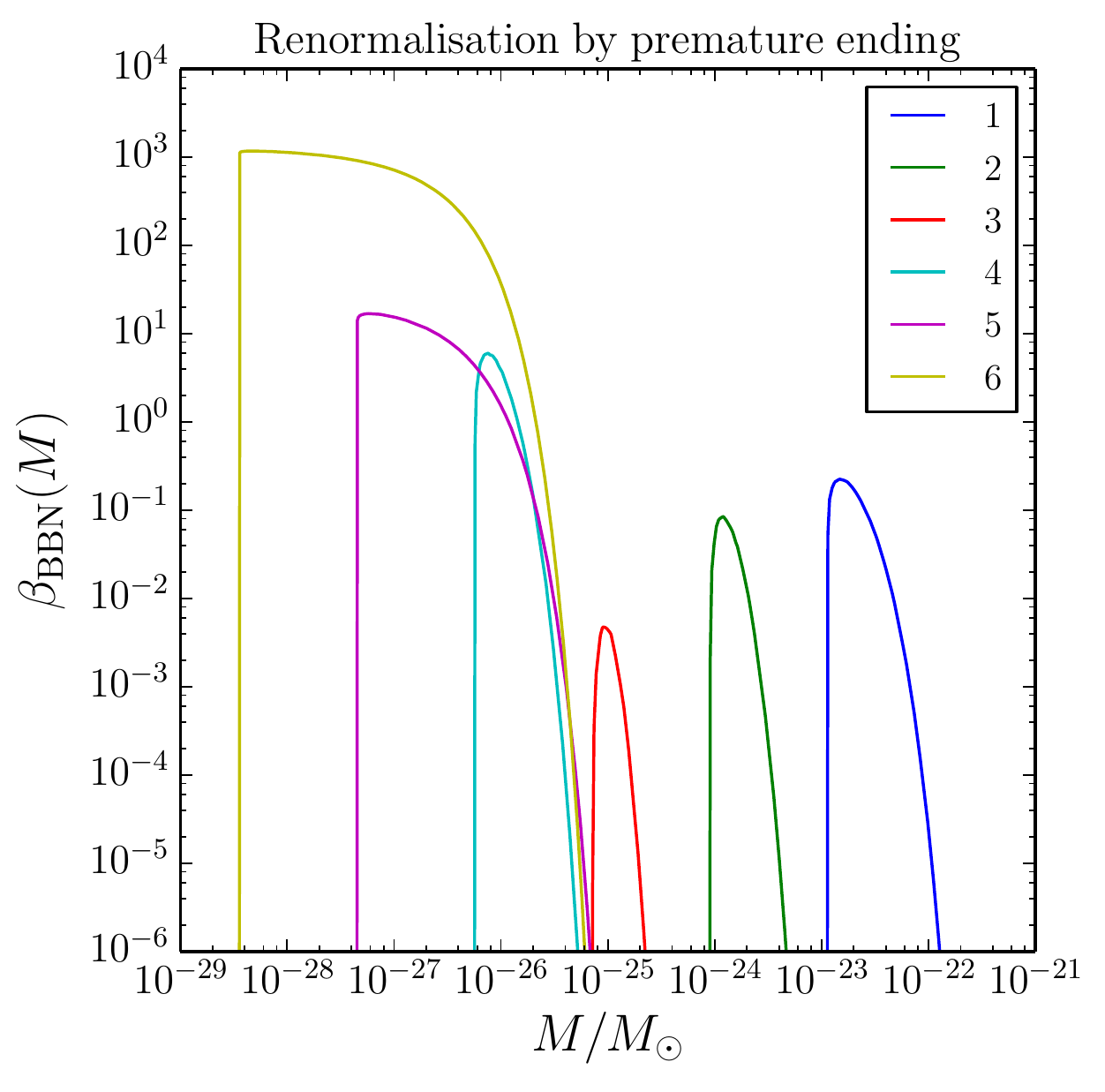}
\caption{Same as in \Fig{fig:BBN_constraints_inclusion}, when the mass fraction is renormalised by premature ending.}
\label{fig:BBN_constraints_premature_ending}
\end{center}
\end{figure}

Let us now discuss observational constraints from the predicted
abundance of PBHs. The amount of DM made of PBHs is constrained by
various astrophysical and cosmological probes, through their
evaporation or gravitational effects (for a recent review, see \eg
\Ref{Carr:2009jm, Carr:2017jsz}). The earliest constraint, \ie the one limiting
black holes with the smallest mass, is BBN. This is why in the left
panels of \Figs{fig:BBN_constraints_inclusion}
and~\ref{fig:BBN_constraints_premature_ending}, the fraction of the
universe made of PBHs at BBN is displayed, as a function of
$\rho_\uinf$ and $\rho_\Gamma$.

As before, the model is defined only when $\rho_\Gamma<\rho_\uinf$,
\ie outside the grey region. The parameter space is otherwise
essentially divided into two main regions: in the dark blue region,
\ie for large values of $\rho_\Gamma$,
$\Omega_{\mathrm{PBH}}^{\mathrm{BBN}}\lesssim 10^{-30}$, and all
observational constraints are easily passed. This corresponds to
situations where PBHs are either not substantially produced, or
evaporate before BBN. In the dark red region, \ie for smaller values
of $\rho_\Gamma$, $\Omega_{\mathrm{PBH}}^{\mathrm{BBN}}\simeq 1$ and
the universe is not radiation dominated at the time of BBN, which is
not allowed at more than the few percents level~\cite{RefToAdd}. A substantial fraction
of the reheating parameter space can therefore be excluded from the
considerations presented in this work, which is our second main result. For
instance, for the typical value $\rho_\uinf^{1/4}\simeq 10^{15}\GeV$,
$\Omega_{\mathrm{PBH}}^{\mathrm{BBN}}\gtrsim 0.1$ if
$\rho_\Gamma^{1/4}\lesssim 1.6\times 10^7 \GeV$ when renormalisation is
performed by inclusion.

The location of the boundary between the excluded and the allowed
regions can be worked out as follows. Requiring that the evaporation time, estimated in \Eq{eq:rho:evap:estimate}, is later than BBN leads to
$\rho_\mathrm{bc}/\Mp^4<(9\times
6^{2/3})\xi^2(10240/g)^{2/3}(\rho_\mathrm{BBN}/\Mp^4)^{1/3}$. In
addition, we must also make sure that the corresponding PBHs have been
produced in a non-negligible quantity which leads to the upper bound
on $\rho_\mathrm{instab}$ derived in the text above
\Eq{eq:reheating:via:PBH:evap:criterion}. Combining these two
expressions, one obtains $\rho_{\mathrm{instab}}/\Mp^4< (36\times
6^{2/3}\xi^2)/(125 \sqrt{3}\pi^5)
({10240}/{g})^{2/3}({\rho_{\mathrm{BBN}}}/{\Mp^4})^{1/3}({\rho_\uinf}/{\Mp^4})^{3/2}
\sim 2.5\times 10^{-29} ({\rho_\uinf}/{\Mp^4})^{3/2}$. Combined with
\Eq{eq:reheating:via:PBH:evap:criterion}, this gives rise to
\begin{align}
\frac{\rho_\mathrm{{instab}}}{\Mp^4}<\mathrm{min}\left[6.0\times
  10^{-5}\left(\frac{\rho_\uinf}{\Mp^4}\right)^{5/2},\,2.5\times
  10^{-29}\left(\frac{\rho_\uinf}{\Mp^4}\right)^{3/2} \right]\, .
\end{align}
One can check that this rough estimate indeed provides a good enough
description of the boundary between the blue and the red regions in
\Fig{fig:BBN_constraints_inclusion} where one simply has
$\rho_{\mathrm{instab}}=\rho_\Gamma$ (the situation in
\Fig{fig:BBN_constraints_premature_ending} is more complicated since
those are two different quantities).

In between the excluded and the allowed regions, there is a
fine-tuned, thin line along which
$\Omega_{\mathrm{PBH}}^{\mathrm{BBN}}$ can take fractional
values. There, the details of the mass fraction, \ie the value of
$\beta$ and the range of masses it covers, matter. For this reason,
both in \Figs{fig:BBN_constraints_inclusion}
and~\ref{fig:BBN_constraints_premature_ending}, we have sampled $6$
points along this thin line, for which
$\OmegaPBH^{\mathrm{BBN}}=10^{-2}$, and we show the corresponding mass
fraction in the right panels, as a function of $M/M_\odot$. We have checked that fixing
$\OmegaPBH^{\mathrm{BBN}}$ to values different than $10^{-2}$ does not
qualitatively change the following remarks.

First, one may be surprised that some values of $\beta$ are larger
than one. This is because, although $\beta$ at the end of the
instability is smaller than one by definition, see \Eq{eq:beta:erf},
it is then redshifted by $\mathfrak{b}$, see \Eq{eq:frakb:beta}, which
can be much larger than one. The integrated mass fraction,
$\Omega_{\mathrm{PBH}}$, does always remain smaller than one.

\begin{figure}[t]
\begin{center}
\includegraphics[width=0.496\textwidth, clip=true]{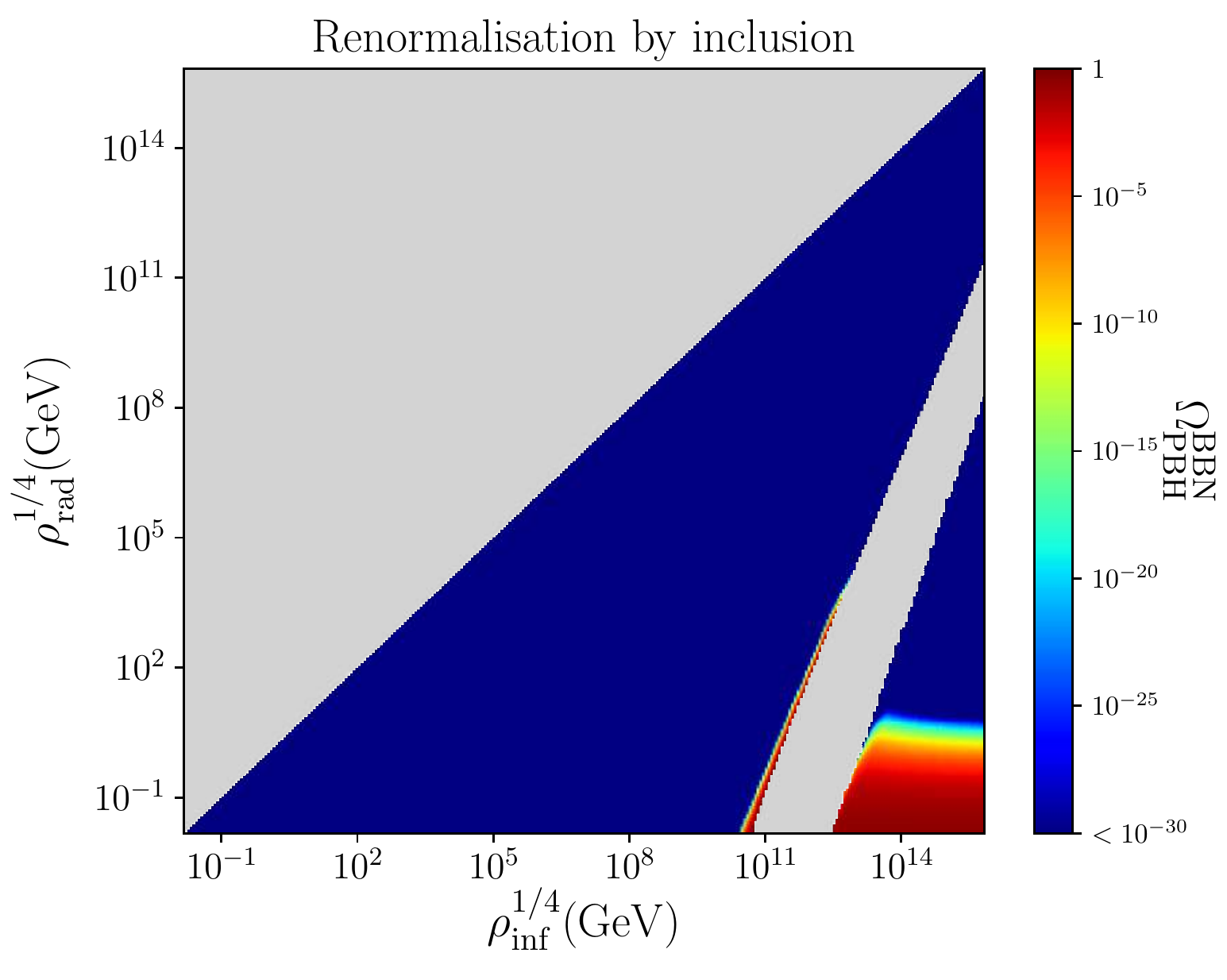}
\includegraphics[width=0.496\textwidth, clip=true]{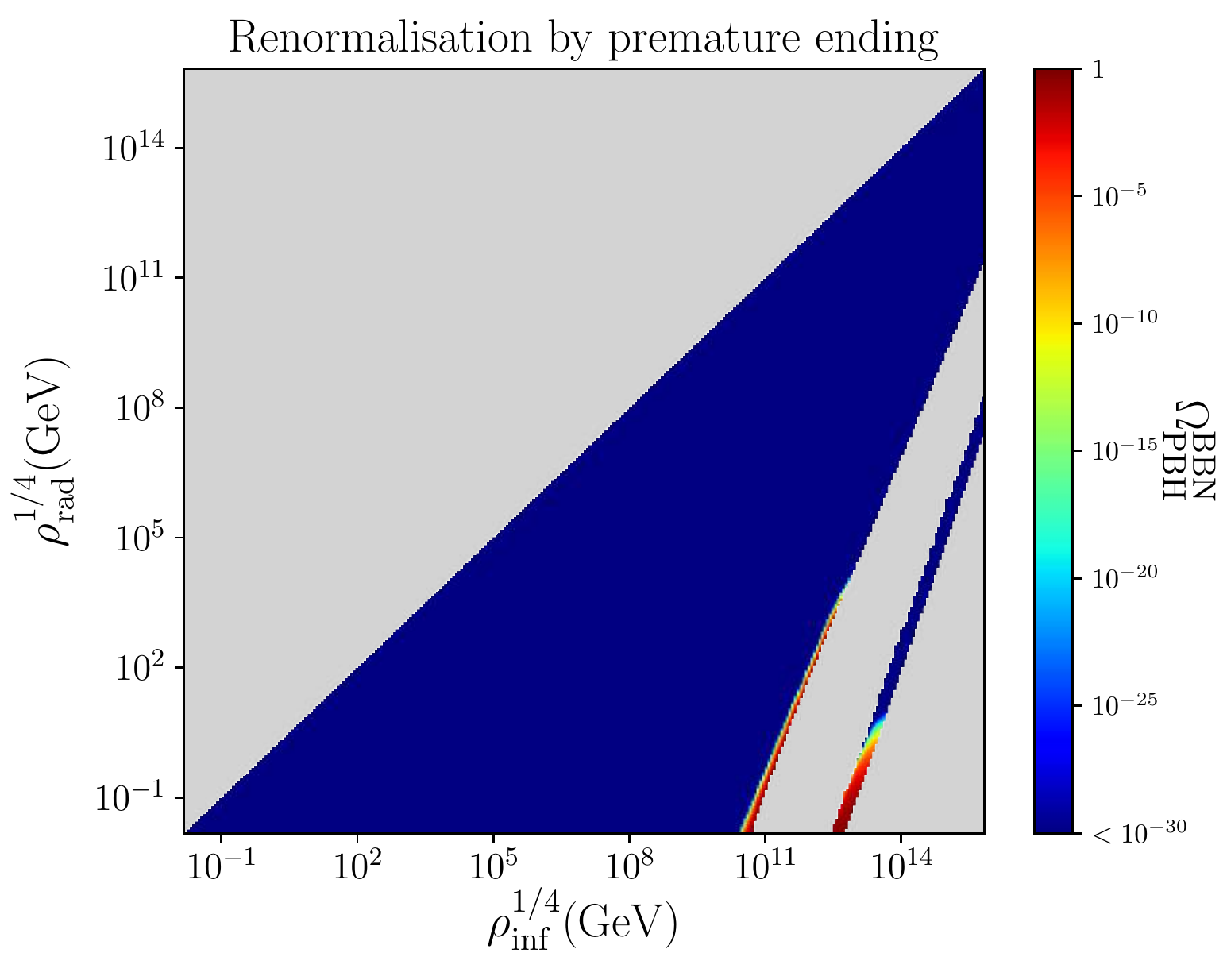}
\caption{Fraction of the universe made of PBHs at BBN, as a function
  of $\rho_\uinf$ and $\rho_\urad$, when the mass fraction is
  renormalised by inclusion (left panel) and premature ending (right
  panel). The grey region is not realised either because
  $\rho_\urad>\rho_\uinf$, or because the corresponding value of
  $\rho_\urad$ is never realised, see the discussion around
  \Fig{fig:rho:rad}.}
\label{fig:Omega_BBN_maps}
\end{center}
\end{figure}

Second, the observational constraints on the value of $\beta$ depend
on whether the mass distribution is monochromatic (\ie all black holes
have the same mass) or extended. In our case, it is clearly extended,
and the constraints then depend on its precise profile. Let us however
note~\cite{Carr:2009jm} that the smallest mass being constrained is of
the order $10^{-24}M_{\odot}$. Only the points labeled $1$ and $2$ in
\Figs{fig:BBN_constraints_inclusion}
and~\ref{fig:BBN_constraints_premature_ending}, \ie the ones with
$\rho_\uinf\sim 10^{-30}\Mp^{4}\simeq (7.7\times 10^{10}\GeV)^4$ and
very small values of $\rho_\Gamma$, can therefore be constrained. More
precisely, for monochromatic mass distributions, one
has\footnote{Observational constraints are usually quoted at the time
  of formation, assuming that PBHs form in the radiation era. In the
  present setup, PBHs form in a matter-dominated phase, so it is more
  convenient to express BBN constraints at the time of BBN itself. In
  terms of the mass fraction $\tilde{\beta}_{\mathrm{form}}$ at the
  time of formation \emph{in the case} where the universe is radiation
  dominated between PBH formation and BBN (\ie the quantity quoted in
  most reports on observational constraints), it is given by
  \begin{align}
    \label{beta_BBN}
    \beta_{\mathrm{BBN}} = 3^{1/4}\sqrt{4\pi
      \xi}\left(\frac{\Mp^{6}}{M^{2}\rho_{\mathrm{BBN}}}\right)^{1/4}
    \tilde{\beta}_{\mathrm{form}}\, .  \end{align}}
$\beta_{\mathrm{BBN}}(10^{-24}M_{\odot}<M<10^{-23}M_{\odot}) <
10^{-7}$ and
$\beta_{\mathrm{BBN}}(10^{-23}M_{\odot}<M<10^{-19}M_{\odot}) <
10^{-12}$. Although this would have to be adapted to the extended mass
distributions we are dealing with, this confirms that the points
labeled $1$ and $2$ are probably excluded. This however does not
change the main shape of the excluded region.

Third, no black hole with masses larger than $10^{-20} M_{\odot}$ are
produced unless they are too abundantly produced. This implies that
the present scenario cannot account for merger progenitors as
currently seen in gravitational-wave detectors such as LIGO/VIRGO, nor can
it explain dark matter since such black holes have all evaporated by
now.

In \Fig{fig:Omega_BBN_maps}, we finally display
$\Omega_{\mathrm{PBH}}$ at BBN as a function of $\rho_\uinf$ and
$\rho_\urad$, in order to derive constraints in that parameter space
too. As above, the upper-left grey triangle corresponds to
$\rho_\urad>\rho_\uinf$ and is therefore to be discarded. There are
however additional grey regions corresponding to values of
$\rho_\urad$ that are not realised: an intermediate grey band that
stands for the discontinuity gap commented on around
\Fig{fig:rho:rad}, and in the case of renormalisation by premature
ending, a lower right grey triangle that arises from the saturation
effect discussed around \Fig{fig:rho:rad} as well.
\subsection{Constraints  from  the abundance of Planckian relics}
\label{subsec:result:relics}
\begin{figure}[t]
\begin{center}
\includegraphics[width=0.496\textwidth, clip=true]{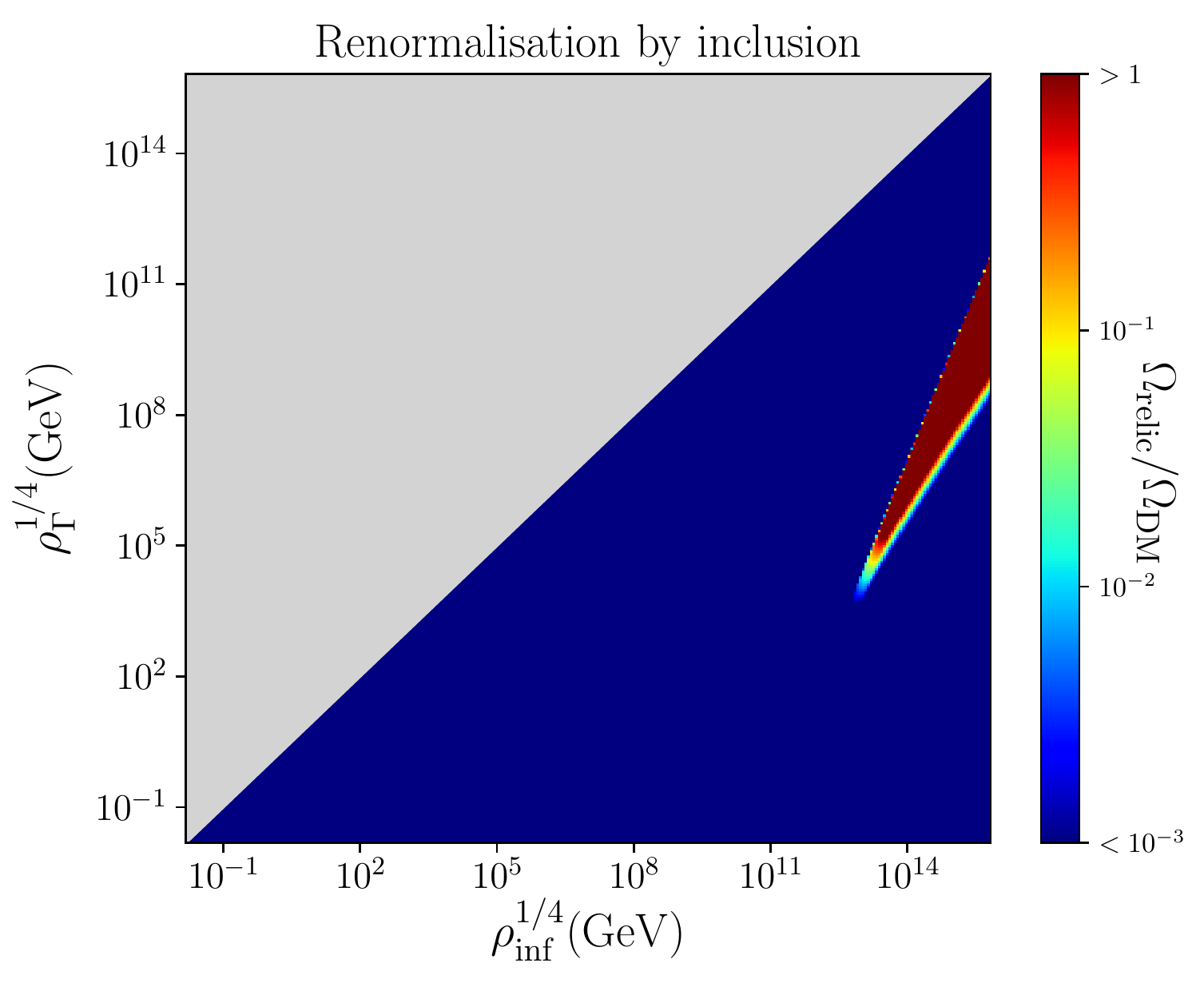}
\includegraphics[width=0.496\textwidth, clip=true]{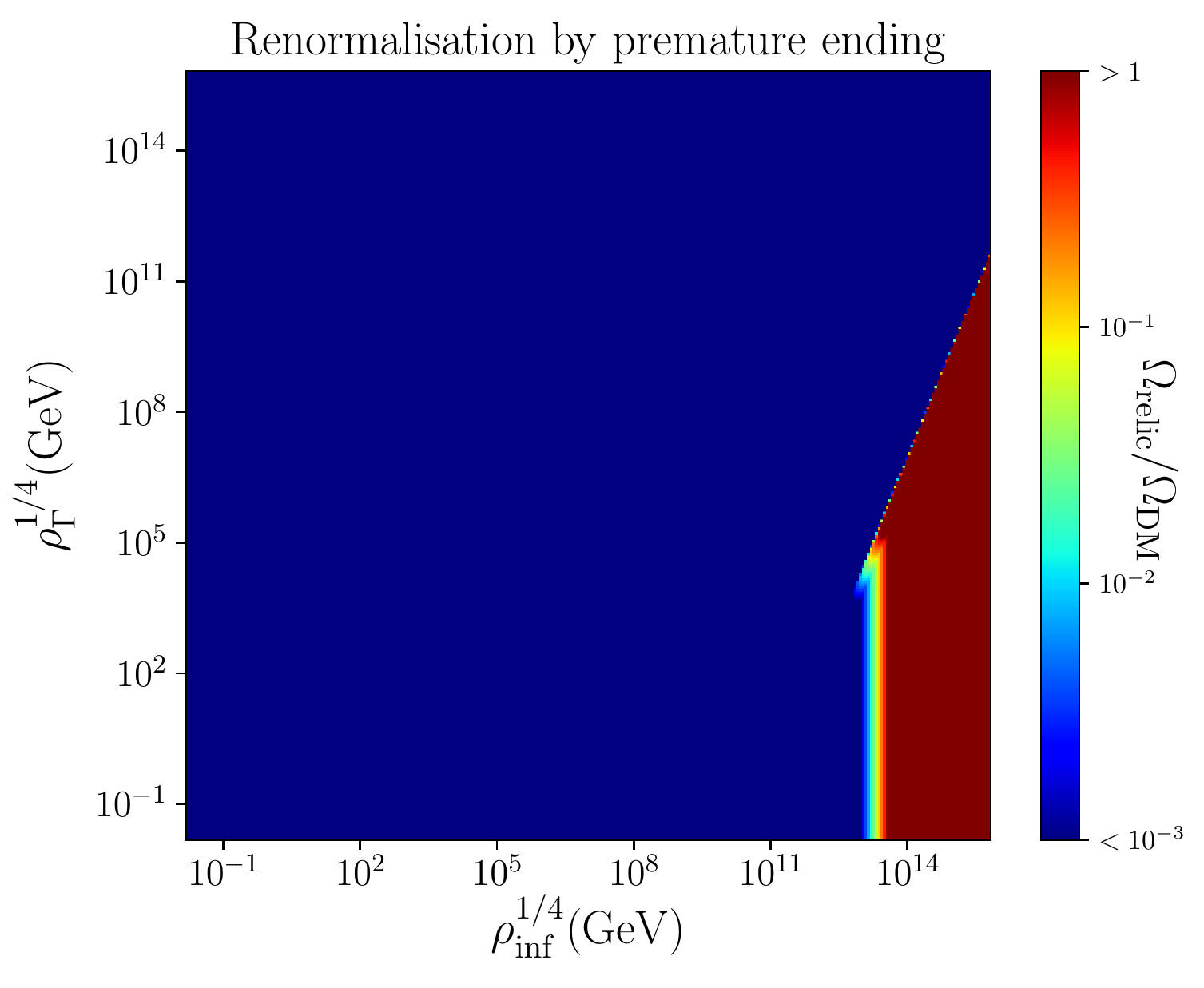}
\caption{Abundance of Planckian relics normalised to the one of dark
  matter, as a function of $\rho_\uinf$ and $\rho_\Gamma$, when the
  mass fraction is renormalised by inclusion (left panel) and
  premature ending (right panel).}
\label{fig:constraints_relics:rhoGamma}
\end{center}
\end{figure}

\begin{figure}[t]
\begin{center}
\includegraphics[width=0.496\textwidth, clip=true]{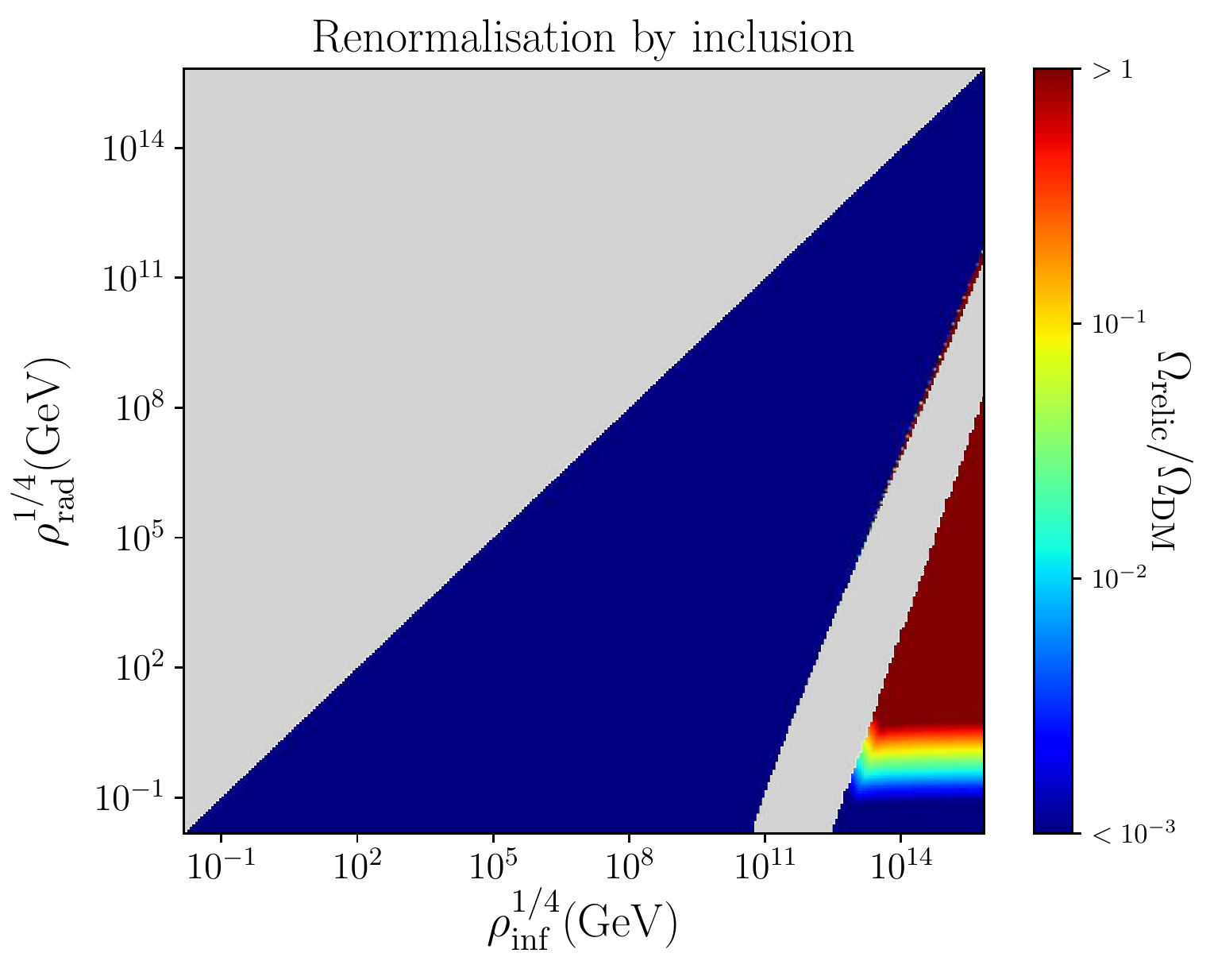}
\includegraphics[width=0.496\textwidth, clip=true]{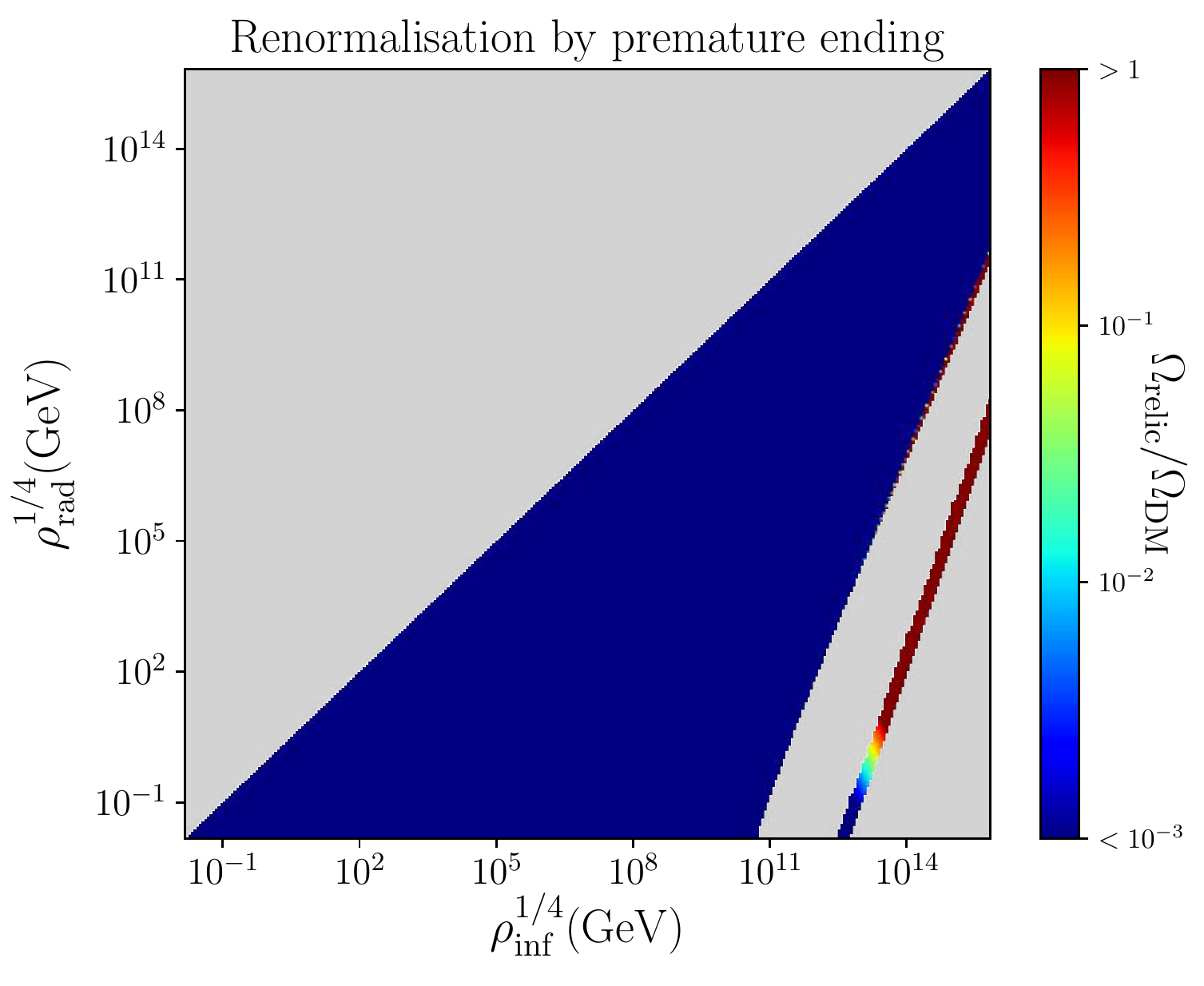}
\caption{Abundance of Planckian relics normalised to the one of dark
  matter, as a function of $\rho_\uinf$ and $\rho_\urad$, when the
  mass fraction is renormalised by inclusion (left panel) and
  premature ending (right panel).}
\label{fig:constraints_relics}
\end{center}
\end{figure}

In \Sec{subsec:Planckian:Relics}, we discussed the possibility that
evaporated PBHs leave Planckian relics behind, \ie objects of mass
$\sim \Mp$ that do not further evaporate. If they exist, their density
is expressed in \Eq{eq:Omega_Relics}, and it should be smaller than
the one of dark matter. This is why in
\Fig{fig:constraints_relics:rhoGamma}, the ratio
$\Omega_{\mathrm{relic}}/\Omega_{\mathrm{DM}}$ is displayed, as a
function of $\rho_\uinf$ and $\rho_\Gamma$, and in
\Fig{fig:constraints_relics}, as a function $\rho_\uinf$ and
$\rho_\urad$. Similarly to \Fig{fig:Omega_BBN_maps}, one can see that
parameter space is essentially divided into two regions: one (dark
blue) where the amount of Planckian relics left over from PBHs is
negligible, and one (dark red) that is excluded since Planckian relics
overtake the dark matter abundance. From
\Fig{fig:constraints_relics:rhoGamma}, we see that, if
$\rho_\uinf^{1/4}\simeq 10^{15}\GeV$, then
$\Omega_\mathrm{relics}>\Omega_\mathrm{DM}$ if 
$4.1\times 10^7 \mathrm{GeV} \lesssim \rho_\Gamma^{1/4}\lesssim 4.0\times 10^9\GeV$ 
and renormalisation is
performed by inclusion. If it is performed by premature ending, then $\Omega_\mathrm{relics}>\Omega_\mathrm{DM}$ if
$\rho_\Gamma^{1/4}\lesssim 4.0 \times 10^9 \mathrm{GeV}$. Further regions of
parameter space can thus be excluded from the predicted abundance of
relics, if they exist. In between the excluded and allowed regions, there is a
fine-tuned boundary where Planckian relics could constitute a
substantial fraction of the dark matter.

\section{Discussion and conclusions}
\label{sec:conclusion}
In this work, we have shown how the coherent oscillations of the
inflaton field around a local minimum of its potential at the end of
inflation can lead to the resonant amplification of its fluctuations
at small scales, that can then collapse and form PBHs. We have shown how the abundance and mass distribution of these
PBHs can be calculated from the spectrum of fluctuations as predicted
by inflation. In some cases, it was found that the production
mechanism is so efficient that one needs to account for possible
inclusion effects, and/or for the possibility that PBHs backreact and prematurely
terminate the preheating instability. In such cases, the universe undergoes
a phase where it is dominated by a gas of PBHs, that later reheats the
universe by Hawking evaporation. This happens when
\Eq{eq:reheating:via:PBH:evap:criterion} is satisfied.

A first result obtained in the present paper is therefore that, in the most simple models
of inflation, reheating does not necessarily occur via inflaton decay,
but for a large fraction of parameter space, it rather proceeds from the evaporation of
PBHs produced during preheating. For the iconic value
$\rho_\uinf^{1/4}\simeq 10^{15}\GeV$ (corresponding to a tensor-to-scalar ratio $r\sim 10^{-3}$), this is the case provided
$\rho_\Gamma^{1/4}\lesssim 2\times 10^9\GeV$. This deeply modifies our
view of how the universe is reheated in the context of the
inflationary theory: the radiation in our
universe could well originate from Hawking radiation rather than from inflaton
decay as usually thought.

\begin{figure}[t]
\begin{center}
\includegraphics[width=0.496\textwidth, clip=true]{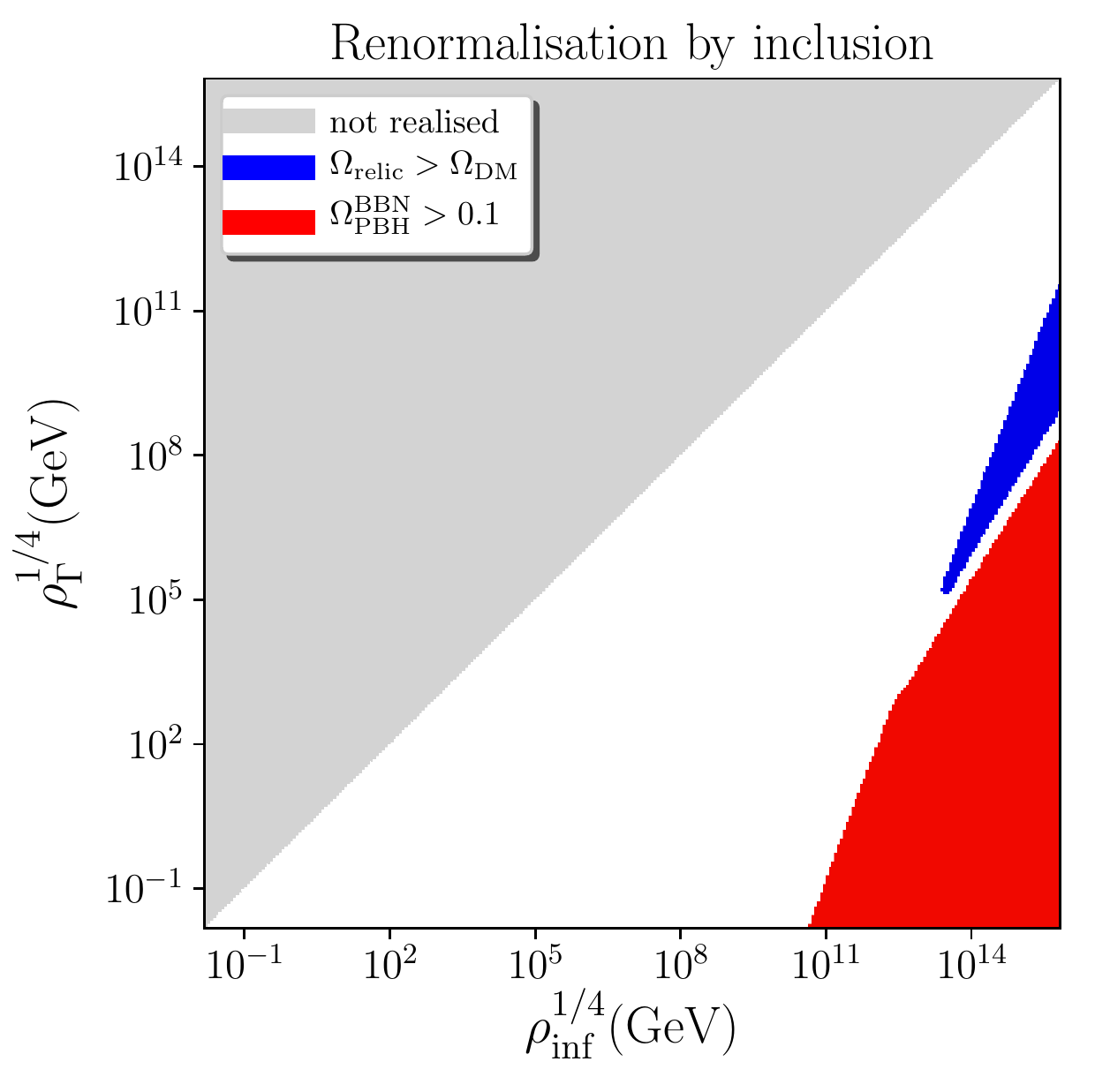}
\includegraphics[width=0.496\textwidth, clip=true]{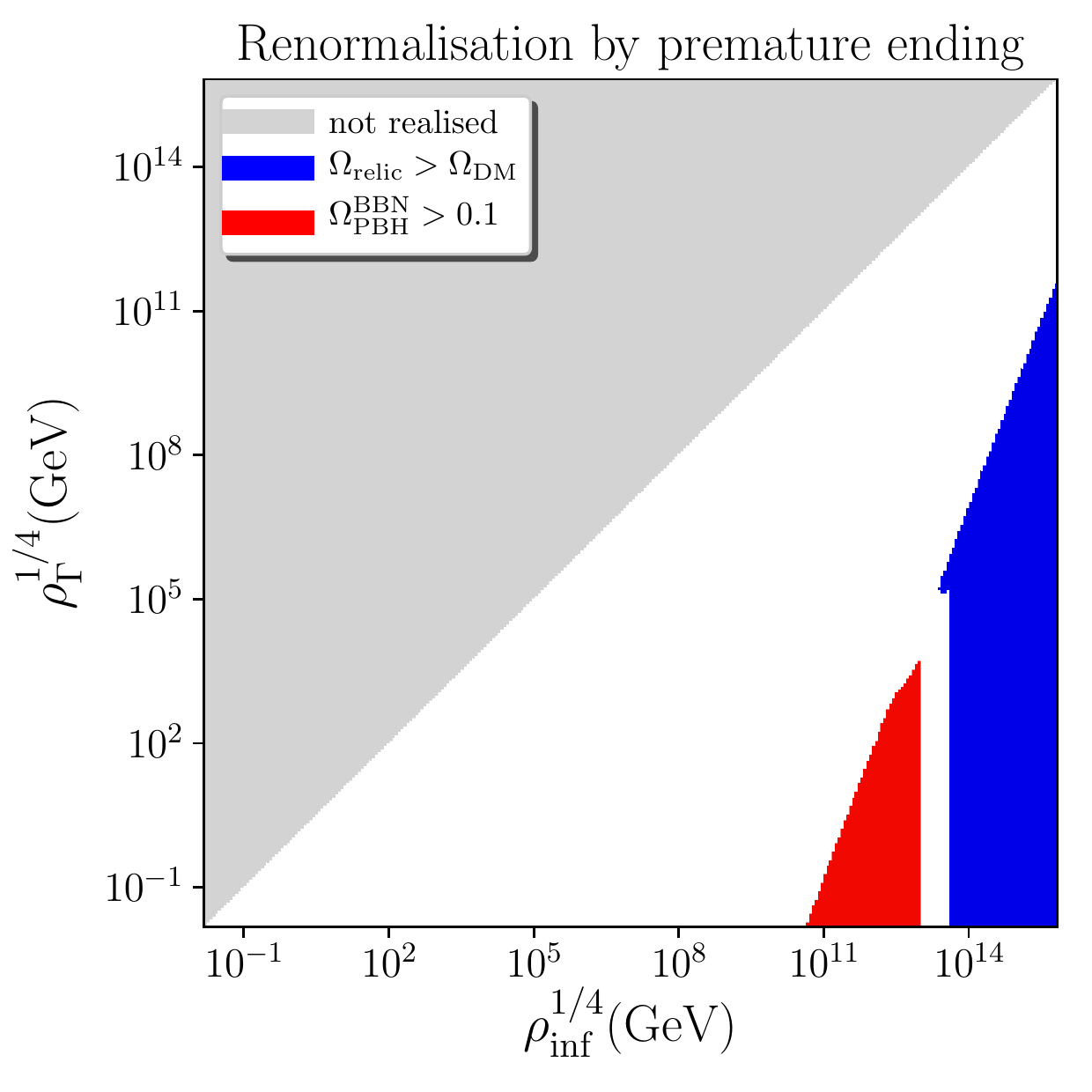}
\caption{Combined constraints in the the space $(\rho_\uinf,
  \rho_\Gamma)$, when the mass fraction is renormalised by inclusion
  (left panel) and premature ending (right panel). Red regions are
  excluded since they yield a too large abundance of primordial black
  holes. If black holes leave Planckian relics behind after
  evaporation, the blue regions are also excluded since they lead to
  too many of them. The remaining region, displayed in white, is the
  allowed one.}
\label{fig:combined_constraints:rhoGamma}
\end{center}
\end{figure}

\begin{figure}[t]
\begin{center}
\includegraphics[width=0.496\textwidth, clip=true]{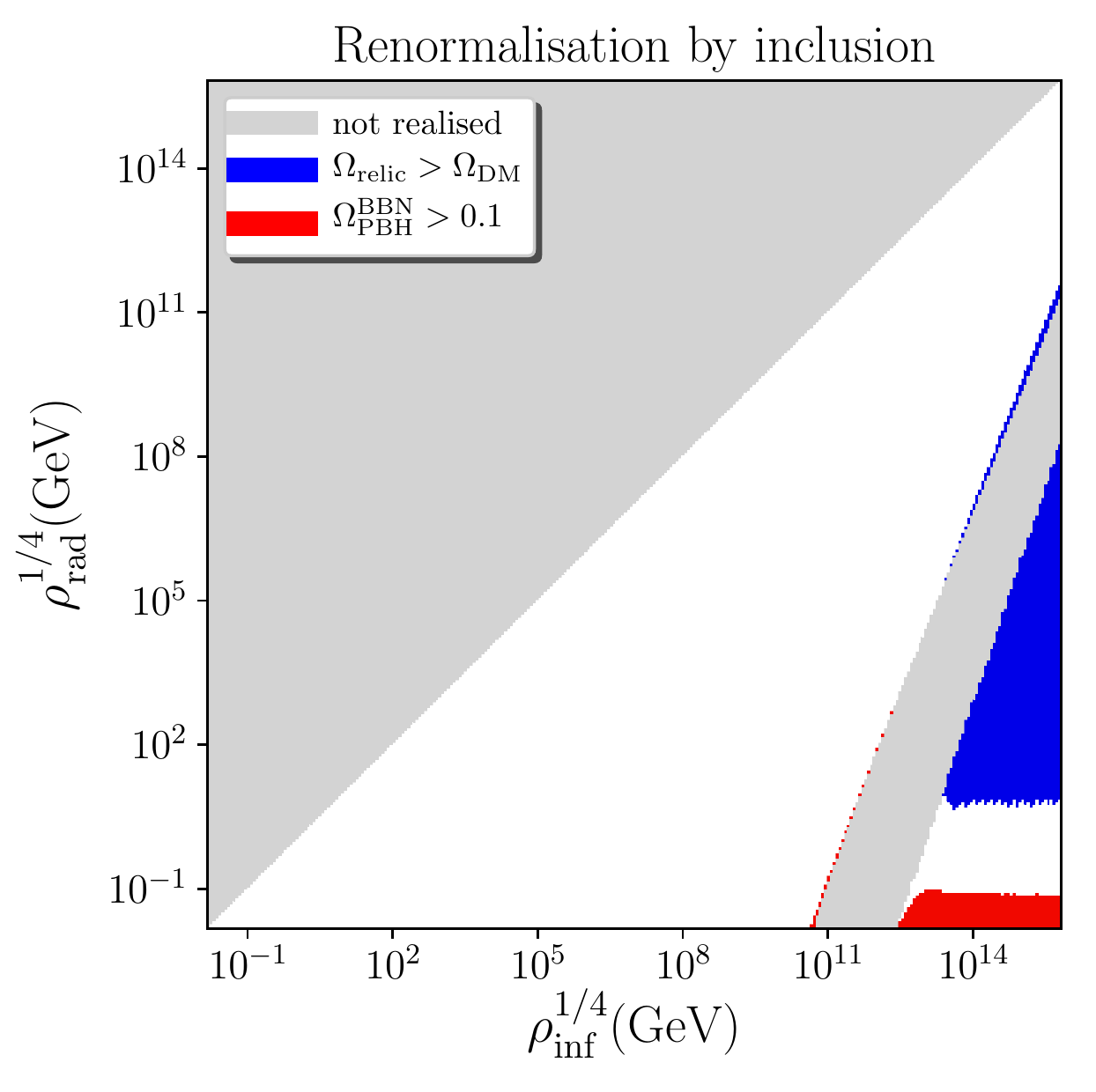}
\includegraphics[width=0.496\textwidth, clip=true]{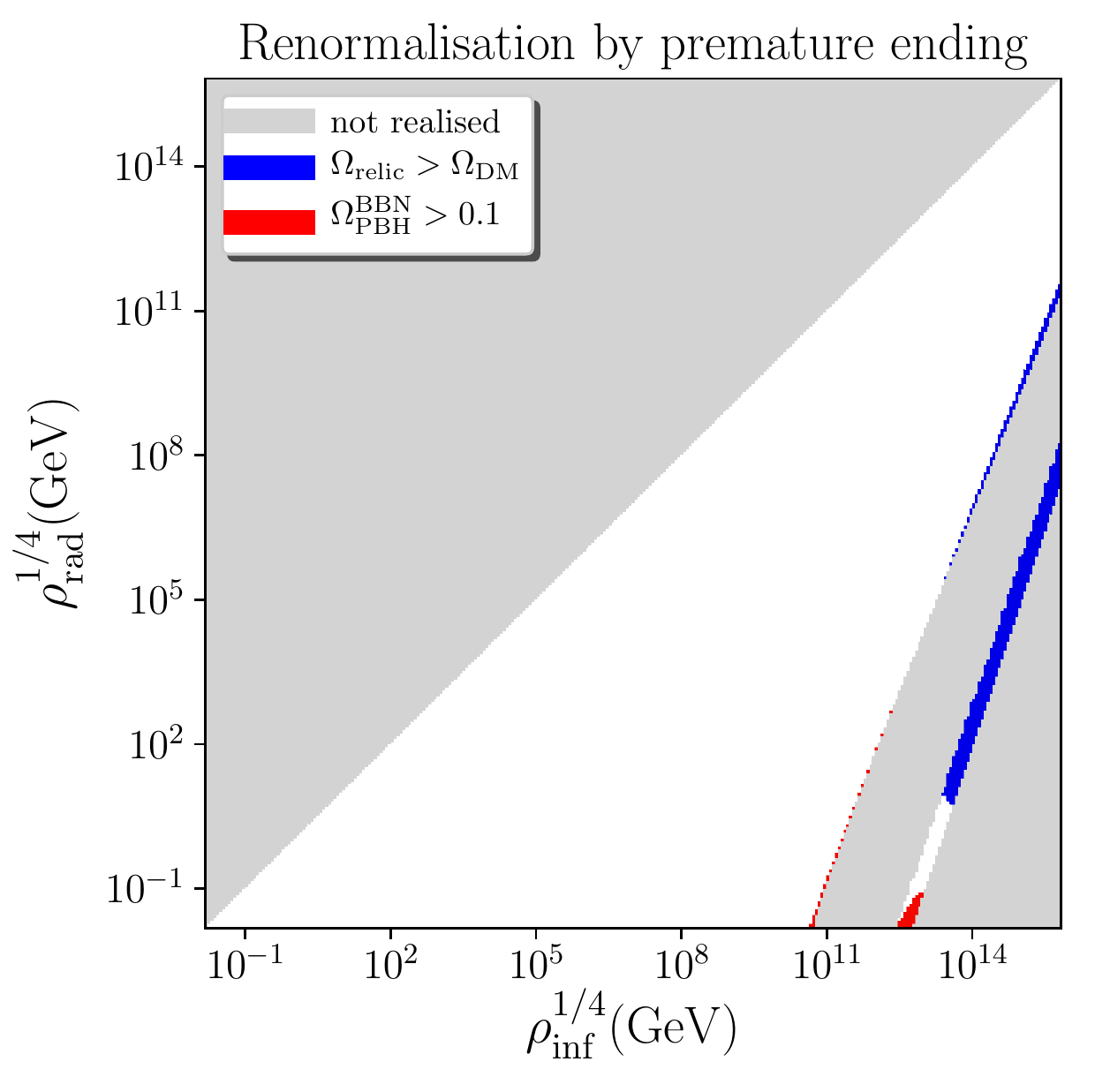}
\caption{Combined constraints in the the space $(\rho_\uinf,
  \rho_\urad)$, when the mass fraction is renormalised by inclusion
  (left panel) and premature ending (right panel). The grey regions
  are excluded since they correspond to values of $\rho_\urad$, the
  energy density at the onset of the radiation epoch, that cannot be
  realised. Red regions are excluded since they yield too large
  abundance of primordial black holes. If black holes leave Planckian
  relics behind after evaporation, the blue regions are also excluded
  since they lead to too many of them. The remaining region, displayed
  in white, is the one allowed.}
\label{fig:combined_constraints}
\end{center}
\end{figure}

A second result concerns the constraints on the energy scale of
inflation and the energy at the onset of the radiation-dominated epoch
that follow from the above-described mechanism. These
combined constraints on the two parameters describing our setup,
either $\rho_\uinf$ and $\rho_\Gamma$ or $\rho_\uinf$ and $\rho_\urad$,
are given in \Fig{fig:combined_constraints:rhoGamma} and
\Fig{fig:combined_constraints} respectively. All coloured regions are
excluded: the grey one since it corresponds to values of $\rho_\Gamma$
and/or $\rho_\urad$ that cannot be realised; the red one since it
leads to an overproduction of PBHs that is excluded by observations;
and, if evaporated black holes leave Planckian relics behind, the blue
one since it yields more relics than the measured abundance of dark
matter. Only the white region remains, which strongly constrains the
energy scale of inflation and reheating. For $\rho_\uinf^{1/4}\simeq
10^{15}\GeV$, if renormalisation is performed by inclusion, values
such that $\rho_\Gamma^{1/4}\lesssim 2\times 10^9\GeV$ and $4.1\times
10^7\GeV\lesssim \rho_\Gamma^{1/4}\lesssim 4.0\times 10^9\GeV$ are
excluded. If renormalisation is performed by premature ending, then
values such that $\rho_\Gamma^{1/4}\lesssim 4.0\times 10^9\GeV$ are
excluded. The constraints on $\rho_\urad$ are also relevant
since, as already mentioned, they correspond to constraints on
the reheating temperature. For $\rho_\uinf^{1/4}\simeq 10^{15}\GeV$,
if renormalisation is performed by inclusion, we find that only values
such that $10^2\MeV\lesssim \rho_\urad^{1/4}\lesssim 6 \GeV$
and $\rho_\urad^{1/4}\gtrsim 4\times10^{9}\GeV$ are allowed. If renormalisation is
performed by premature ending, then only $\rho_\urad^{1/4}\gtrsim
4\times 10^{9}\GeV$ is possible. 

This has very important implications. For
instance, the Starobinsky model and the Higgs inflation models, which are among
the best models of inflation~\cite{Martin:2013tda,Martin:2013nzq} and
yield a tensor-to-scalar ratio of $r\simeq 10^{-3}$, share the
same potential but have different reheating temperatures. More precisely, the Starobinsky model is usually associated with low reheating temperatures (typically $T_\ureh\sim 10^8\GeV$ in supergravity embeddings, see \Ref{Terada:2014uia}), and Higgs inflation with large reheating temperatures such as $T_\ureh\simeq
10^{12}\GeV$~\cite{Bezrukov:2007ep, GarciaBellido:2008ab, Figueroa:2015rqa}, see \eg Fig.~2 in
\Ref{Martin:2016iqo}. Using $\rho_\urad^{1/4}\simeq
(\pi^2g_*/30)^{1/4}T_\ureh$ with $g_*\simeq 1000$, this leads to
$\rho_\urad^{1/4}\simeq 4.2\times 10^8\GeV$ for the Starobinsky model
and $\rho_\urad^{1/4}\simeq 4.2\times 10^{12}\GeV$ for Higgs
inflation. According to the constraints obtained here, the reheating
temperatures typically associated with the Starobinsky model are therefore excluded.

Finally, let us comment on
the robustness of our results. One should note that in the case where PBHs are abundantly produced,
the use of the Press-Schechter formalism, or of the peak theory, might
be questionable since those typically assume PBHs to be rare
events. The precise way in which the mass fraction needs to be
renormalised is also an open question in that case. By considering two
extreme possibilities, \ie black hole inclusion and premature ending
of the instability, we have tried to cover the range of the possible
outcomes from that renormalisation procedure, but it would be clearly
more satisfactory to have a better description of PBHs production in
the dense regime.

Let us also stress that over the course of the present analysis,
conservative assumptions have been made, which tend to underestimate
the predicted abundance of PBHs, in order to make safe the statement
that the coloured regions in \Figs{fig:combined_constraints:rhoGamma}
and~\ref{fig:combined_constraints} are excluded. It would however be
interesting to go beyond these assumptions and make the constraints
even tighter.

Another effect we have neglected is black-hole accretion and
merging. Since the evaporation time of PBHs scales as their masses
cubed, see \Eq{eq:Hawking:time}, accretion and merging make them live
longer and modelling these effects would therefore render our bounds
tighter. This might be of little importance when the abundance of PBHs
is tiny, but it may play a bigger role in the case where PBHs
transiently dominate the universe content. In that case, one may also
expect that substantial amounts of gravitational waves are emitted by
PBH mergers, which provides another channel through which the
preheating instability could be constrained.

Let us also mention that in the analysis of \Apps{sec:ltb} and~\ref{sec:deltacri}, for simplicity, the over-density has been assumed to be initially spherically symmetric. The impact of deviations from spherically symmetric configurations on the production of PBHs has been studied in dust-like environment, \eg in \Ref{Harada:2016mhb}, where it has been shown that it can play an important role, and substantially decrease the abundance of the black holes. One may expect similar effects for black holes forming from a massive scalar field inhomogeneity~\cite{Hidalgo:2017dfp, Carr:2018nkm}, although the situation is different\footnote{A massive scalar field differs from a pressure-less perfect fluid in different ways~\cite{Pinol:2019}: at the background level, the equation-of-state parameter oscillates, and vanishes only when averaged over time, and at the perturbative level, the scalar field perturbations have a different dynamics than those of a perfect fluid, as revealed for instance by the fact that the density contrast grows proportionally with the scale factor only inside the instability band for a scalar field, while it takes place at all scales for a perfect fluid. This explains why standard techniques investigating the formation of PBHs in a pressure-less dustlike universe cannot be directly applied to the situation discussed in this paper.} and techniques developed to track perfect fluid inhomegeneities cannot be directly applied here. One would thus have to generalise the analysis of \Apps{sec:ltb} and~\ref{sec:deltacri} to non-spherically symmetric configurations, which may require a numerical approach and which we leave for future work. This may lead to less black holes than what we have estimated in the present work, although it is worth mentioning that non-spherically symmetric configurations should also provide spins to the resulting black holes. Since the Hawking evaporation time depends on the spin~\cite{Unruh:1974bw}, PBHs forming from non-spherically symmetric over-densities have different life times, and are affected by observational constraints differently.

It is also worth stressing that the preheating instability has here
been discussed in the context of a quadratic potential, since most
inflationary potentials are quadratic close to their minimum, but it
also takes place for quartic potentials~\cite{Jedamzik:2010dq}. In
that case, the instability is even more pronounced, but it is
restricted to a narrower range of modes, and it would be
interesting to study its consequences for PBH formation.

Finally, let us mention that CMB predictions are also affected by our results. As explained in \Sec{sec:intro}, for a fixed inflationary single-field
potential, the only theoretical uncertainty in observational
predictions is on the number of e-folds elapsed between the time when
the CMB pivot scale exits the Hubble radius and the end of
inflation. This number depends~\cite{Martin:2006rs} on the energy
scale of inflation, which is given by the inflationary model under
consideration, the energy density at which the radiation era starts,
and the averaged equation-of-state parameter between the end of
inflation and that time. By restricting these values, the present work
allows one to make inflationary predictions more focused, and this
will be the topic of a separate article.

\begin{acknowledgments}
T.~P. acknowledges support from the Fondation CFM pour la
Recherche in France, the Alexander S. Onassis
Public Benefit Foundation in Greece, the Foundation for Education
and European Culture in Greece and the A.G. Leventis
Foundation. V.~V. acknowledges funding from the European Union's
Horizon 2020 research and innovation programme under the Marie Sk\l
odowska-Curie grant agreement N${}^0$ 750491.
\end{acknowledgments}
%
\appendix

\section{Black holes formation from scalar field collapse}
\label{sec:ltb}
In this first appendix, we review (and correct a few typos in the work of) \Ref{Goncalves:2000nz}, that studies black hole formation from massive scalar field collapse. Let us consider an inhomogeneous massive scalar field $\phi(t,r)$
living in an inhomogeneous but isotropic (spherically symmetric) space
time endowed with the metric
\begin{align}
\label{eq:metric}
\dd s^2=-\dd t^2+e^{-2 \Lambda(t,r)}\dd r^2 +R^2(t,r)\left(\dd \theta^2
+\sin ^2 \theta \dd \varphi^2\right).
\end{align}
In order to follow the evolution of the scalar field, we must solve
the corresponding Einstein equations $G_{\mu \nu}=T_{\mu \nu}/\Mp^2$
where
$T_{\mu \nu}=\partial_{\mu }\phi \partial_{\nu }\phi-\frac12g_{\mu
  \nu}\left(g^{\alpha \beta}\partial_{\alpha
  }\phi \partial_{\beta}\phi+m_\phi^2 \phi^2\right)$
is the stress energy tensor of the scalar field ($m_\phi$ is the mass of
the field) and the Klein-Gordon equation
$\left(g^{\mu \nu}\nabla_{\mu }\nabla_{\nu}-m^2_\phi\right)\phi=0$. This
last equation takes the following form
\begin{align}
  \label{eq:kg}
  \ddot{\phi}-e^{2\Lambda}\phi''+\left(2\frac{\dot{R}}{R}-\dot{\Lambda}\right)
  \dot{\phi}-e^{2\Lambda}\left(2\frac{R'}{R}+\Lambda'\right)\phi'
  +m_\phi^2\phi=0,
\end{align}
where a dot denotes a derivative with respect to time and a prime a
derivative with respect to the radial coordinate $r$. This
Klein-Gordon equation should be compared to Eq.~(7) of
\Ref{Goncalves:2000nz}. The two formula are nearly identical but there
are sign differences. As can be seen on the above expression, in
Eq.~(7) of \Ref{Goncalves:2000nz}, $2\dot{R}/R+\dot{\Lambda}$ should
read $2\dot{R}/R-\dot{\Lambda}$ and $-2R'/R+\Lambda '$ should read
$2R'/R+\Lambda '$.

Then, the components of the Einstein tensor are given by
\begin{align}
G_{tt} &= \frac{1}{R^2}\left[1+\dot{R}^2-2\dot{\Lambda}\dot{R}R
-Re^{2\Lambda}\left(2\Lambda'R'+2R''+\frac{R'^2}{R}\right)\right],
\\
G_{tr} &=-\frac{2}{R}\left(\dot{R}'+\dot{\Lambda}R'\right),
\\
G_{rr} &= \frac{1}{R^2}\left[R'^2-e^{-2\Lambda}\left(\dot{R}^2
+2R\ddot{R}+1\right)\right], 
\\
G_{\theta \theta} & =\sin^{-2}\theta \, G_{\varphi \varphi}
=R\left(\dot{R}\dot{\Lambda}+\Lambda' R' e^{2\Lambda}
+R''e^{2\Lambda}-\ddot{R}+\ddot{\Lambda}R-R\dot{\Lambda}^2\right).
\end{align}
These equations exactly correspond to Eqs.~(2)-(5) of
\Ref{Goncalves:2000nz}. On the other hand, the components of the
stress-energy tensor can be expressed as
\begin{align}
\label{eq:stresstt}
T_{tt} &= \frac12 \dot{\phi}^2+\frac12 e^{2\Lambda}\phi'^2
+\frac12 m_\phi^2\phi^2, 
\\
\label{eq:stressrt}
T_{rt}&= \dot{\phi}\phi', 
\\
\label{eq:stressrr}
T_{rr} &=\frac12 e^{-2\Lambda}\dot{\phi}^2+\frac12 \phi'^2-
\frac12 e^{-2\Lambda}m_\phi^2\phi^2,
\\
\label{eq:stressthetatheta}
T_{\theta \theta}&=\sin ^{-2}\theta \, T_{\varphi \varphi}=\frac{R^2}{2}
\left(\dot{\phi}^2-e^{2\Lambda}\phi'^2-m_\phi^2\phi^2\right).
\end{align}
These formulas are identical to Eqs.~(9)-(12) in
\Ref{Goncalves:2000nz}. 

Having the components of the Einstein and stress-energy tensors, we are 
in a position to write down Einstein equations. However, these ones can 
be greatly simplified by introducing two auxiliary functions $k(t,r)$
and $m(t,r)$ defined by the following relations
\begin{align}
  \label{eq:defk&m}
k(t,r)=1-R'^2 e^{2\Lambda}, \quad m(t,r)
=\frac{R}{2}\left(\dot{R}^2+k\right).
\end{align}
These definitions correspond to Eqs.~(13) and (14) in
\Ref{Goncalves:2000nz}. Notice, however, the misprint in Eq.~(13)
where the factor $R'^2$ in front of the term $e^{2\Lambda}$ is
absent. Then, Einstein equations take the form
\begin{align}
\label{eq:15}
k'&= 8\pi RR'\left(T_{tt}+T^r{}_r\right)+2R'\left(\ddot{R}+
\dot{\Lambda}\dot{R}\right), \\
\label{eq:16}
\dot{k}&=8\pi RR'T^r{}_t, \\
\label{eq:17}
m'&= 4\pi R^2R'T_{tt}-4\pi R^2\dot{R}T_{rt}, \\
\label{eq:18}
\dot{m}&=4\pi R^2R'T^r{}_t-4\pi \dot{R}R^2T^r{}_r.
\end{align}
Notice that, in order to compare our results to
\Ref{Goncalves:2000nz}, we have used $\Mp^{-2}=8\pi G$ with $G=1$ ($G$
is the Newton constant). The above equations are Eqs.~(15)-(18) of
\Ref{Goncalves:2000nz} and agree with our results, except Eq.~(15) for
which the sign of the right-hand sign is incorrect.

Despite their apparent simplicity, the above equations remain
difficult to solve. As discussed in \Ref{Goncalves:2000nz}, they can
nevertheless be solved by expanding the scalar field in inverse powers
of its mass. For this purpose we write
\begin{align}
\label{eq:defcapphi}
\phi(t,r)=\frac{1}{m_\phi}\Phi(t,r)\cos \left(m_\phi t\right).
\end{align}
Then we insert this expression in Eqs.~(\ref{eq:stresstt}),
(\ref{eq:stressrt}), (\ref{eq:stressrr})
and~(\ref{eq:stressthetatheta}). This leads to
\begin{align}
\label{eq:stressttmod}
T_{tt} & \kern-0.1em = \kern-0.1em \frac12 \Phi^2-\frac{1}{2m_\phi}\Phi 
\dot{\Phi}\sin \left(2m_\phi t\right)
+\frac{1}{4m_\phi^2}\left(\dot{\Phi}^2+e^{2\Lambda}\Phi'^2\right)
\left[1+\cos\left(2 m_\phi t\right)\right],
\\
\label{eq:stressrtmod}
T_{rt} &  \kern-0.1em = \kern-0.1em \frac{1}{2m_\phi ^2}\dot{\Phi}\Phi'
\left[1+\cos \left(2m_\phi t\right)\right]
-\frac{1}{2m_\phi}\Phi \Phi'\sin\left(2m_\phi t\right),
\\
\label{eq:stressrrmod}
T^r{}_r &  \kern-0.1em = \kern-0.1em -\frac{1}{2} \Phi^2 \cos\left(2m_\phi t\right)
-\frac{1}{2m_\phi}
\Phi \dot{\Phi}\sin \left(2 m_\phi t\right)+
\frac{1}{4m_\phi^2}\left(\dot{\Phi}^2+e^{2\Lambda}\Phi'^2\right)
\left[1+\cos\left(2 m_\phi t\right)\right],
\\
\label{eq:stressthetathetamod}
T_{\theta \theta}&  \kern-0.1em = \kern-0.1em \frac{R^2}{2}\left\{  \kern-0.1em  \frac{1}{2m_\phi^2}\left(\dot{\Phi}^2-e^{2\Lambda}\Phi'^2\right)
\left[1+\cos\left(2 m_\phi t\right)\right] -\Phi^2\cos\left(2m_\phi t\right)
-\frac{1}{m_\phi}\Phi \dot{\Phi}\sin \left(2m_\phi t\right) \right\}.
\end{align}
These equations correspond to Eqs.~(21)-(24) in
\Ref{Goncalves:2000nz}.  We notice that Eq.~(\ref{eq:stressttmod})
differs from Eq.~(21) for two reasons: firstly, our third term is
proportional to $m_\phi^{-2}$, while Eq.~(21) in \Ref{Goncalves:2000nz}
does not contain this factor, and, secondly, our term
$\dot{\Phi}^2+e^{2\Lambda}\Phi'^2$ reads
$\dot{\Phi}^2+e^{4\Lambda}\Phi'^2$ in \Ref{Goncalves:2000nz}. Since $\Phi$ has dimension two,
see \Eq{eq:defcapphi}, it is clear that the
$m_\phi^{-2}$ factor must be present in that term in order for the equation
to be dimensionally correct. On the other hand, \Eq{eq:stressrtmod}
coincides with Eq.~(22) in
\Ref{Goncalves:2000nz}. \Eq{eq:stressrrmod}, however, is again
different from Eq.~(23) in \Ref{Goncalves:2000nz}, exactly for the
same reasons as Eq.~(21) differs from our
Eq.~(\ref{eq:stressttmod}). Finally, \Eq{eq:stressthetathetamod} is
also different from Eq.~(24) of \Ref{Goncalves:2000nz}: our term
$e^{2\Lambda}$ reads $e^{4\Lambda}$ in that paper.

The Einstein equations and the Klein-Gordon equation are non-linear
partial differential equations and, therefore, are complicated to
solve. Following \Ref{Goncalves:2000nz}, it is useful to perform an
expansion in inverse powers of the mass $m_\phi$ for the field and the free
functions appearing in the metric tensor. Concretely, one writes
\begin{align}
\Phi(t,r)&=\Phi_0(t,r)+\sum_{i=1}^{+\infty}\sum_{j=1}^{+\infty}
\frac{1}{m_\phi^i}\left[\Phi_{ij}^\mathrm{c}\cos\left(jm_\phi t\right)
+\Phi_{ij}^\mathrm{s}\sin\left(jm_\phi t\right)\right], 
\\
k(t,r)&=k_0(t,r)+\sum_{i=1}^{+\infty}\sum_{j=1}^{+\infty}
\frac{1}{m_\phi^i}\left[k_{ij}^\mathrm{c}\cos\left(jm_\phi t\right)
+k_{ij}^\mathrm{s}\sin\left(jm_\phi t\right)\right],
\\
m(t,r)&=m_0(t,r)+\sum_{i=1}^{+\infty}\sum_{j=1}^{+\infty}
\frac{1}{m_\phi^i}\left[m_{ij}^\mathrm{c}\cos\left(jm_\phi t\right)
+m_{ij}^\mathrm{s}\sin\left(jm_\phi t\right)\right],
\\
R(t,r)&=R_0(t,r)+\sum_{i=1}^{+\infty}\sum_{j=1}^{+\infty}
\frac{1}{m_\phi^i}\left[R_{ij}^\mathrm{c}\cos\left(jm_\phi t\right)
+R_{ij}^\mathrm{s}\sin\left(jm_\phi t\right)\right].
\end{align}
Inserting these expansions into the equations of motion leads, at
leading order (namely order $m_\phi^0$ for the equations of motion),
to the following expressions (restricting ourselves to a sum from
$j=1$ to $j=2$ which, for the leading order, will be fully justified
below): \Eq{eq:16} implies that
$k_{11}^\mathrm{c}=k_{12}^\mathrm{c}=k_{11}^\mathrm{s}=k_{12}^\mathrm{s}=0$
and $\dot{k}_0=0$. The definition of $m(t,r)$ in \Eqs{eq:defk&m}
reduces to
$R_{11}^\mathrm{c}=R_{12}^\mathrm{c}=R_{11}^\mathrm{s}=R_{12}^\mathrm{s}=0$
and $-R_0k_0/2+m_0-R_0\dot{R}_0^2/2=0$. \Eq{eq:18} leads to
$\dot{m}_0=0$,
$m_{11}^\mathrm{c}=m_{21}^\mathrm{c}=m_{11}^\mathrm{s}=0$ and
$m_{12}^\mathrm{s}=\pi R_0^2\dot{R}_0\Phi_0^2$. Notice that this last
formula coincides with Eq.~(35) of \Ref{Goncalves:2000nz}. The
Klein-Gordon equation~(\ref{eq:kg}) implies that
$-R_0'{}^3(1-k_0)(2\phi_0\dot{R}_0+2\dot{\phi}_0R_0+\Phi_0R_0\dot{R}_0'/R_0')=0$
and $\Phi_{12}^\mathrm{s}=0$. Finally \Eq{eq:17} reduces to $m_0'=2\pi
R_0^2R_0'\Phi_0^2=0$, which is also Eq.~(33) of
\Ref{Goncalves:2000nz}. Notice that by time differentiating the last
expression of $m_0'$, leading to zero since we have already shown that
$\dot{m}_0=0$, one demonstrates that the expression obtained
before, namely
$-R_0'{}^3(1-k_0)(2\phi_0\dot{R}_0+2\dot{\phi}_0R_0+\Phi_0R_0\dot{R}_0'/R_0')=0$,
is identically satisfied and, therefore, does not lead to additional
constraints. We wee that, at leading order, it is consistent to assume
that all coefficients of the above expansions vanish but
$m_{12}^\mathrm{s}$. This means that, at leading order, the solution
to the Einstein equations reads
\begin{align}
\Phi(t,r)&=\Phi_0(t,r)+{\cal O}\left(m_\phi^{-2}\right), 
\\
m(t,r)&=m_0(t,r)+\frac{1}{m_\phi}m_{12}^\mathrm{s} \sin\left(2m_\phi t\right)
+{\cal O}\left(m_\phi^{-2}\right),
\\
k(t,r)&=k_0(t,r)
+{\cal O}\left(m_\phi^{-2}\right),
\\
R(t,r)&=R_0(t,r)
+{\cal O}\left(m_\phi^{-2}\right).
\end{align}

In \Ref{Goncalves:2000nz}, it is claimed that one can go to next-to-leading order (namely order $m_\phi^{-1}$ for the equations of
motion), the solution at this order being given by the following
expressions
\begin{align}
\Phi(t,r)&=\Phi_0(t,r)+{\cal O}\left(m_\phi^{-3}\right), 
\\
m(t,r)&=m_0(t,r)+\frac{1}{m_\phi}m_{12}^\mathrm{s} \sin\left(2m_\phi t\right)
+\frac{1}{m_\phi^2}m_{22}^\mathrm{c} \cos\left(2m_\phi t\right)
+{\cal O}\left(m_\phi^{-3}\right),
\\
k(t,r)&=k_0(t,r)
+\frac{1}{m_\phi^2}k_{22}^\mathrm{c} \cos\left(2m_\phi t\right)
+{\cal O}\left(m_\phi^{-3}\right),
\\
R(t,r)&=R_0(t,r)
+\frac{1}{m_\phi^2}R_{22}^\mathrm{c} \cos\left(2m_\phi t\right)
+{\cal O}\left(m_\phi^{-3}\right).
\end{align}
However, if one repeats the above analysis, one finds the
following. At next-to-leading order, \Eq{eq:16} implies that
$k_{21}^\mathrm{c}=k_{21}^\mathrm{s}=k_{22}^\mathrm{s}=0$ and
$k_{22}^\mathrm{c}=2\pi R_0(1-k_0)\Phi_0\Phi_0'/R_0'$. This last
formula is identical to Eq.~(34) of \Ref{Goncalves:2000nz} [In
  Eq.~(34), there is a misprint: $R_0'/R_0$ should read $R_0/R_0'$]. At
next-to-leading order, the definition of $m(t,r)$ in \Eqs{eq:defk&m}
reduces to $m_{11}^\mathrm{c}=R_0\dot{R}_0R_{21}^\mathrm{s}$,
$m_{11}^\mathrm{s}=-R_0\dot{R}_0R_{21}^\mathrm{c}$,
$m_{12}^\mathrm{c}=2R_0\dot{R}_0R_{22}^\mathrm{s}$ which, given what
has been established at leading order, implies that
$R_{21}^\mathrm{c}=R_{21}^\mathrm{s}=R_{22}^\mathrm{s}=0$. Moreover,
one also has $m_{12}^\mathrm{s}=-2R_0\dot{R}_0R_{22}^\mathrm{c}$,
which, given that $m_{12}^\mathrm{s}$ has already been determined,
implies that $R_{22}^\mathrm{c}=-\pi R_0\Phi_0^2/2$ in accordance with
Eq.~(36) of \Ref{Goncalves:2000nz}. Let us now turn to
\Eq{eq:18}. This leads to
$\Phi_{11}^\mathrm{c}=\Phi_{11}^\mathrm{s}=\Phi_{12}^\mathrm{c}=0$ and
$m_{21}^\mathrm{c}=m_{21}^\mathrm{s}=0$, $m_{22}^\mathrm{c}=\pi
R_0^2\Phi_0\Phi_0'/R_0'$. One also obtains an equation for the
derivative of $m_{12}^\mathrm{s}$, namely
$\dot{m}_{12}^\mathrm{s}=2m_{22}^\mathrm{c}-2\pi
R_0^2\Phi_0\Phi_0'/R_0'+2\pi R_0^2k_0\Phi_0\Phi_0'/R_0'+2\pi
R_0^2\dot{R}_0\Phi_0\dot{\Phi}_0$, and an equation for
$R_{22}^\mathrm{c}$ that reads
$R_{22}^\mathrm{c}=2\dot{R}_0\Phi_{12}^\mathrm{s}$. But we have seen
that the Klein-Gordon equation at leading order implies
$\Phi_{12}^\mathrm{s}=0$ and, therefore, $R_{22}^\mathrm{c}=0$. This
result is inconsistent with the result established above, namely
$R_{22}^\mathrm{c}=-\pi R_0\Phi_0^2/2$. We interpret this
inconsistency as an indication that, if one works at next-to-leading
order, it is impossible to truncate the expansions of $\Phi$, $R$, $k$
and $m$ to second harmonics. Since this is what was done in
\Ref{Goncalves:2000nz}, we conclude that the next-to-leading order
solution presented in this article is not correct. In the present
article, we therefore restrict ourselves to the leading order.

It follows from the previous considerations that, as long as the above
perturbative solution remains valid, the metric
tensor~(\ref{eq:metric}) takes the form
\begin{align}
\label{eq:tb}
  \dd s^2\simeq -\dd t^2+\frac{R_0^2{}'(t,r)}{1-k_0(r)}\dd r^2
+R_0^2(t,r)\dd \Omega ^2,
\end{align}
where
\begin{align}
\label{eq:fried}
\frac{\dot{R_0}^2(t,r)}{R_0^2(t,r)}&=\frac{2m_0(r)}{R_0^3(t,r)}
-\frac{k_0(r)}{R_0^2(t,r)},
\\
\label{eq:mass}
\frac{\dd m_0(r)}{\dd r}&=4\pi \frac{\Phi_0^2}{2}R_0^2R_0'.
\end{align}
One recognises the Tolman-Bondi solution which corresponds to an
inhomogeneous solution of the Einstein equations for a pressureless
fluid. The corresponding energy density is given by $\Phi_0^2/2$ which
is consistent since, at leading order,
$T_{tt}=\rho(t,r)=\Phi_0^2/2+{\cal O}(m_\phi^{-1})$. Therefore, we
reach the conclusion that, as long as the above described
approximation is valid, a scalar field overdensity behaves as the one of a pressureless fluid
and, as a consequence, unavoidably evolves into a black hole. This
solution is the equivalent for a scalar field of the spherical
collapse model and allows us to follow the evolution of the system
beyond the perturbative regime.

Let us now study how an overdensity made of scalar field can proceed
to a black hole. For convenience, in the following, we write
$\rho(t,r)$ as
\begin{align}
  \label{eq:defcontrast}
  \rho(t,r)=\rho_{\mathrm b}(t)\left[1+\Delta(t,r)\right]
  =\rho_{\mathrm b}(t)\left[1+\delta(t,r)\Theta(r_{\mathrm{c}}-r)\right],
\end{align}
where $\rho_\mathrm{b}(t)$ represents the homogeneous background
energy density outside the overdensity and $\Delta
(t,r)=[\rho(t,r)-\rho_{\mathrm{b}}]/\rho_\mathrm{b}$ the density
contrast. The quantity $r_{\mathrm{c}}$ represents the comoving radius
of the overdensity and $\delta(t,r)$ is its profile. Notice that we
need to know $\delta(t,r)$ only for $r<r_\mathrm{c}$ since this term
does not contribute to $\rho(t,r)$ outside the overdensity, thanks to
the Heaviside function $\Theta(r_{\mathrm{c}}-r)$. The line
element~(\ref{eq:tb}) describes the evolution of spherical dust shells
labelled by $r$. Notice that $r$ is a comoving radial coordinate and
that each shell has surface area $4\pi R^2_0(t,r)$. As a consequence,
the total mass $M$ of the overdensity is given by
\begin{align}
  M=\int _0^{r_\uc}\frac{\dd m_0(r)}{\dd r}=
  \int _0^{r_\uc} \rho(t,r) 4\pi R_0^2 \dd R_0
  =\int _0^{r_\uc} \rho(t,r) 4\pi R_0^2 R_0' \dd r.
  \end{align}
The conservation equation,
$\dot{\rho}+(\dot{R}_0'/R_0'+2\dot{R}_0/R_0)\rho=0$, guarantees that
this mass is conserved, namely $\dot{M}=0$.

To proceed further and study the dynamics of the collapse, we need to
choose initial conditions, in particular the initial profile for the
overdensity. At this stage, let us recall that there is a gauge
freedom that can be be fixed by using the gauge condition
$R_0(t_\mathrm{ini},r)=r$. This condition will be used in the rest of
these appendices. Once the initial conditions have been chosen, one can
calculate the behaviour of the functions characterising the model. In
particular, using \Eq{eq:mass}, the function $m_0(r)$ can be expressed
as
\begin{align}
m_0(r)=\frac{4\pi}{3}\rho_\mathrm{b}(t_\mathrm{ini})r^3\left[1
  +\frac{3}{r^3}\int _0^r\Delta (t_\mathrm{ini},x) x^2\dd x\right]
=\frac{4\pi}{3}\rho_\mathrm{b}(t_\mathrm{ini})r^3\left[1
  +\left\langle \Delta(t_\uini,r)\right \rangle\right].
\end{align}
Of course, many different choices for the initial density profile are
a priori possible. The important point is that, once a choice is made,
the function $m_0(r)$ is uniquely specified thanks to the above
equation (explicit examples are given below). The mass of the
overdensity is nothing but $M=m_0(r_\uc)$, which implies that the
function $m_0(r)$ can also be rewritten as
\begin{align}
  m_0(r)=M\left(\frac{r}{r_\uc}\right)^3\frac{1
  +\left\langle \Delta(t_\uini,r)\right \rangle}{1
  +\left\langle \Delta(t_\uini,r_\uc)\right \rangle}\, .
\end{align}

Another initial data that needs to be provided is the value of
$\dot{R}_0(t_\mathrm{ini},r)$. For this purpose, we define the
``inhomogeneous'' Hubble parameter by
\begin{align}
H(t,r)\equiv \frac{\dot{R}_0(t,r)}{R_0(t,r)}.
\end{align}
Then, one just needs to provide the function
$H(t_\mathrm{ini},r)\equiv H_\mathrm{ini}$. A natural choice is to
simply assume that the initial value of $H(t,r)$ is determined by the
initial background energy density (and, therefore, does not depend on
$r$), that is to say
\begin{align}
\label{eq:Hini}
H^2(t_\mathrm{ini},r)\equiv H^2_\mathrm{ini}
=\frac{8\pi}{3}\rho_\mathrm{b}(t_\mathrm{ini})
=\frac{2M}{r_\uc^3}\frac{1}{1
  +\left\langle \Delta(t_\uini,r_\uc)\right \rangle}
.
\end{align}

Finally, $k_0(r)$ remains to be calculated. In order to concretely
determine this function, one needs to integrate \Eq{eq:fried}. This
can be easily done and the solution reads
\begin{align}
\label{eq:radius}
R_0(\eta,r) &=\frac{2m_0(r)}{k_0(r)}\cos ^2\frac{\eta}{2}, \\
\label{eq:time}
t_0(\eta,r) &= t_\mathrm{BB}(r)+\frac{m_0(r)}{k^{3/2}_0(r)}
\left(\eta+\sin \eta\right),
\end{align}
where $\eta$ is, a priori, a parameter in the range $[-\pi,\pi]$ [not to be confused with the conformal time introduced below \Eq{eq:MS}]. The
radial dependent integration constant $t_\mathrm{BB}(r)$ is usually
called the big-bang time function since, in a cosmological context, it
allows for inhomogeneous Big Bangs. Using the gauge condition
$R_0(\eta_\mathrm{ini},r)=r$, \Eq{eq:radius} implies that
\begin{align}
\label{eq:kinter}
k_0(r)=\frac{2m_0(r)}{r}\left[1-\sin^2\left(
\frac{\eta_\mathrm{ini}}{2}\right)\right].
\end{align}
However, $\sin(\eta_\mathrm{ini}/2)$ remains to be found. In fact, it
can be evaluated in terms of $H(t_\mathrm{ini},r)$. Indeed, from the
above parametric solution~(\ref{eq:radius})-(\ref{eq:time}), the Hubble parameter reads
\begin{align}
\label{eq:Hubble}
H(t,r)=\frac{k^{3/2}_0(r)}{2m_0(r)}\frac{\sin(\eta/2)}{\cos^3(\eta/2)}.
\end{align}
Then, using the expression of $k_0$ already derived above, namely
$k_0(r)=2m_0(r)\cos^2(\eta_\mathrm{ini}/2)/r$, one has
\begin{align}
  H^2(t_\mathrm{ini},r)=\frac{2m_0(r)}{r^3}\sin^2\left(
  \frac{\eta_\mathrm{ini}}{2}\right).
\end{align}
Inserting this formula back into \Eq{eq:kinter}, one finally
obtains
\begin{align}
  \label{eq:k0model}
  k_0(r)=\frac{2m_0(r)}{r}-r^2H^2(t_\mathrm{ini},r)=M\frac{2
  \left\langle \Delta(t_\uini,r)\right \rangle}{1
  +\left\langle \Delta(t_\uini,r_\uc)\right \rangle}\frac{r^2}{r_\uc^3},
\end{align}
where one has used \Eq{eq:Hini}. Everything is now known and,
therefore, from the knowledge of the initial density profile, we have
completely characterised the model, in particular the functions
$m_0(r)$ and $k_0(r)$.

\section{Calculation of the critical density contrast}
\label{sec:deltacri}

We now focus on the fate of the overdensity and, as a
consequence, we restrict ourselves to $r\leq r_\mathrm{c}$. One can
re-write the parametric solution using the expression of $m_0(r)$ and
$k_0(r)$ that we have established. Inside the overdensity, namely for
$r\leq r_\mathrm{c}$, one finds
\begin{align}
\label{eq:radius2}
\frac{R_0(\eta,r)}{r} &=\frac{1+\left\langle \Delta(t_\uini,r)\right \rangle}
     {\left\langle \Delta(t_\uini,r)\right \rangle}
\cos ^2\frac{\eta}{2}, \\
\label{eq:time2}
t_0(\eta,r) &= t_\mathrm{BB}(r)+\frac{1}{2H_{\rm ini}}
\frac{1+\left\langle \Delta(t_\uini,r)\right \rangle}
     {\left\langle \Delta(t_\uini,r)\right \rangle^{3/2}}
\left(\eta+\sin \eta\right).
\end{align}
Let us now discuss the initial condition for this model. We start from
a value of $R_0(t,r)$ which is non-vanishing but in the linear
regime. The wavelength of the Fourier mode under consideration is
related to the radius of the overdensity by
$R_0(\eta_\mathrm{ini},r_\mathrm{c})=r_\mathrm{c}=\lambda$. The value of
$\eta_\mathrm{ini}$ depends on $\left\langle
\Delta(t_\uini,r)\right \rangle$ since using \Eq{eq:radius2}
together with the gauge condition, one has
\begin{align}
\label{eq:initialtime}
\sin^2\left(\frac{\eta_\mathrm{ini}}{2}\right)=
\frac{1}{1+\left\langle \Delta(t_\uini,r)\right \rangle}.
\end{align}
This expression implies that $\eta_\uini$ is a function of the radial
coordinate $r$. We notice that, if we change the value of
$\left\langle \Delta(t_\uini,r)\right \rangle$, then we change the
initial value of the parameter $\eta_\uini$. However, one can always
ensure that $t_\mathrm{ini}=0$ by properly choosing the big-bang function
$t_\mathrm{BB}(r)$, concretely
\begin{align}
\label{eq:bbtime}
t_\mathrm{BB}(r)=-\frac{1}{2H_{\rm ini}}
\frac{1+\left\langle \Delta(t_\uini,r)\right \rangle}
     {\left\langle \Delta(t_\uini,r)\right \rangle^{3/2}}
\left(\eta_\uini+\sin \eta_\uini\right).
\end{align}
The fact that $t_\mathrm{BB}$ can depend on $r$ plays an important
role and allows us to treat a situation where $\eta_\uini$ is itself
dependent on $r$. In the present context, $t_\mathrm{ini}$ is in fact
the band crossing (bc) time. So times calculated in this way should in
fact be interpreted as $t-t_\mathrm{bc}$. The question is now which
values of $\left\langle \Delta(t_\uini,r)\right \rangle$ lead to black
hole formation.  There are in fact two conditions for black hole
formation: first, the approximation leading to a Tolman-Bondi solution
should be valid until the spherical overdensity becomes smaller than
the Schwarzschild horizon and, second, this should happen before the
inflaton decay. This last condition can easily be worked out. Using
\Eqs{eq:time2} and~\eqref{eq:bbtime}, one obtains that the time at which
black hole formation occurs is given by
\begin{align}
t_\mathrm{coll}-t_\mathrm{bc}=
\frac{1}{2H_\mathrm{bc}}
\frac{1+\left\langle \Delta(t_\uini,r)\right \rangle}
     {\left\langle \Delta(t_\uini,r)\right \rangle^{3/2}}
\left(\pi-\eta_{\rm ini}-\sin \eta_{\rm ini}\right),
\end{align}
with, using \Eq{eq:initialtime},
\begin{align}
\eta_\mathrm{ini}=-2\, \mathrm{arcsin}
\left(\frac{1}{\sqrt{1+\left\langle \Delta(t_\uini,r)\right \rangle}}\right).
\end{align}
Expanding $t_\mathrm{coll}-t_\mathrm{bc}$ in terms of
$\left\langle \Delta(t_\uini,r)\right \rangle$, one finds
\begin{align}
  \label{eq:collapsetime}
t_\mathrm{coll}-t_\mathrm{bc}=\frac{1}{2H_\mathrm{bc}}
\left\{\frac{2\pi}{\left\langle \Delta(t_\uini,r)\right \rangle^{3/2}}
  +\frac{\pi}{\left\langle \Delta(t_\uini,r)\right \rangle}
  -\frac43
  +{\cal O}\left[\left\langle \Delta(t_\uini,r)
    \right \rangle^{1/2}\right]\right\}.
\end{align}
On the other hand, since $a\propto t^{2/3}$ during the
phase where the scalar field oscillates around its quadratic minimum, cosmic time at the end of the instability phase is given by
\begin{align}
t_\mathrm{instab}-t_\mathrm{bc}=\frac{2}{3H_\mathrm{bc}}
\left[e^{3(N_\mathrm{instab}-N_\mathrm{bc})/2}-1\right].
\end{align}
Then, the requirement that black hole formation occurs before the end of the instability phase implies that $t_\mathrm{instab}-t_\mathrm{bc}>t_\mathrm{coll}-t_\mathrm{bc}$,
which amounts to a lower bound on the initial value of the density
contrast, namely
\begin{align}
  \label{eq:criterion}
  \left\langle \Delta(t_\uini,r)\right \rangle
  >\delta_\mathrm{c}\equiv \left(\frac{3\pi}{2}\right)^{2/3}
\left[e^{3(N_\mathrm{instab}-N_\mathrm{bc})/2}-1\right]^{-2/3}\, .
\end{align}
One checks that in the absence of an instability phase, namely when
$N_\mathrm{instab}=N_\mathrm{bc}$, the initial overdensity should be
infinite.

Let us now see how the criterion~(\ref{eq:criterion}) depends on the
profile of the overdensity. The first example we consider, most
certainly the simplest one, is such that
$\Delta(t_\uini,r)=\delta_\uini\Theta(r_\uc-r)$, namely a top hat
profile. In that case, it is straighforward to show that $\left\langle
\Delta(t_\uini,r)\right \rangle=\delta_\uini$. Moreover, it is also
easy to show that, for $r<r_\mathrm{c}$,
\begin{align}
m_0(r)=M\left(\frac{r}{r_\mathrm{c}}\right)^3,
\end{align}
while, for $r>r_\mathrm{c}$,
\begin{align}
m_0(r)=M+\frac{M}{1+\delta_\mathrm{ini}}\left(\frac{r^3}
{r_\mathrm{c}^3}-1\right).
\end{align}
The function $m_0(r)$ is continuous everywhere but its derivative is
discontinuous at the boundary of the overdensity. On the other hand,
the function $k_0(r)$ is obtained from \Eq{eq:k0model} and one
obtains
\begin{align}
  \label{eq:k0inside1}
  k_0(r)=2M\frac{\delta_\mathrm{ini}}{1+\delta_\mathrm{ini}}
  \frac{r^2}{r_\mathrm{c}^3},
\end{align}
if $r<r_\mathrm{c}$ and, if $r>r_\mathrm{c}$, one has
\begin{align}
  \label{eq:k0outside1}
k_0(r)=2M\frac{\delta_\mathrm{ini}}{1+\delta_\mathrm{ini}}\frac{1}{r}.
\end{align}
In particular, one can check that, outside the overdensity, the
spacetime is asymptotically Einstein-de Sitter. Therefore, the
model correctly captures the idea of an overdensity embedded into
a cosmological spacetime.

Let us now consider another example: instead of a top hat profile as
before, one chooses a non flat profile defined by
\begin{align}
  \Delta (t_\uini,r)=\delta _\uini\left(1-\frac{1}{e}\right)^{-1}
  \left(e^{-r/r_\uc}-\frac{1}{e} \right)\Theta(r_\uc-r).
    \end{align}
In this case, $\delta_\uini$ represents the value of
$\Delta(t_\uini,r)$ at the center of the overdensity [this is the origin
  of the presence of the factor $(1-1/e)^{-1}$]. In that case, one has
\begin{align}
\left\langle
\Delta(t_\uini,r)\right \rangle
&= 3\frac{\delta_\uini}{1-1/e}\biggl[2\left(\frac{r_\uc}{r}\right)^3
    \left(1-e^{-r/r_\uc}\right)
    -2\left(\frac{r_\uc}{r}\right)^2
    e^{-r/r_\uc}-\frac{r_\uc}{r}e^{-r/r_\uc}-\frac{1}{3e}\biggr].
\end{align}
From this formula, one can determine the functions $m_0(r)$ and
$k_0(r)$. However, we do not give them here since their expression is
not especially illuminating. It is more interesting to study the form of the
criterion~(\ref{eq:criterion}) in that case. Since
$\left\langle \Delta(t_\uini,r)\right \rangle$ now depends on $r$, one
can imagine different scenarios such as, for instance, a case where
only a fraction of the overdensity collapses to form a black
hole. However, the simplest case is when the entire overdensity
proceeds to a black hole. In that situation, it seems reasonable to
interpret the criterion~(\ref{eq:criterion}) as being valid for the
radius of the overdensity, that is to say for $r=r_\uc$. It is easy to
show that $\left\langle \Delta(t_\uini,r_\uc)\right
\rangle=3[2-16/(3e)]/(1-1/e)\delta_\uini\simeq 0.18 \, \delta _\uini$. As
a consequence, the criterion becomes
$0.18\, \delta_\uini>\delta_\mathrm{c}$, where $\delta_\mathrm{c}$
has been defined in Eq.~(\ref{eq:criterion}). Up to a factor of order
one, this is very similar to the criterion obtained from a top-hat
profile, and one concludes that our formation criterion
is rather independent of the profile details.

\bibliographystyle{JHEP}
\bibliography{PBHreh}
\end{document}